\documentclass[aps,twocolumn,superscriptaddress,prd,nofootinbib,10pt]{revtex4-2} %
\usepackage{graphicx}%
\usepackage[caption=false,singlelinecheck=false]{subfig}%
\usepackage{amsmath}%
\usepackage{amssymb}%
\usepackage{amsfonts}%
\usepackage{amsthm}%
\usepackage{mathrsfs}%
\usepackage{txfonts}%
\usepackage{paralist}%
\usepackage{ulem}%
\usepackage{tikz}%
\usepackage[colorlinks,urlcolor=black,citecolor=black,link
color=black]{hyperref}

\def \Re{\text{Re}}
\def \Im{\text{Im}}
\newcommand{\GammaDB}{\ensuremath{\Gamma^D_{\!\leftrightarrow}}}
\newcommand{\GammaMB}{\ensuremath{\Gamma^M_{\!\leftrightarrow}}}

\newcommand{\modulus}[1]{\left| #1 \right|}

\newcommand{\braketSq}[1]{\Big\langle \left\lvert #1
\right\rvert^2 \Big\rangle}

\allowdisplaybreaks

\begin{document}

\title{Inferring the nature of active neutrinos: Dirac or Majorana?}

\author{C.~S.~Kim}%
\email[Email at: ]{cskim@yonsei.ac.kr}%
\affiliation{Department of Physics and IPAP, Yonsei University, Seoul
03722, Korea}%
\affiliation{Institute of High Energy Physics, Dongshin University,
Naju 58245, Korea}%

\author{M.~V.~N.~Murthy}%
\email[Email at: ]{murthy@imsc.res.in}%
\affiliation{The Institute of Mathematical Sciences, Taramani, Chennai
600113, India}%

\author{Dibyakrupa~Sahoo}%
\email[Email at: ]{Dibyakrupa.Sahoo@fuw.edu.pl}%
\affiliation{Faculty of Physics, University of Warsaw, Pasteura 5,
02-093 Warsaw, Poland}
\affiliation{Institute of Physics, Sachivalaya Marg, Bhubaneswar 751005, India}
\affiliation{Homi Bhabha National Institute, BARC Training School Complex,
Anushaktinagar, Mumbai 400094, India}


\begin{abstract}
The nature of a neutrino, whether it is a Dirac type or Majorana type,
may be comprehensively probed using their quantum statistical
properties. If the neutrino is a Majorana fermion, then by definition
it is identical and  indistinguishable from the corresponding
antineutrino. When a Majorana neutrino and antineutrino are pair
produced, the corresponding state has to obey the Pauli principle
unlike in the Dirac case. We use this property to distinguish between
the two cases using the process $B^0 \to
\mu^-\,\mu^+\,\nu_\mu\,\bar{\nu}_\mu$. We show that the two cases
differ dramatically in a special kinematic scenario where, in the rest
frame of the parent $B$ meson, the muons fly away back-to-back (i.e.\
fly with 3-momenta of equal magnitudes but opposite directions), and
so do the neutrino and antineutrino. Unlike any other scenario, we
know the energies and magnitudes of $3$-momenta of both the neutrino
and the antineutrino in this back-to-back configuration without even
directly measuring them. This provides a way of avoiding the
constraint imposed by the `practical Dirac-Majorana confusion
theorem', as one need not fully integrate over neutrino and
antineutrino in this case. As a true signature of the universal
principle of quantum statistics which does not depend on the size of
the mass of the particle but its spin, the difference between Dirac
and Majorana cases in this special kinematic configuration does
survive independent of the neutrino mass as long as neutrino mass is
nonzero. The analysis presented here is applicable immediately to
several other processes with the same final state as in the case of
$B^0$ decay without any major change.%
\end{abstract}
\maketitle


\section{Introduction}

Neutrinos are the most ubiquitous elementary particles after the
photon in the universe. Nevertheless they are also one of the least
understood in terms of their properties. We know  that the active
neutrinos in the standard model (SM) come in three flavors: electron
neutrino, muon neutrino and tau neutrino each associated with a
corresponding charged lepton. From neutrino oscillation experiments
\cite{Fukuda:1998mi,Ahmad:2002jz} it has been established that the
neutrinos $\nu_\ell$ with $\ell=e,\mu,\tau$ can oscillate from one
flavor to another. This is usually explained by considering the flavor
neutrinos as linear combinations of three different neutrino mass
eigenstates. The oscillation experiments suggest that at least two of
these mass eigenstates must have tiny but non-zero masses, whereas in
the SM neutrinos are regarded as massless. As neutrinos are charge
neutral and have non-zero mass, they could in principle be their own
antiparticles. In that case they are called Majorana fermions
\cite{Majorana:1937vz,Racah:1937qq, Furry:1938zz}. Since a Majorana
neutrino is quantum mechanically identical  to its antiparticle, any
state having a Majorana neutrino antineutrino pair must obey the
Fermi-Dirac statistics, a fact that is independent of the magnitude of
the neutrino mass. This means that the probability amplitude must be
totally antisymmetric under exchange. There is no such requirement if
neutrinos are of Dirac type. Thus the main difference between the
Dirac or Majorana nature of the neutrino arises from its quantum
statistical properties. We exploit this connection to construct a
novel way of probing the nature of neutrinos.

Such a connection between the statistics and the nature of neutrino
and antineutrino has been studied previously by using
antisymmetrization of amplitude for final states having Majorana
neutrino antineutrino pair. However, the effect of antisymmetrization
gets lost when the unobservable neutrino and antineutrino get fully
integrated out. This then leads to the ``practical Dirac-Majorana
confusion theorem'' (DMCT) \cite{Kayser:1981nw, Kayser:1982br}. The
theorem, which still lacks a rigorous, process independent, general
proof, as far as we are aware, states that the difference between
Dirac and Majorana neutrinos is proportional to some power of the
neutrino mass. This poses a challenge since the neutrino masses are
not known precisely, except that at least two of them must have
non-zero masses as indicated by neutrino oscillation experiments and
the masses are very small ($< 1$~eV) compared to other mass scales in
the SM \cite{Fukuda:1998mi,Ahmad:2002jz}. Thus, any proposal
conforming to DMCT depends on this tiny neutrino mass and as a result
carries this mass uncertainty apart from the probability being small.
It is therefore necessary and important to explore whether there are
any SM allowed processes that can directly probe the quantum
statistics of Majorana neutrinos avoiding this DMCT constraint.

Both experimentally as well as theoretically, an important proposal to
probe the Majorana nature of the neutrino is through the neutrino-less
double beta decay ($0\nu\beta\beta$) \cite{Furry:1939qr,
Schechter:1981bd, Nieves:1984sn, Takasugi:1984xr, Hirsch:2006yk,
Doi:1985dx, Engel:1988au, Haxton:1985am, Tomoda:1990rs,
Suhonen:1998ck, Faessler:1999zg, Vergados:2002pv, Bilenky:1987ty,
Elliott:2004hr, Aalseth:2004hb, Avignone:2007fu, Rodin:2009yc,
Barabash:2008dj, Giuliani:2010zz, Rodejohann:2011mu, Giuliani:2012zu,
Vergados:2012xy, Simkovic:2013kna, Engel:2016xgb, Ejiri:2019ezh}.
Since the proposal looks at a lepton number violating (LNV) process,
it is beyond SM. While there are many on-going experiments
\cite{Albanese:2021dhn, Agostini:2018tnm, Aalseth:2017btx,
Albert:2017owj, Abgrall:2017syy, Arnold:2016ezh, Xue:2017qbf,
Ni:2019kms, Alduino:2017ehq, KlapdorKleingrothaus:2000sn,
Bakalyarov:2003jk, KlapdorKleingrothaus:2006ff, Agostini:2013mzu,
Pocar:2015ota, Gando:2016cqd, KamLAND-Zen:2016pfg}, there is no
conclusive evidence experimentally as yet or from any other LNV
decays. Another LNV process, the neutrinoless double-electron capture
\cite{Winter:1955zz, Bernabeu:1983yb, Sujkowski:2003mb,
Karpeshin:2008zz, Krivoruchenko:2010ng, Eliseev:2011zza,
Eliseev:2011zzc, Eliseev:2012ih, Faessler:2012ku, Suhonen:2012zzb,
Bernabeu:2017ape, Alduino:2017mnz, Blaum:2020ogl, Abe:2018gyq} has
also been studied experimentally and is yet to be observed. Both these
processes involve a single Majorana neutrino as a propagator. One can
also consider processes mediated by exchange of a pair of virtual
neutrinos, as done in Ref.~\cite{Grifols:1996fk, Segarra:2020rah,
Costantino:2020bei}, as a way to distinguish Dirac and Majorana
neutrinos by observing the resulting potential. In this method DMCT
also holds except for distances that are of the same order or larger
than the inverse of the unknown neutrino mass. Also the process of
coherent scattering of neutrino on nucleus with bremsstrahlung
radiation has been explored and shown to be consistent with DMCT
\cite{Millar:2018hkv}. Therefore it is worthwhile exploring other
possibilities, especially those that do not involve LNV, or include
Majorana neutrino(s) as propagator(s).

In Refs.~\cite{Yoshimura:2006nd, Dinh:2012qb, Fukumi:2012rn,
Yoshimura:2013wva, Song:2015xaa, Tanaka:2017juo, Berryman:2018qxn,
Tashiro:2019ghs} SM allowed process of radiative emission of neutrino
pair was considered as a probe of Dirac or Majorana neutrino. Since
the final state involves $\nu\,\bar\nu$, statistics was accounted in
the Majorana case by explicit antisymmetrisation. Earlier, Nieves and
Pal~\cite{Nieves:1985ir} analysed the decay $K^+\rightarrow
\pi^+\,\nu\,\bar\nu$ as  a  test of Majorana neutrinos. They pointed
out that while the Dirac case involves both vector and axial vector
contribution, the Majorana case involves pure axial vector current.
This is due to the explicit antisymmetrisation of the final state of
two identical particles. In all these cases the difference between
Dirac and Majorana type appeared in the different event rates.
Crucially,  in all the above analyses, the neutrino and antineutrino
variables were integrated out since they are not observable. The
results for the rates was found to be directly proportional to some
power of the neutrino mass as required by DMCT. As far as we know it
is only in the analysis by Chhabra and Babu \cite{Chhabra:1992be} that
an effect independent of $m_\nu$ has been found between Dirac and
Majorana type neutrinos. Chabra and Babu considered the process
$e^+e^-\rightarrow \nu\,\bar{\nu}\,\gamma$. Their result is in
conformity with DMCT when they integrate over all the neutrino
variables. However, they also point out that the difference between
Dirac and Majorana nature can be ascertained independent of the mass
of the neutrinos provided their momenta are \textit{not} integrated
out.

It is important to note that when one considers massless neutrinos,
i.e.\ $m_\nu = 0$, both the Dirac and Majorana neutrinos can be
described as Weyl fermions. The reduction of neutrino degrees of
freedom from 4 to 2 for $m_\nu = 0$ is a discrete jump, and not a
continuous change. So the massless neutrino is an entirely different
species than the massive one even with extremely tiny mass. Therefore,
the presumed smooth transitional difference between Majorana and Dirac
neutrinos at $m_\nu \to 0$ is only a misperception.

In this paper, we show that the difference between Dirac and Majorana
neutrino persists independent of the magnitude of neutrino mass
provided the neutrino/antineutrino momenta are either measured
directly or indirectly fixed. As shown by Chabra and Babu, this is not
in violation of DMCT. We elaborate on this theme in this paper. In
particular we consider the decay $B^0 \,(\text{or }\bar{B}^0)
\rightarrow \mu^+\,\mu^-\,\nu\,\bar\nu$, for example, and discuss the
rates and branching ratios in a chosen kinematic scenario in which we
may indirectly discern the $\nu\,\bar\nu$ variables without the need
of any explicit observation of the neutrinos which is extremely
difficult any way at present. The method may be adopted to many other
such processes simply by replacing the appropriate parameters like
mass etc. We discuss in detail the dramatic differences between Dirac
and Majorana scenarios in differential distributions in such SM
allowed processes. {\it Most importantly this difference mainly
involves well known and measured quantities and is independent of the
unknown neutrino mass as long as it is non-zero.} We do not consider
massless neutrinos in this paper. Moreover, we would like to emphasize
that our work is not dependent on specific details of any neutrino
mass generation mechanism.

The paper is organised as follows. In
section~\ref{sec:previous-studies}, we provide a brief overview of the
previous studies using SM allowed processes to put things in
perspective. In section~\ref{sec:previous-studies} we also lay down
the basic issues that we  address in this paper. This is followed by
section~\ref{sec:expectation} in which we provide a broad outline of
our approach showing the main differences between Dirac and Majorana
neutrinos. In section~\ref{sec:B-decay}, we look at the decay of
$B^0\rightarrow \mu^+\mu^-\nu\bar\nu$ in detail. In this section we
make a case study with its experimental feasibility and future
prospects. This is followed by a discussion of other possible decay
modes in section~\ref{sec:other-decay}. Finally we conclude in
section~\ref{sec:conclusion} emphasizing the salient features of our
approach.

\section{A brief overview of previous studies}\label{sec:previous-studies}

First a note about the convention here: In general a neutrino flavour
is denoted by $\nu_\ell=\nu$ with $\ell=e,\mu,\tau$, where we drop the
subscript which is already implicit in the process. Same for
antineutrinos. When we explicitly denote the Majorana neutrinos in
the Feynman diagrams or elsewhere, we use the convention $\nu \equiv
\nu_\ell = \bar{\nu} \equiv \bar{\nu}_\ell \equiv \nu^M$.

\subsection{Processes with 2-body final states}%

As noted earlier the practical Dirac-Majorana confusion theorem (DMCT)
states that any difference between Dirac and Majorana neutrinos must
vanish in the limit of neutrino mass  going to zero. The DMCT was
first discussed by Kayser in Ref.~\cite{Kayser:1982br}. The loop
induced process $\gamma^* \to \nu\,\bar{\nu}$ was discussed. Angular
momentum analysis shows that $\nu\,\bar{\nu}$ final state can exist in
any one of the four possible $J=1$ states: ${}^3S_1$, ${}^3P_1$,
${}^3D_1$ and ${}^1P_1$. In the case of Dirac neutrinos all the four
states are possible where as for Majorana neutrinos only the ${}^3P_1$
state is allowed, since this is the only antisymmetric state. This
also fixes the parity of the Majorana neutrino relative to the photon
while leaving it undetermined in the Dirac case. Using this
information it was proposed that the angular distribution of neutrinos
in the decay $\psi(J^P=2^+) \to \nu\,\bar{\nu}$ could be different for
Dirac and Majorana neutrinos. While this has not been realised
experimentally, this remains the first application of the quantum
statistics apart from proposing DMCT.

A more direct application, instead of a loop induced process, is the
tree-level decay $Z^0\to\nu\,\bar{\nu}$. This was discussed in
Ref.~\cite{Shrock:1982jh}. In the Dirac case both vector and
axial-vector currents contribute where as in the Majorana case it is a
pure axial vector, due to antisymmetrisation taking into account the
statistics. The decay width, more appropriately called the missing
width, is given by
\begin{equation}
\Gamma\left(Z^0\to\nu\,\bar{\nu}\right) =
\frac{G_F\,m_Z^3}{12\,\pi\sqrt{2}} \times
\begin{cases}
\left(1-r\right)\left(1-4\,r\right)^{1/2}, & \text{Dirac}\\[2mm]
\left(1-4\,r\right)^{3/2}, & \text{Majorana}
\end{cases}
\end{equation}
where $r=\left(m_\nu/m_Z\right)^2$ with $m_\nu$, $m_Z$ being the
masses of neutrino and $Z$ boson respectively, and $G_F$ denotes the
Fermi coupling constant. Thus the difference between Dirac and
Majorana cases is directly proportional to $r$ or $m_\nu^2$ as
expected from DMCT. Alternatively, one could also study the process
$e^+\,e^-\to\nu\,\bar{\nu}$ \cite{Ma:1989jpa}. While spin dependent
and spin-independent cross sections for Dirac and Majorana cases show
substantial difference near threshold, the results are consistent with
DMCT once the spins are summed over. This example comes close to the
conclusions of this paper as we shall see later.

\subsection{Processes with 3-body final states}

The main difficulty with just $\nu\bar\nu$ in the final state is that
it can not be observed; in the case of $Z$ decay this corresponds to
the invisible width of the $Z$ boson as the final state can not be
directly observed. One way to improve upon this situation is to look
at 3- and  4- body finals states which contain the $\nu\bar\nu$ pair.
Nieves and Pal~\cite{Nieves:1985ir} analysed the decay
$K^+\to\pi^+\,\nu\,\bar{\nu}$. Because the final state pion is a
pseudo-scalar, the process still involves only the axial vector
current in the Majorana case as in the two body decays. However, the
presence of the pion allows for a differential distribution, even
after integrating over the $\nu,\bar\nu$ variables. Once again, while
the rates are different for Dirac and Majorana scenarios, the
difference in pion energy distributions is proportional to the
neutrino mass in accordance with DMCT. On the other hand, Chabra and
Babu in Ref.~\cite{Chhabra:1992be} analysed in detail the scattering
process $e^+\,e^- \to \nu\,\bar{\nu}\,\gamma$. Because of the presence
of $\gamma$ in the final state this process has a richer spin
structure. Most importantly, it is shown that when there is no
integration over $\nu,\bar\nu$ variables, the difference between Dirac
and Majorana cases does not vanish even if the neutrino mass is set to
zero. However, upon integration, the result is proportional to the
neutrino mass in accordance with DMCT. This is a clear demonstration
of both conformity and an exception to DMCT but suffers from the fact
that it is still not possible to observe any neutrino related
variables experimentally.

More recently, radiative emission of neutrino pair has been attracting
some attention~\cite{Yoshimura:2006nd, Dinh:2012qb, Fukumi:2012rn,
Yoshimura:2013wva, Song:2015xaa, Tanaka:2017juo, Tashiro:2019ghs}. In
this proposal, one looks at atomic transition from an excited state to
a ground state as in $\left|es\right\rangle \to
\left|gs\right\rangle+\gamma+\nu\,\bar{\nu}$. The photon energy
spectrum is sensitive to the absolute masses of the neutrino mass
eigenstates. The Dirac or Majorana cases may be probed by looking at
the decay rate near the threshold for neutrino pair production. Since
the momenta of neutrino (antineutrino) are integrated out, the
difference between the two cases is always proportional to the
neutrino mass, again in agreement with DMCT. Complimentary to the
studies on radiative emission of neutrino pair in atomic experiments,
the authors of Ref.~\cite{Berryman:2018qxn} studied the stimulated
emission of neutrino pair via the process $e^-\gamma \to
e^-\nu\,\bar{\nu}$. Here also they consider the difference between
Dirac and Majorana neutrinos close to the kinematic threshold of pair
production and compare the decay rates for the two cases which is
extremely small.

\subsection{Summary of results from previous studies}

The common features in all of the above studies are the following:
\begin{enumerate}
\item All the processes considered, have a neutrino and an
antineutrino in the final state, are SM allowed and do not violate
lepton number.%
\item The amplitude is antisymmetrised in the case of Majorana
neutrinos as required by statistics.%
\item The `observable' difference between Dirac and Majorana neutrinos
is a direct consequence of the antisymmetrization in the Majorana
case. It is proportional to the neutrino mass \text{when} the neutrino
and antineutrino momenta are integrated out.%
\item However, exceptions to DMCT constraint occur under some special
conditions, e.g.\ near kinematic threshold of pair production, when
the spin sum is not done, or when the neutrino and antineutrino
momenta are not integrated out.%
\end{enumerate}

We continue along the theme considered in many of the references cited
above and show that the the difference between Dirac and Majorana
cases may be seen more clearly under certain kinematical conditions
especially with 4-body fermion final states in an SM allowed process
without lepton number violation.

In particular we choose processes in which we have a final state given
by $\mu^+\,\mu^-\,\nu_\mu\,\bar\nu_\mu$. Of course, we could have
chosen either $e^+e^-$ or $\tau^+\tau^-$ instead of the muon pair. The
analysis remains the same though experimentally muon pair production
is preferred.

The initial state could be either a symmetric collision of $e^- e^+$
or decay of some resonance such as neutral $B$ or $D$ mesons or even
the SM Higgs, the main criteria being which initial state offers the
best ability to measure the total missing 4-momentum of the escaping
neutrino and antineutrino pair. In this work we specifically focus on
the decay $B^0(\bar{B}^0) \rightarrow \mu^-\,\mu^+\,\nu_\mu\,\bar{\nu}_\mu$. Even though
$\nu_\mu$ is strictly not a mass eigenstate, for simplicity we denote
its effective mass by $m_\nu$.

\section{General formalism}\label{sec:expectation}%

Consider the SM allowed decay,
$$B^0(p_B)\to\mu^-(p_-)\,\mu^+(p_+)\,\bar\nu_\mu(p_1)\,\nu_\mu(p_2),$$
where the corresponding 4-momenta are shown in parentheses. There are various other allowed initial states one could also consider, such as $\bar{B}^0$, $D^0$, $\bar{D}^0$, or neutral kaons, or even Higgs. The following analysis holds for all such decays
with appropriate changes in the form factors or vertex factors
as well as the allowed phase space due to the mass of the
parent particle. The amplitude for Dirac case is denoted as
\begin{equation}\label{eq:amp-D}
\mathscr{M}^D = \mathscr{M}(p_1,p_2),
\end{equation}
where for brevity we have not shown any other dependencies in the
amplitude. For Majorana case the amplitude is antisymmetrized with
respect to the exchange of $p_1,p_2$ and is given by,
\begin{equation}\label{eq:amp-M}
\mathscr{M}^M = \frac{1}{\sqrt{2}} \Big( \mathscr{M}(p_1,p_2) -
\mathscr{M}(p_2,p_1) \Big).
\end{equation}
The difference between amplitude squares for the two cases after
summing over final spins is given by
\begin{align}
\left| \mathscr{M}^D \right|^2 - \left| \mathscr{M}^M \right|^2 &=
\frac{1}{2} \bigg( \underbrace{\left| \mathscr{M}(p_1,p_2)
\right|^2}_{\text{Direct term}} - \underbrace{\left|
\mathscr{M}(p_2,p_1) \right|^2}_{\text{Exchange term}} \bigg) \nonumber\\%
&\quad + \underbrace{\text{Re}\Big(\mathscr{M}(p_1,p_2)^*\,
\mathscr{M}(p_2,p_1)\Big)}_{\text{Interference term}}
.\label{eq:DmM-DTETIT}
\end{align}
Consistent with the prior studies in the literature as mentioned in
Sec.~\ref{sec:previous-studies}, we observe the following.
\begin{enumerate}
\item The antisymmetrization in Majorana amplitude gives rise to the
three terms: direct, exchange and interference terms, which are
identified in Eq.~\eqref{eq:DmM-DTETIT}. The Dirac case involves only
the direct term.

\item The interference term is \textit{always} (except for $p_1 =
p_2$) directly proportional to $m_\nu^2$ as it involves helicity
flips,
\begin{equation}
\text{Re}\Big(\mathscr{M}(p_1,p_2)^*\, \mathscr{M}(p_2,p_1)\Big)
\propto m_\nu^2.
\end{equation}

\item Neither the direct nor the exchange terms is proportional to
$m_\nu$. In general,
\begin{equation}\label{eq:DTneqET}
\underbrace{\left| \mathscr{M}(p_1,p_2) \right|^2}_{\text{Direct
term}} \neq \underbrace{\left| \mathscr{M}(p_2,p_1)
\right|^2}_{\text{Exchange term}}.
\end{equation}
The difference between direct and exchange terms is, in general, not
proportional to $m_\nu$. However, this difference  vanishes after
integration over the neutrino momenta, i.e.\
\begin{equation}
\iint \underbrace{\left| \mathscr{M}(p_1,p_2) \right|^2}_{\text{Direct
term}}\, \mathrm{d}^4 p_1 \, \mathrm{d}^4 p_2 = \iint
\underbrace{\left| \mathscr{M}(p_2,p_1) \right|^2}_{\text{Exchange
term}}\, \mathrm{d}^4 p_1 \, \mathrm{d}^4 p_2,
\end{equation}
since the amplitude squared is symmetric under exchange of $p_1,p_2$
even though the amplitude is antisymmetric. Therefore,
\begin{align}
&\iint \left( \left| \mathscr{M}^D \right|^2 - \left| \mathscr{M}^M
\right|^2 \right)\, \mathrm{d}^4 p_1 \, \mathrm{d}^4 p_2 \nonumber\\%
&= 2\iint \underbrace{\text{Re}\Big(\mathscr{M}(p_1,p_2)^*\,
\mathscr{M}(p_2,p_1)\Big)}_{\text{Interference term}}\, \mathrm{d}^4
p_1 \, \mathrm{d}^4 p_2 
\propto m_\nu^2. \label{eq:general-DMCT}
\end{align}
This is consistent with DMCT once the integration over neutrino and
antineutrino momenta are done.
\end{enumerate}

\subsection{A thought experiment highlighting an exception to DMCT}\label{subsec:tht-expt}

In order to show that there exist exceptions to DMCT we consider a
simple thought experiment for illustration only. Let us assume, for
arguments sake, that the 4-momenta of both neutrino and anti-neutrino
are individually measured. Consider the special case when neutrino and
antineutrino are collinear, i.e.\ their 4-momenta are equal, $p_1 =
p_2$. Due to antisymmetrization the amplitude for Majorana case in
Eq.~\eqref{eq:amp-M} vanishes for such collinear events
($\mathscr{M}^M_\text{collinear} = 0$). However, the amplitude for the
Dirac case is non-zero, $\mathscr{M}^D_\text{collinear} \neq 0$. This
is a dramatic illustration of the difference between the Dirac and
Majorana cases. Furthermore, as we show later in the specific example
of the $B$ decay in Sec.~\ref{sec:B-decay}, the
$\mathscr{M}^D_\text{collinear}$ is in fact not proportional to
$m_\nu^2$. Hence, the difference between the Dirac and the Majorana
cases does not vanish when we neglect terms proportional to $m_\nu$.
This starkly contradicts the DMCT. The kinematics chosen here is only
for the purpose of illustration. The collinear $\nu\,\bar{\nu}$
scenario has never been probed experimentally. On the contrary, there
exists another kinematic scenario, the back-to-back $\nu\,\bar{\nu}$
configuration, using which the exception to the DMCT may be easily
explored. As we will show, this scenario is experimentally accessible.
Unless otherwise mentioned, we focus on this new specific back-to-back
kinematic scenario in our discussions ahead.

\subsection{Back-to-back neutrino antineutrino configuration: an experimentally observable exception to DMCT}

Before we discuss the detailed structure of the amplitudes, we can
make certain statements based on angular momentum analysis and quantum
statistics. In a frame where the neutrino and antineutrino are
back-to-back, i.e.\ flying with $3$-momenta of equal magnitude but
opposite direction, this reduces to the helicity analysis\footnote{In
this work we have $V-A$ interaction which fixes the helicity of all
the particles involved. One could, in principle, consider mass
dependent contributions, which are negligible for neutrino and
antineutrino due to their tiny mass.}. This is the kinematic situation
that we are interested. The transition from a left-handed neutrino to
the right-handed antineutrino is achieved by the combined
transformation of charge conjugation (C) and parity (P). Thus,
\begin{equation}\label{eq:CP}
\textrm{C~P} \left\lvert \nu_\ell (\vec{s},E_\nu,\vec{p}_\nu)
\right\rangle = \eta_P \left\lvert \bar{\nu}_\ell
(\vec{s},E_\nu,-\vec{p}_\nu) \right\rangle,
\end{equation}
where $\vec{s},E_\nu,\vec{p}_\nu$ denote the spin, energy and
$3$-momentum of the neutrino respectively, and $\eta_P$ is the parity
phase factor which is arbitrary for Dirac neutrinos but takes the
values $\pm i$ for Majorana neutrinos \cite{Kayser:1982br}.
Disregarding this phase factor for the time being, we can
schematically express Eq.~\eqref{eq:CP} as follows,
\begin{equation}\label{eq:B2B-CP}
\raisebox{-1cm}{\includegraphics[scale=1]{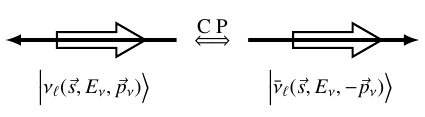}}
\end{equation}
where the long thin arrows represent the $3$-momenta of the neutrino
and antineutrino, and the short thick arrows represent their spins. It
is clear from Eqs.~\eqref{eq:CP} and \eqref{eq:B2B-CP} that if the
Majorana neutrino and antineutrino are back-to-back we can consider
the consequence of their exchange as a proper signature of the quantum
statistics.

\subsection{Helicity considerations}

This back-to-back configuration has one important consequence. If in the rest
frame of the parent $B^0$ meson the neutrino antineutrino pair is
found to fly away back-to-back, the muon pair must also fly away
back-to-back since $3$-momentum is conserved. This is a much simpler
kinematic configuration than the general kinematics for any $4$-body
decay. Instead of the usual five independent variables one needs to
describe any $4$-body decay, we only need two independent variables to
describe the back-to-back configuration. In this case, the energies of the two
muons are the same and let us denote them by $E_\mu$. Similarly, the
energies of the back-to-back neutrino and antineutrino are the same
and let us denote them by $E_\nu$. Either $E_\mu$ or $E_\nu$ is
independent, because from conservation of energy we get,
\begin{equation}\label{eq:B2B-energies-relation}
E_\nu = m_B/2 - E_\mu,
\end{equation}
where $m_B$ is the mass of the $B^0$ meson. Let us choose $E_\mu$ as
one independent variable. The other independent variable would then be
the angle, say $\theta$, between the muon direction and the neutrino
direction.

\begin{figure}[hbtp]
\centering%
\subfloat[Helicity configuration involving Dirac neutrinos,
$\nu_\mu\equiv \nu^D$, $\bar{\nu}_\mu \equiv
\bar{\nu}^D$.\label{fig:B2B-helicity-configuration-D}]{\includegraphics[scale=.9]{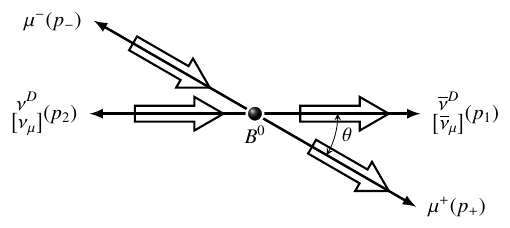}}\hfill%
\subfloat[Helicity configuration involving Majorana neutrinos,
$\nu_\mu=\bar{\nu}_\mu \equiv
\nu^M$.\label{fig:B2B-helicity-configuration-M}]{\includegraphics[scale=.9]{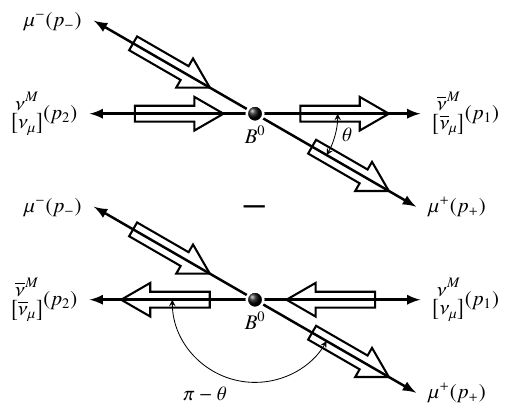}}%
\caption{The helicity configuration for back-to-back muons in the rest
frame of $B^0$ in the decay
$B^0\to\mu^-\,\mu^+\,\nu_\mu\,\bar{\nu}_\mu$. Here the second diagram
in Majorana case is a result of antisymmetrisation and is not related
to any helicity-flip. We show the $\big[ \nu_\mu \big]$ and $\big[
\bar{\nu}_\mu \big]$ labels just for bookkeeping.}%
\label{fig:B2B-helicity-configuration}%
\end{figure}

\begin{figure}[hbtp]
\centering%
\includegraphics[scale=.9]{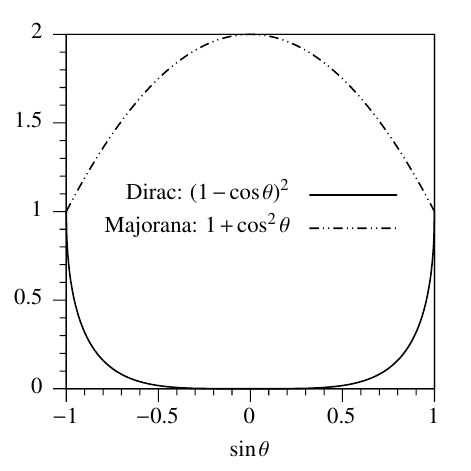}%
\caption{Comparison of the angular distributions as given in
Eqs.~\eqref{eq:B2B-helicity-dist-D} and
\eqref{eq:B2B-helicity-dist-M}. The reason for taking $\sin\theta$ as
the independent variable instead of $\cos\theta$ would be clear from
the detailed discussion later.}%
\label{fig:B2B-angular-distribution}%
\end{figure}

\begin{figure*}[hbtp]
\centering%
\subfloat[For Dirac neutrinos: $\nu_\mu \equiv \nu^D$, $\bar{\nu}_\mu
\equiv \bar{\nu}^D$.
\label{fig:Feynman-diagram-D}]{\includegraphics[width=.9\linewidth,keepaspectratio]{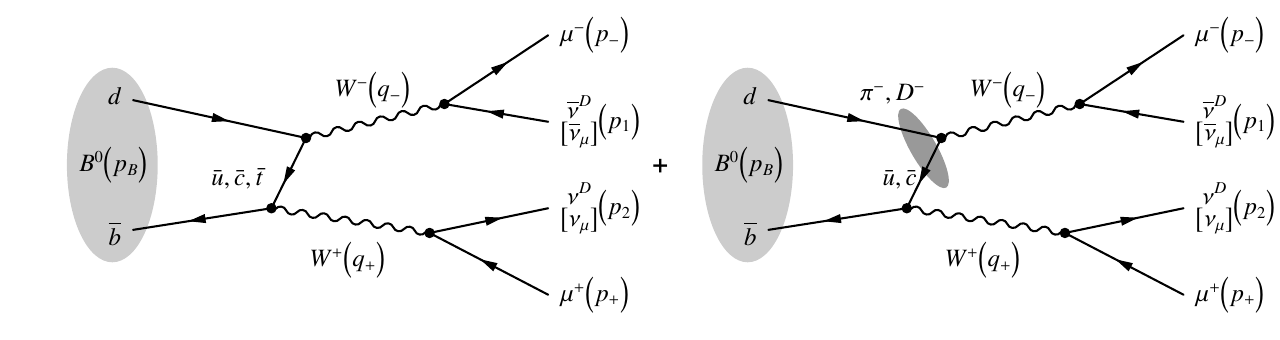}}\\%
\subfloat[For Majorana neutrinos: $\nu_\mu = \bar{\nu}_\mu \equiv
\nu^M$.
\label{fig:Feynman-diagram-M}]{\includegraphics[width=.9\linewidth,keepaspectratio]{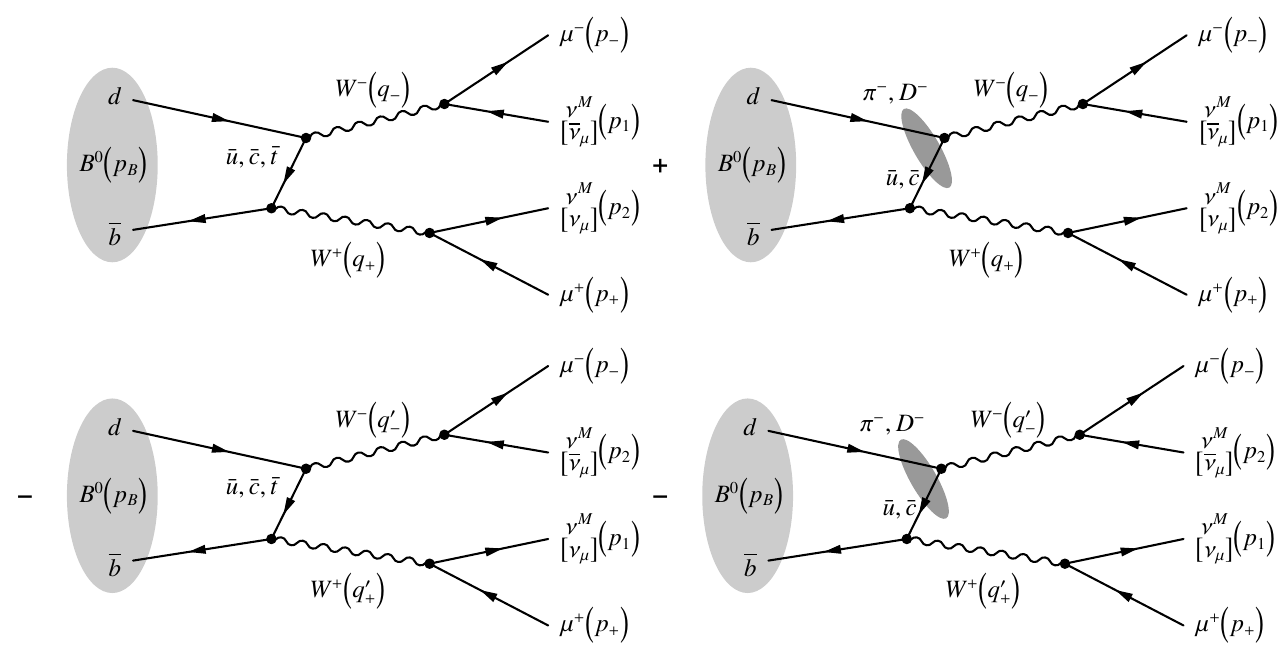}}%
\caption{The Feynman diagrams for $B^0 \to \mu^- \mu^+ \nu_\mu
\bar{\nu}_\mu$ for both Dirac and Majorana cases. Here the internal
4-momenta are denoted by $q_- = p_- + p_1$, $q_+ = p_+ + p_2$, $q'_- =
p_- + p_2$ and $q'_+ = p_+ + p_1$. Here one can consider $\pi^-
\left(d\overline{u}\right)$ and $D^- \left(d\overline{c}\right)$ as
the possible resonances. Due to the identical nature of Majorana
neutrino and antineutrino, we have two probable resonant diagrams
involving intermediate $\pi^-$ or $D^-$. We show the $\big[ \nu_\mu
\big]$ and $\big[ \bar{\nu}_\mu \big]$ labels just for bookkeeping.}%
\label{fig:Feynman-diagram}
\end{figure*}

Let us analyze the helicity configuration of this back-to-back muons
(and back-to-back neutrino antineutrino) case as shown in
Fig.~\ref{fig:B2B-helicity-configuration}, where the long arrows
represent particle momenta and the short thick arrows represent their
spins. Let us denote the decay amplitude describing the back-to-back
configuration by $\mathscr{M}_\leftrightarrow^{D/M}$ for
Dirac/Majorana neutrinos. In the case of Dirac neutrinos, it is clear
from Fig.~\ref{fig:B2B-helicity-configuration-D} that for $\theta=0$
we have a net final spin $\neq 0$. This violates conservation of
angular momentum, since the parent $B^0$ meson has spin-0. Therefore,
for the Dirac case we have,
\begin{equation}\label{eq:B2B-helicity-dist-D}
\left| \mathscr{M}^D_\leftrightarrow \right|^2 \propto
\underbrace{\Big(1-\cos\theta\Big)^2}_{\textrm{Direct term}}.
\end{equation}
However for Majorana neutrinos, it is clear from
Fig.~\ref{fig:B2B-helicity-configuration-M} that both the $\theta$ and
$\pi-\theta$ configurations are indistinguishable since $\nu_\mu$ and
$\bar{\nu}_\mu$ are quantum mechanically identical\footnote{The
antisymmetrization for Majorana case gives the exchange term (via $p_1
\leftrightarrow p_2$ exchange) and is not associated with any helicity
flip, as shown in Fig.~\ref{fig:B2B-helicity-configuration} and
Fig.~\ref{fig:Feynman-diagram}. However, helicity flip is present in
the interference term making it proportional to $m_\nu^2$.}. The
interference term which is proportional to $m_\nu^2$ can be neglected.
Thus,
\begin{align}
\left| \mathscr{M}^M_\leftrightarrow \right|^2 &\propto \frac{1}{2}
\Bigg[ \underbrace{\Big(1-\cos\theta\Big)^2}_{\text{Direct term}} +
\underbrace{\Big(1-\cos\left(\pi-\theta\right)\Big)^2}_{\text{Exchange
term}} -
\underbrace{\mathcal{O}\left(m_\nu^2\right)}_{\text{Interference
term}}\Bigg] \nonumber\\%
&\simeq 1 + \cos^2\theta.\label{eq:B2B-helicity-dist-M}
\end{align}
Thus, the Dirac and Majorana cases have completely different angular
distributions in the back-to-back configuration, see
Fig.~\ref{fig:B2B-angular-distribution}. We would like to emphasize
that this difference is simply a result of the antisymmetrization of
the amplitude for Majorana neutrinos, and we have already neglected
the interference term which is proportional to $m_\nu^2$. Therefore,
it can be considered as a proper test of the quantum statistics of the
Majorana neutrinos.

The distinct signature between Dirac and Majorana cases as shown in
Fig.~\ref{fig:B2B-angular-distribution} appears only in the restricted
kinematic situation of back-to-back muons in the $B^0$ rest frame. The
branching ratio in general is dominated by the non-back-to-back configurations
which dominate the phase space and as we shall see later the branching
ratio for back-to-back configuration is small but significant for
distinguishing Dirac and Majorana cases. Therefore back-to-back configuration
provides an exception to DMCT. Of course, once the full phase space
integration over $\nu,\bar\nu$ variables is carried out, the
difference between Dirac and Majorana cases is proportional to $m_\nu$
and we are back to DMCT domain.

In the next section a detailed analysis of the decay
$B^0\to\mu^-\,\mu^+\,\nu_\mu\,\bar{\nu}_\mu$ is presented covering all
the nuances in the differences between Dirac and Majorana cases.

\section{A detailed study of the decay \texorpdfstring{$B^0\to\mu^-\,\mu^+\,\nu_\mu\,\bar{\nu}_\mu$}{B0 ---> mu- mu+ nu nubar}}\label{sec:B-decay}%

In Fig.~\ref{fig:Feynman-diagram} the Feynman diagrams that contribute
to the decay $B^0\to\mu^-\,\mu^+\,\nu_\mu\,\bar{\nu}_\mu$ are shown.
This is a doubly weak decay. The branching ratio of this mode will
also have contributions from intermediate resonances such as $\pi^-$
and $D^-$ which tend to enhance the total branching ratio.

\subsection{Structures in the decay amplitude}

In order to present both the resonant and non-resonant contributions
to the decay amplitudes in a uniform form, we note that the hadronic
part will involve the following factors:%
\begin{enumerate}
\item product of Cabibbo-Kobayashi-Maskawa (CKM) matrix elements,
$V_{\!ub}^* V_{\!ud}^{}$ or $V_{\!cb}^* V_{\!cd}^{}$ or $V_{\!tb}^*
V_{\!td}^{}$ depending on whether $u$ or $c$ or $t$ quark is being
considered as the propagating quark,%
\item product of coupling constants and the virtual $W$ propagators,
which gives an overall factor of $\dfrac{g_w^4}{64\, m_W^4} =
\dfrac{G_F^2}{2}$, as the Fermi constant ($G_F$) is related to the
weak coupling constant ($g_w$) and the mass of $W$ boson ($m_W$) by
the relation $G_F = \dfrac{\sqrt{2}}{8}
\left(\dfrac{g_w}{m_W}\right)^2$,%
\item the effective vertex factors for the contribution from $B^0(p_B)
\to W^{+*} \left(q_+^{(\prime)}\right) \; W^{-*}
\left(q_-^{(\prime)}\right)$, which are different for resonant and
non-resonant channels (see Fig.~\ref{fig:Feynman-diagram} for the
definitions of $q_\pm^{(\prime)}$ and more details about the vertex
factors are given in Sec.~\ref{subsec:resonant} and
\ref{subsec:nonresonant} below).%
\end{enumerate}
There are two combinations of product of leptonic currents in our case,
\begin{subequations}
\begin{align}
L_{\alpha\beta} &= \left[\overline{\it u}(p_-) \, \gamma_\alpha
\left(1-\gamma^5\right) \, \varv(p_1)\right] \left[\overline{u}(p_2) \,
\gamma_\beta \left(1-\gamma^5\right) \, \varv(p_+)\right],
\label{eq:lepton-current-1}\\%
L_{\alpha\beta}^{\prime} &= \left[\overline{u}(p_-) \, \gamma_\alpha
\left(1-\gamma^5\right) \, \varv(p_2)\right] \left[\overline{u}(p_1) \,
\gamma_\beta \left(1-\gamma^5\right) \, \varv(p_+)\right].
\label{eq:lepton-current-2}
\end{align}
\end{subequations}
It is easy to see that $L_{\alpha\beta}$ and
$L_{\alpha\beta}^{\prime}$ are related to one another by $p_1
\leftrightarrow p_2$ exchange.

The decay amplitudes for Dirac and Majorana neutrinos can be written
as,
\begin{subequations}\label{eq:amplitude-full}
\begin{align}
\mathscr{M}^D &= \frac{G_F^2}{2} H^{\alpha\beta} L_{\alpha\beta}
\equiv \mathscr{Q}_{12} + \mathscr{R}_{12},\label{eq:Dirac-amplitude-full}\\%
\mathscr{M}^M &= \frac{G_F^2}{2\sqrt{2}} \left( H^{\alpha\beta}
L_{\alpha\beta} - H^{\prime\alpha\beta} L_{\alpha\beta}^\prime \right)
\nonumber\\%
&\equiv \frac{1}{\sqrt{2}} \left(
\mathscr{Q}_{12} - \mathscr{Q}_{21} + \mathscr{R}_{12} - \mathscr{R}_{21}
\right),\label{eq:Majorana-amplitude-full}
\end{align}
\end{subequations}
where $H^{(\prime)\alpha\beta}$ denote the hadronic currents which
contain the combination of products of CKM matrix elements and
effective vertex factors, and we discuss about the structure of the
hadronic currents below in detail leading to its final expression in
Eq.~\eqref{eq:hadronic-current}, $\mathscr{Q}_{12}$ and
$\mathscr{R}_{12}$ are respectively the non-resonant and resonant
parts of the decay amplitude which are the sole contributors in case
of Dirac neutrinos, and the non-resonant amplitude $\mathscr{Q}_{21}$,
the resonant amplitude $\mathscr{R}_{21}$ which appear in Majorana
case are obtained from $\mathscr{Q}_{12}$, $\mathscr{R}_{12}$
respectively by $p_1 \leftrightarrow p_2$ exchange. Below we look at
the content of the hadronic currents, the resonant and non-resonant
amplitudes in more detail.

\subsection{Resonant amplitude and the hadronic current}\label{subsec:resonant}

For the resonant case, we can have both $\pi^-$ and $D^-$ as
intermediate resonances depending on whether $q_-^{(\prime)2} =
m_\pi^2$ or $m_D^2$ in the decay $B^0 \to \mu^- \mu^+ \nu_\mu
\bar{\nu}_\mu$. If we were to consider the conjugate process
$\overline{B}^0 \to \mu^- \mu^+ \nu_\mu \bar{\nu}_\mu$, then the
resonances would be $\pi^+$ and $D^+$, both associated with the
4-momentum $q_+^{(\prime)}$ instead of $q_-^{(\prime)}$. Thus, knowing
the flavor of the initial neutral $B$ meson, i.e.\ whether it is $B^0$
or $\overline{B}^0$, the allowed resonances get fixed distinguishing
the 4-momenta $q_+^{(\prime)}$ and $q_-^{(\prime)}$. This is in fact
easily discernible from the expression for the effective vertex
factors for $B^0 \to W^{+*}\left(q_+^{(\prime)}\right) \;
W^{-*}\left(q_-^{(\prime)}\right)$ from the resonant channel,
\begin{equation}\label{eq:resonant-vertex-factors}
\mathbf{V}^{(\prime) \alpha\beta}_R = \frac{f_R}{q_-^{(\prime) 2} - m_R^2
+ i m_R \Gamma_R} q_-^{(\prime)\alpha} \Big(F^{(\prime)}_{R+} \, q_+^{(\prime)\beta}
+ F^{(\prime)}_{R-} \, q_-^{(\prime)\beta} \Big),
\end{equation}
where $F_{R\pm}^{(\prime)} \equiv F_{R\pm}
\left(q_+^{(\prime)2}\right)$ are known form factors and $f_R$ is the
known decay constant of the resonance $R$ ($R=\pi,D$) that has mass
$m_R$ and total decay rate $\Gamma_R$. Therefore, for the process $B^0
\to \mu^- \mu^+ \nu_\mu \bar{\nu}_\mu$, the resonant part of the decay
amplitude for Dirac case is given by
\begin{equation}\label{eq:R12}
\mathscr{R}_{12} =\frac{G_F^2}{2}  \mathbf{H}^{\alpha\beta}
L_{\alpha\beta},
\end{equation}
where the resonant hadronic current is given by,
\begin{equation}
\mathbf{H}^{\alpha\beta} \equiv V^*_{ub} V_{ud}
\mathbf{V}^{\alpha\beta}_\pi + V^*_{cb} V_{cd}
\mathbf{V}^{\alpha\beta}_D =  \left(\mathbf{F}_+ q_+^\beta +
\mathbf{F}_- q_-^\beta \right) q_-^\alpha,
\label{eq:resonant-hadronic-current-1}
\end{equation}
with the combined form factors $\mathbf{F}_\pm$ (``\textit{resonant
transition form factors}'') being given by
\begin{align}
\mathbf{F}_\pm \equiv \mathbf{F}_\pm\left(q_+^2,q_-^2\right) &=
\frac{V^*_{ub} V_{ud} \; f_\pi}{q_-^2 - m_\pi^2 + i m_\pi \Gamma_\pi}
\, F_{\pi \pm}\left(q_+^2\right) \nonumber\\%
&\quad + \frac{V^*_{cb} V_{cd} \; f_D}{q_-^2 - m_D^2 + i m_D \Gamma_D}
\, F_{D \pm}\left(q_+^2\right).
\label{eq:resonant-combined-form-factors-D}
\end{align}
It is important to reiterate that the vertex factor for resonant case
as defined in Eq.~\eqref{eq:resonant-vertex-factors} and the related
form factors of Eq.~\eqref{eq:resonant-combined-form-factors-D} are
specific to the decay mode $B^0 \to \mu^- \mu^+ \nu_\mu
\bar{\nu}_\mu$.

For Majorana case, in addition to $\mathscr{R}_{12}$ we have
$\mathscr{R}_{21}$ which is given by
\begin{equation}\label{eq:R21}
\mathscr{R}_{21} \equiv \frac{G_F^2}{2} \mathbf{H}^{\prime
\alpha\beta} L'_{\alpha\beta},
\end{equation}
with the resonant hadronic current $\mathbf{H}^{\prime \alpha\beta}$
being given by,
\begin{equation}
\mathbf{H}^{\prime \alpha\beta} \equiv V^*_{ub} V_{ud}
\mathbf{V}^{\prime \alpha\beta}_\pi + V^*_{cb} V_{cd}
\mathbf{V}^{\prime \alpha\beta}_D =  \left(\mathbf{F}_+^\prime
q_+^{\prime\beta} + \mathbf{F}_-^\prime q_-^{\prime \beta} \right)
q_-^{\prime\alpha}, \label{eq:resonant-hadronic-current-2}
\end{equation}
which includes the combined form factors $\mathbf{F}_\pm^\prime$ that
can be easily obtained by substituting $q_\pm^{2}$ by $q_\pm^{\prime
2}$ in Eq.~\eqref{eq:resonant-combined-form-factors-D}.

\subsection{Non-resonant amplitude and the hadronic current}\label{subsec:nonresonant}

Unlike the resonant case, in the non-resonant case neither
$q_+^{(\prime)}$ nor $q_-^{(\prime)}$ has any preferred role over the
other. Hence, following Lorentz covariance, the effective vertex
factors for $B^0(p_B) \to W^{+*} \left(q_+^{(\prime)}\right)\; W^{-*}
\left(q_-^{(\prime)}\right)$ for non-resonant case (involving
intermediate quark $Q=u,c,t$) can be written as
\begin{equation}
\mathbb{V}^{(\prime)\alpha\beta}_Q =  F_a^{(\prime)Q} \,
g^{\alpha\beta} + F_b^{(\prime)Q} \, p_B^\alpha \, p_B^\beta + i \,
F_c^{(\prime)Q} \, \epsilon^{\alpha\beta\mu\nu} \, q_{+\mu}^{(\prime)}
\, q_{-\nu}^{(\prime)},
\end{equation}
where $F_a^{(\prime)Q} \equiv
F_a^{(\prime)Q}\left(q_+^{(\prime)2},q_-^{(\prime)2}\right)$,
$F_b^{(\prime)Q} \equiv
F_b^{(\prime)Q}\left(q_+^{(\prime)2},q_-^{(\prime)2}\right)$,
$F_c^{(\prime)Q} \equiv
F_c^{(\prime)Q}\left(q_+^{(\prime)2},q_-^{(\prime)2}\right)$ are the
relevant ``\textit{non-resonant transition form
factors}\footnote{These \textit{transition form factors} are functions
of two different $q^2$. An example where similar situation occurs is
while considering the pion transition form factors for
$\gamma^*\,\gamma^*\,\pi^0$ vertex, see Ref.~\cite{Lichard:2010ap}.}''
and $p_B = q_+^{(\prime)} + q_-^{(\prime)}$. Currently, the exact
expressions for the form factors $F_a^{(\prime)Q}, F_b^{(\prime)Q}$
and $F_c^{(\prime)Q}$ are unknown and we consider them to be complex,
in general. Thus, the non-resonant decay amplitude for Dirac case
neutrinos is given by,
\begin{equation}\label{eq:Q12}
\mathscr{Q}_{12} = \frac{G_F^2}{2} \left( \sum_{Q=u,c,t} V^*_{Qb}
V_{Qd} \mathbb{V}^{\alpha\beta}_Q \right) L_{\alpha\beta} =
\frac{G_F^2}{2} \; \mathbb{H}^{\alpha\beta} \; L_{\alpha\beta},
\end{equation}
where the non-resonant hadronic current is given by
\begin{equation}\label{eq:nonresonant-hadronic-current-1}%
\mathbb{H}^{\alpha\beta} = \mathbb{F}_a \; g^{\alpha\beta} +
\mathbb{F}_b \; p_B^\alpha \, p_B^\beta + i \, \mathbb{F}_c \;
\epsilon^{\alpha\beta\mu\nu} \, q_{+\mu} \, q_{-\nu},
\end{equation}
with the combined form factors being,
\begin{equation}\label{eq:nonresonant-combined-form-factors-D}
\mathbb{F}_i \equiv \mathbb{F}_i \left(q_+^2,q_-^2\right) =
\sum_{Q=u,c,t} V^*_{Qb} V_{Qd} \, F^Q_i\left(q_+^2,q_-^2\right),
\end{equation}
with $i=a,b,c$. For Majorana case, in addition to $\mathscr{Q}_{12}$
we have $\mathscr{Q}_{21}$ which is given by,
\begin{equation}\label{eq:Q21}
\mathscr{Q}_{21} = \frac{G_F^2}{2} \mathbb{H}^{\prime\alpha\beta} \,
L^{\prime}_{\alpha\beta}
\end{equation}
with
\begin{equation}\label{eq:nonresonant-hadronic-current-2}%
\mathbb{H}^{\prime\alpha\beta} = \mathbb{F}_a^\prime \;
g^{\alpha\beta} + \mathbb{F}_b^\prime \; p_B^\alpha \, p_B^\beta + i
\, \mathbb{F}_c^\prime \; \epsilon^{\alpha\beta\mu\nu} \, q'_{+\mu} \,
q'_{-\nu},
\end{equation}
and the combined form factors $\mathbb{F}_i^\prime$ with $i=a,b,c$ can
be easily obtained by substituting $q_\pm^{2}$ by $q_\pm^{\prime 2}$
in Eq.~\eqref{eq:nonresonant-combined-form-factors-D}.

\subsection{Complete expressions for the hadronic currents}

Taking both resonant and non-resonant contributions, the decay
amplitudes for Dirac and Majorana cases for the decay $B^0 \to \mu^-
\mu^+ \nu_\mu \bar{\nu}_\mu$ are given by
Eq.~\eqref{eq:amplitude-full} with the hadronic currents having both
resonant and non-resonant components,
\begin{equation}\label{eq:hadronic-current}
H^{(\prime)\alpha\beta} = \mathbf{H}^{(\prime)\alpha\beta} +
\mathbb{H}^{(\prime)\alpha\beta},%
\end{equation}
where the expressions for $\mathbf{H}^{\alpha\beta}$,
$\mathbf{H}^{\prime\alpha\beta}$, $\mathbb{H}^{\alpha\beta}$ and
$\mathbb{H}^{\prime\alpha\beta}$ are shown in
Eqs.~\eqref{eq:resonant-hadronic-current-1},
\eqref{eq:resonant-hadronic-current-2},
\eqref{eq:nonresonant-hadronic-current-1} and
\eqref{eq:nonresonant-hadronic-current-2} respectively. For brevity
the primed and unprimed hadronic currents are written in the same
equation above.

It should be noted that the form factors $\mathbf{F}_\pm$ and
$\mathbf{F}_\pm^{\prime}$, as well as $\mathbb{F}_{a,b,c}$ and
$\mathbb{F}_{a,b,c}^{\prime}$ are the same functions with different
arguments since the hadronic structure is independent of the process.
They are simply related by the exchange  $p_1 \leftrightarrow p_2$.
Furthermore, currently we do not know the exact functional forms of
the various non-resonant transition form factors. On the other hand
the individual resonant form factors for any given resonance are
known, but there could be relative phase difference between resonant
and non-resonant form factors. Though the resonance contribution is
substantial for the total branching ratio, as we show later they are
not important for the back-to-back kinematic configuration which is the focus
here.

\begin{figure}[ht!]
\centering%
\includegraphics[width=\linewidth,keepaspectratio]{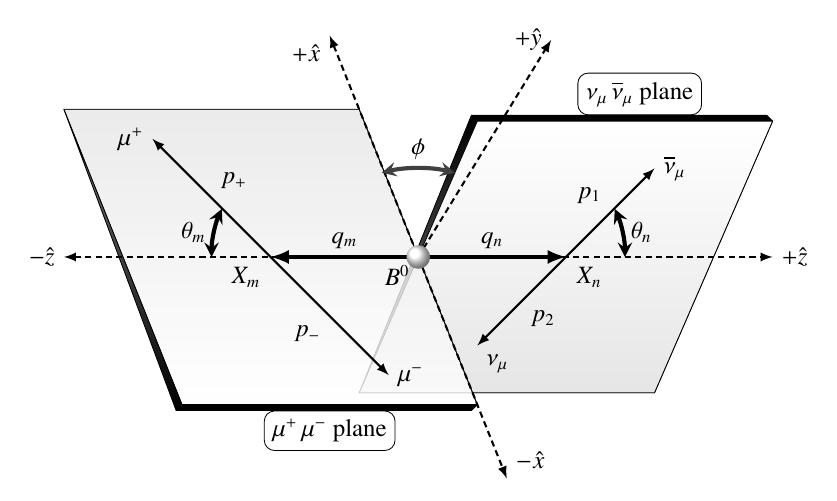}%
\caption{The kinematics of $B^0 \to \mu^- \mu^+ \nu_\mu \bar{\nu}_\mu$
in the rest frame of $B$, showing the polar angles $\theta_m$ and
$\theta_n$, as well as the azimuthal angle $\phi$. Here $X_m$ and
$X_n$ denote the muon pair and the neutrino pair.}%
\label{fig:kinematics}
\end{figure}

\subsection{General kinematics and differential decay rates}

It is convenient to visualize the decay $B^0 \to \mu^- \mu^+ \nu_\mu
\bar{\nu}_\mu$, in the rest frame of the $B^0$ meson, as a two-body
decay into a ``di-muon'' $X_m$ of mass $m_{\mu\mu}$ and a
``di-neutrino'' $X_n$ of mass $m_{\nu\nu}$. The subsequent decay of
each of these two subsystems is considered in its own
center-of-momentum frame as shown in Fig.~\ref{fig:kinematics}. The
$4$-momentum of the di-muon is denoted by $q_m$ and that of the
di-neutrino is denoted by $q_n$. The process is then described by the
following five variables:
\begin{enumerate}
\item $m_{\mu\mu}^2 \equiv q_m^2= \left(p_+ + p_-\right)^2$, the
invariant effective mass squared of the di-muon system,%
\item $m_{\nu\nu}^2 \equiv q_n^2 = \left(p_1 + p_2\right)^2$, the
invariant effective mass squared of the di-neutrino system,%
\item $\theta_m$, the angle between the direction of flight of the
$\mu^+$ in the center-of-momentum frame of the di-muon and the
direction of flight of the di-muon in the $B^0$ rest frame,%
\item $\theta_n$, the angle between the direction of flight of the
$\bar{\nu}_\mu$ in the center-of-momentum frame of the di-neutrino and
the direction of flight of the di-neutrino in the $B^0$ rest frame,
and%
\item $\phi$, the angle between the plane formed by the muons in the
$B^0$ rest frame and the corresponding plane formed by the neutrino
and antineutrino.
\end{enumerate}
The angles $\theta_m$ and $\theta_n$ are polar; $\phi$ is azimuthal.

The differential decay rate for the decay $B \to \mu^- \mu^+ \nu_\mu
\bar{\nu}_\mu$ is given by,
\begin{equation}\label{eq:diff-decay-rate-1}
\frac{\mathrm{d}^5\Gamma^{D/M}}{\mathrm{d}m_{\mu\mu}^2 \; \mathrm{d}m_{\nu\nu}^2 \; \mathrm{d}\cos\theta_m
\; \mathrm{d}\cos\theta_n \; \mathrm{d}\phi} = \frac{Y \,Y_m
\,Y_n\,\braketSq{\mathscr{M}^{D/M}}}{(4\,\pi)^6\, m_B^2 \, m_{\mu\mu}
\, m_{\nu\nu}},
\end{equation}
where $m_B$ is the mass of the $B$ meson, the magnitude of
$3$-momentum of $X_m$ or $X_n$ in the $B$ rest frame is $Y$, the
magnitude of $3$-momentum of $\mu^-$ or $\mu^+$ in the rest frame of
the di-muon is $Y_m$, the magnitude of $3$-momentum of $\nu_\mu$ or
$\bar{\nu}_\mu$ in the rest frame of the di-neutrino is $Y_n$ and
these are given by
\begin{subequations}
\begin{align}
Y &= \frac{\sqrt{\lambda\left(m_B^2, m_{\mu\mu}^2, m_
{\nu\nu}^2\right)}}{2m_B},\\%
Y_m &= \frac{\sqrt{m_{\mu\mu}^2-4m_\mu^2}}{2},\\%
Y_n &= \frac{\sqrt{m_ {\nu\nu}^2-4m_\nu^2}}{2},
\end{align}
\end{subequations}
with $m_\mu$, $m_\nu$ being the masses of muon and neutrino
respectively, and $\lambda\left(x,y,z\right) = x^2 + y^2 + z^2 - 2
\left(xy+yz+zx\right)$ is the K\"{a}ll\'{e}n function. The expression
for the square of the modulus of the decay amplitude with average over
initial spins and sum over final spins,
$\braketSq{\mathscr{M}^{D/M}}$, is a complicated function of
$\theta_m$, $\theta_n$, $\phi$, $m_{\mu\mu}$ and $m_{\nu\nu}$ for both
Dirac and Majorana cases, and can be written as,
\begin{align}
&\braketSq{\mathscr{M}^{D}} = G_F^4 \nonumber\\%
&\times \bigg( \modulus{\mathbb{F}_a}^2
S_{aa}^D + \modulus{\mathbb{F}_b}^2 S_{bb}^D +
\modulus{\mathbb{F}_c^2} S_{cc}^D + \modulus{\mathbf{F}_+}^2 S_{pp}^D
+ \modulus{\mathbf{F}_-}^2 S_{mm}^D \nonumber\\*%
& \quad + \Re\left(\mathbb{F}_a
\mathbb{F}_b^*\right) R_{ab}^D + \Re\left(\mathbb{F}_a
\mathbb{F}_c^*\right) R_{ac}^D + \Re\left(\mathbb{F}_a
\mathbf{F}_+^*\right) R_{ap}^D \nonumber\\%
& \quad + \Re\left(\mathbb{F}_a \mathbf{F}_-^*\right) R_{am}^D
+ \Im\left(\mathbb{F}_a \mathbb{F}_b^*\right) I_{ab}^D +
\Im\left(\mathbb{F}_a \mathbb{F}_c^*\right) I_{ac}^D \nonumber\\*%
& \quad +
\Im\left(\mathbb{F}_a \mathbf{F}_+^*\right) I_{ap}^D +
\Im\left(\mathbb{F}_a \mathbf{F}_-^*\right) I_{am}^D +
\Re\left(\mathbb{F}_b \mathbb{F}_c^*\right) R_{bc}^D \nonumber\\%
& \quad + \Re\left(\mathbb{F}_b \mathbf{F}_+^*\right) R_{bp}^D
+ \Re\left(\mathbb{F}_b \mathbf{F}_-^*\right) R_{bm}^D +
\Re\left(\mathbb{F}_c \mathbf{F}_+^*\right) R_{cp}^D \nonumber\\*%
& \quad +
\Im\left(\mathbb{F}_b \mathbb{F}_c^*\right) I_{bc}^D +
\Im\left(\mathbb{F}_b \mathbf{F}_+^*\right) I_{bp}^D +
\Im\left(\mathbb{F}_b \mathbf{F}_-^*\right) I_{bm}^D \nonumber\\%
& \quad + \Im\left(\mathbb{F}_c \mathbf{F}_+^*\right) I_{cp}^D
+ \Re\left(\mathbb{F}_c \mathbf{F}_-^*\right) R_{cm}^D +
\Re\left(\mathbf{F}_+ \mathbf{F}_-^*\right) R_{pm}^D \nonumber\\*%
& \quad +
\Im\left(\mathbb{F}_c \mathbf{F}_-^*\right) I_{cm}^D +
\Im\left(\mathbf{F}_+ \mathbf{F}_-^*\right) I_{pm}^D \bigg),
\label{eq:DAmpSq} \\%
&\braketSq{\mathscr{M}^{M}} = \frac{G_F^4}{2} \nonumber\\%
&\times \bigg(
\modulus{\mathbb{F}_a}^2 S_{aa}^M + \modulus{\mathbb{F}_b}^2 S_{bb}^M
+ \modulus{\mathbb{F}_c^2} S_{cc}^M + \modulus{\mathbf{F}_+}^2
S_{pp}^M + \modulus{\mathbf{F}_-}^2 S_{mm}^M \nonumber\\*%
& \quad +
\modulus{\mathbb{F}'_a}^2 S_{a'a'}^M + \modulus{\mathbb{F}'_b}^2
S_{b'b'}^M + \modulus{\mathbb{F}'_c}^2 S_{c'c'}^M +
\modulus{\mathbf{F}'_+}^2 S_{p'p'}^M \nonumber\\%
& \quad + \modulus{\mathbf{F}'_-}^2 S_{m'm'}^M +
\Re\left(\mathbb{F}_a \mathbb{F}_b^*\right) R_{ab}^M +
\Re\left(\mathbb{F}_a \mathbb{F}_c^*\right) R_{ac}^M \nonumber\\*%
& \quad +
\Re\left(\mathbb{F}_a \mathbf{F}_+^*\right) R_{ap}^M +
\Re\left(\mathbb{F}_a \mathbf{F}_-^*\right) R_{am}^M +
\Re\left(\mathbb{F}_b \mathbb{F}_c^*\right) R_{bc}^M \nonumber\\%
&\quad + \Re\left(\mathbb{F}_b \mathbf{F}_+^*\right) R_{bp}^M +
\Re\left(\mathbb{F}_b \mathbf{F}_-^*\right) R_{bm}^M +
\Re\left(\mathbb{F}_c \mathbf{F}_+^*\right) R_{cp}^M \nonumber\\*%
& \quad +
\Re\left(\mathbb{F}_c \mathbf{F}_-^*\right) R_{cm}^M +
\Re\left(\mathbb{F}'_a \mathbb{F}_b^{\prime *}\right) R_{a'b'}^M +
\Re\left(\mathbb{F}'_a \mathbb{F}_c^{\prime *}\right) R_{a'c'}^M
\nonumber\\%
& \quad + \Re\left(\mathbb{F}'_a \mathbf{F}_+^{\prime *}\right)
R_{a'p'}^M + \Re\left(\mathbb{F}'_a \mathbf{F}_-^{\prime *}\right)
R_{a'm'}^M + \Re\left(\mathbb{F}'_b \mathbb{F}_c^{\prime *}\right)
R_{b'c'}^M \nonumber\\*%
& \quad + \Re\left(\mathbb{F}'_b \mathbf{F}_+^{\prime *}\right)
R_{b'p'}^M + \Re\left(\mathbb{F}'_b \mathbf{F}_-^{\prime *}\right)
R_{b'm'}^M + \Re\left(\mathbb{F}'_c \mathbf{F}_+^{\prime *}\right)
R_{c'p'}^M \nonumber\\%
& \quad + \Re\left(\mathbb{F}'_c \mathbf{F}_-^{\prime *}\right)
R_{c'm'}^M + \Re\left(\mathbf{F}_+ \mathbf{F}_-^*\right) R_{pm}^M +
\Re\left(\mathbf{F}'_+ \mathbf{F}_-^{\prime *}\right) R_{p'm'}^M \nonumber\\*%
& \quad +
\Im\left(\mathbb{F}_a \mathbb{F}_b^*\right) I_{ab}^M +
\Im\left(\mathbb{F}_a \mathbb{F}_c^*\right) I_{ac}^M +
\Im\left(\mathbb{F}_a \mathbf{F}_+^*\right) I_{ap}^M \nonumber\\%
& \quad + \Im\left(\mathbb{F}_a \mathbf{F}_-^*\right) I_{am}^M
+ \Im\left(\mathbb{F}_b \mathbb{F}_c^*\right) I_{bc}^M +
\Im\left(\mathbb{F}_b \mathbf{F}_+^*\right) I_{bp}^M \nonumber\\*%
& \quad +
\Im\left(\mathbb{F}_b \mathbf{F}_-^*\right) I_{bm}^M +
\Im\left(\mathbb{F}_c \mathbf{F}_+^*\right) I_{cp}^M +
\Im\left(\mathbb{F}_c \mathbf{F}_-^*\right) I_{cm}^M \nonumber\\%
& \quad + \Im\left(\mathbb{F}'_a \mathbb{F}_b^{\prime *}\right)
I_{a'b'}^M + \Im\left(\mathbb{F}'_a \mathbb{F}_c^{\prime *}\right)
I_{a'c'}^M + \Im\left(\mathbb{F}'_a \mathbf{F}_+^{\prime *}\right)
I_{a'p'}^M \nonumber\\*%
& \quad + \Im\left(\mathbb{F}'_a \mathbf{F}_-^{\prime *}\right)
I_{a'm'}^M + \Im\left(\mathbb{F}'_b \mathbb{F}_c^{\prime *}\right)
I_{b'c'}^M + \Im\left(\mathbb{F}'_b \mathbf{F}_+^{\prime *}\right)
I_{b'p'}^M  \nonumber\\%
& \quad + \Im\left(\mathbb{F}'_b \mathbf{F}_-^{\prime *}\right)
I_{b'm'}^M + \Im\left(\mathbb{F}'_c \mathbf{F}_+^{\prime *}\right)
I_{c'p'}^M + \Im\left(\mathbb{F}'_c \mathbf{F}_-^{\prime *}\right)
I_{c'm'}^M \nonumber\\*%
& \quad + \Im\left(\mathbf{F}_+ \mathbf{F}_-^*\right) I_{pm}^M +
\Im\left(\mathbf{F}'_+ \mathbf{F}_-^{\prime *}\right) I_{p'm'}^M
\nonumber\\%
& \quad + m_\nu^2 \bigg( \Re\left(\mathbb{F}_a
\mathbb{F}_a^{\prime *}\right) R_{aa'}^M + \Re\left(\mathbb{F}_a
\mathbb{F}_b^{\prime *}\right) R_{ab'}^M + \Re\left(\mathbb{F}_a
\mathbb{F}_c^{\prime *}\right) R_{ac'}^M \nonumber\\*%
& \qquad\quad + \Re\left(\mathbb{F}_a
\mathbf{F}_+^{\prime *}\right) R_{ap'}^M + \Re\left(\mathbb{F}_a
\mathbf{F}_-^{\prime *}\right) R_{am'}^M + \Re\left(\mathbb{F}_b
\mathbb{F}_a^{\prime *}\right) R_{ba'}^M \nonumber\\*%
& \qquad\quad + \Re\left(\mathbb{F}_b \mathbb{F}_b^{\prime
*}\right) R_{bb'}^M + \Re\left(\mathbb{F}_b \mathbb{F}_c^{\prime
*}\right) R_{bc'}^M + \Re\left(\mathbb{F}_b \mathbf{F}_+^{\prime
*}\right) R_{bp'}^M \nonumber\\*%
& \qquad\quad + \Re\left(\mathbb{F}_b \mathbf{F}_-^{\prime
*}\right) R_{bm'}^M + \Re\left(\mathbb{F}_c \mathbb{F}_a^{\prime
*}\right) R_{ca'}^M + \Re\left(\mathbb{F}_c \mathbb{F}_b^{\prime
*}\right) R_{cb'}^M  \nonumber\\%
& \qquad\quad + \Re\left(\mathbb{F}_c \mathbb{F}_c^{\prime
*}\right) R_{cc'}^M + \Re\left(\mathbb{F}_c \mathbf{F}_+^{\prime
*}\right) R_{cp'}^M + \Re\left(\mathbb{F}_c \mathbf{F}_-^{\prime
*}\right) R_{cm'}^M \nonumber\\*%
& \qquad\quad + \Re\left(\mathbb{F}'_a \mathbf{F}_+^*\right)
R_{a'p}^M + \Re\left(\mathbb{F}'_a \mathbf{F}_-^*\right) R_{a'm}^M +
\Re\left(\mathbb{F}'_b \mathbf{F}_+^*\right) R_{b'p}^M \nonumber\\%
& \qquad\quad + \Re\left(\mathbb{F}'_b \mathbf{F}_-^*\right)
R_{b'm}^M + \Re\left(\mathbb{F}'_c \mathbf{F}_+^*\right) R_{c'p}^M +
\Re\left(\mathbb{F}'_c \mathbf{F}_-^*\right) R_{c'm}^M \nonumber\\*%
& \qquad\quad +
\Im\left(\mathbb{F}_a \mathbb{F}_a^{\prime *}\right) I_{aa'}^M +
\Im\left(\mathbb{F}_a \mathbb{F}_b^{\prime *}\right) I_{ab'}^M +
\Im\left(\mathbb{F}_a \mathbb{F}_c^{\prime *}\right) I_{ac'}^M
\nonumber\\%
& \qquad\quad + \Im\left(\mathbb{F}_a \mathbf{F}_+^{\prime
*}\right) I_{ap'}^M + \Im\left(\mathbb{F}_a \mathbf{F}_-^{\prime
*}\right) I_{am'}^M + \Im\left(\mathbb{F}_b \mathbb{F}_a^{\prime
*}\right) I_{ba'}^M \nonumber\\*%
& \qquad\quad + \Im\left(\mathbb{F}_b \mathbb{F}_b^{\prime
*}\right) I_{bb'}^M + \Im\left(\mathbb{F}_b \mathbb{F}_c^{\prime
*}\right) I_{bc'}^M + \Im\left(\mathbb{F}_b \mathbf{F}_+^{\prime
*}\right) I_{bp'}^M \nonumber\\%
& \qquad\quad + \Im\left(\mathbb{F}_b \mathbf{F}_-^{\prime
*}\right) I_{bm'}^M + \Im\left(\mathbb{F}_c \mathbb{F}_a^{\prime
*}\right) I_{ca'}^M + \Im\left(\mathbb{F}_c \mathbb{F}_b^{\prime
*}\right) I_{cb'}^M \nonumber\\*%
& \qquad\quad + \Im\left(\mathbb{F}_c \mathbb{F}_c^{\prime
*}\right) I_{cc'}^M + \Im\left(\mathbb{F}_c \mathbf{F}_+^{\prime
*}\right) I_{cp'}^M + \Im\left(\mathbb{F}_c \mathbf{F}_-^{\prime
*}\right) I_{cm'}^M \nonumber\\%
& \qquad\quad + \Im\left(\mathbb{F}'_a \mathbf{F}_+^*\right)
I_{a'p}^M + \Im\left(\mathbb{F}'_a \mathbf{F}_-^*\right) I_{a'm}^M +
\Im\left(\mathbb{F}'_b \mathbf{F}_+^*\right) I_{b'p}^M \nonumber\\*%
& \qquad\quad +
\Im\left(\mathbb{F}'_b \mathbf{F}_-^*\right) I_{b'm}^M +
\Im\left(\mathbb{F}'_c \mathbf{F}_+^*\right) I_{c'p}^M +
\Im\left(\mathbb{F}'_c \mathbf{F}_-^*\right) I_{c'm}^M  \bigg) \bigg),
\label{eq:MAmpSq}
\end{align}
where the terms $S_{i^{(\prime)}i^{(\prime)}}^{D/M}$ are associated
with the squares of the form factors
$\modulus{\mathbb{F}^{(\prime)}_i}^2$ or
$\modulus{\mathbf{F}^{(\prime)}_i}^2$, and the terms
$R_{i^{(\prime)}j^{(\prime)}}^{D/M}$ $\left( \text{or }
I_{i^{(\prime)}j^{(\prime)}}^{D/M}\right)$ are associated with the
real (or imaginary) part of the products of form factors
$\mathbb{F}_i^{(\prime)} \mathbb{F}_j^{(\prime)*}$ or
$\mathbb{F}_i^{(\prime)} \mathbf{F}_j^{(\prime)*}$ or
$\mathbf{F}_i^{(\prime)} \mathbb{F}_j^{(\prime)*}$ or
$\mathbf{F}_i^{(\prime)} \mathbf{F}_j^{(\prime)*}$ with $i\neq j$ and
$i,j \in \{a,b,c,p\equiv +,m \equiv -\}$. The total number of possible
terms for Dirac and Majorana cases are $25$ and $92$ respectively. We
have $25$ direct terms, $25$ exchange terms and $42$ interference
terms in the Majorana case which are directly proportional to
$m_\nu^2$ as shown in Eq.~\eqref{eq:MAmpSq}. Some of these terms,
shown in Eq.~\eqref{eq:vanishing-terms} of
Appendix~\ref{appendix:terms}, are zero. The detailed expressions for
the $70$ non-vanishing terms (with Majorana and Dirac cases sharing
$20$ terms) are given explicitly in Appendix~\ref{appendix:terms} from
Eq.~\eqref{eq:term-first} to Eq.~\eqref{eq:term-last}.

\textit{Note:} To briefly illustrate that the difference between Dirac
and Majorana neutrinos need not necessarily be proportional to some
power of $m_\nu$, we reconsider the simple example of collinear
neutrino and antineutrino ($p_1 = p_2$) that was mentioned in
subsec.~\ref{subsec:tht-expt}. We note that the Majorana amplitude for
this collinear case vanishes exactly, while the Dirac amplitude does
not. Considering the leading contribution that comes from the form
factor $\mathbb{F}_a$ alone, and substituting $p_1 = p_2 \equiv p_\nu
(\text{say})$ we obtain
\begin{equation*}
\braketSq{\mathscr{M}^D_\text{collinear}} = 64\,G_F^4\, \modulus{\mathbb{F}_a}^2\,\left(p_\nu \cdot p_+\right)\,\left(p_\nu \cdot p_-\right),
\end{equation*}
which is not proportional to any power of $m_\nu$. This proves that
the collinear $\nu\,\bar{\nu}$ scenario is indeed another exception to
DMCT, albeit being experimentally inaccessible as mentioned in
subsec.~\ref{subsec:tht-expt}.

To analyze the experimentally accessible back-to-back kinematic
configuration which is also capable of distinguishing Dirac and
Majorana neutrinos, we need to first study the differential decay
distribution in detail.

\subsection{Differential decay distribution}

The angles $\theta_n$ and $\phi$ (see Fig.~\ref{fig:kinematics}) are
indeed inaccessible, as the neutrino pair goes missing. Therefore, for
a physically useful differential decay rate we must integrate over
both $\theta_n$ and $\phi$ in Eq.~\eqref{eq:diff-decay-rate-1}, i.e.\
\begin{align}
\frac{\mathrm{d}^3\Gamma^{D/M}}{\mathrm{d}m_{\mu\mu}^2 \; \mathrm{d}m_{\nu\nu}^2 \;
\mathrm{d}\cos\theta_m} &= \frac{Y\,Y_m\,Y_n}{(4\,\pi)^6 m_B^2 \, m_{\mu\mu} \,
m_{\nu\nu}} \nonumber\\%
&\times \int_{-1}^1 \int_0^{2\pi} \braketSq{\mathscr{M}^{D/M}} \,
\mathrm{d}\cos\theta_n \; \mathrm{d}\phi.\label{eq:diff-decay-rate-2}
\end{align}
It is straightforward to show that the difference between Dirac and
Majorana cases is given by
\begin{align}
&\frac{\mathrm{d}^3\Gamma^M}{\mathrm{d}m_{\mu\mu}^2 \; \mathrm{d}m_{\nu\nu}^2 \; \mathrm{d}\cos\theta_m} -
\frac{\mathrm{d}^3\Gamma^D}{\mathrm{d}m_{\mu\mu}^2 \; \mathrm{d}m_{\nu\nu}^2 \; \mathrm{d}\cos\theta_m} =
\dfrac{G_F^4}{2} \nonumber\\%
&\times \frac{Y\,Y_m \,Y_n}{(4\,\pi)^6\, m_B^2 \, m_{\mu\mu} \,
m_{\nu\nu}} \int_{-1}^1 \int_0^{2\pi} \, \mathrm{d}\cos\theta_n \; \mathrm{d}\phi
\nonumber\\%
&\times \Bigg(  -\modulus{\mathbb{F}_a}^2 S_{aa}^M -
\modulus{\mathbb{F}_b}^2 S_{bb}^M - \modulus{\mathbb{F}_c^2} S_{cc}^M
- \modulus{\mathbf{F}_+}^2 S_{pp}^M - \modulus{\mathbf{F}_-}^2
S_{mm}^M \nonumber\\*%
& \quad + \modulus{\mathbb{F}'_a}^2 S_{a'a'}^M +
\modulus{\mathbb{F}'_b}^2 S_{b'b'}^M + \modulus{\mathbb{F}'_c}^2
S_{c'c'}^M + \modulus{\mathbf{F}'_+}^2 S_{p'p'}^M \nonumber\\%
&  \quad +                           
\modulus{\mathbf{F}'_-}^2 S_{m'm'}^M - \Re\left(\mathbb{F}_a \mathbb{F}_b^*\right) R_{ab}^M
- \Re\left(\mathbb{F}_a \mathbb{F}_c^*\right) R_{ac}^M \nonumber\\*%
& \quad  -
\Re\left(\mathbb{F}_a \mathbf{F}_+^*\right) R_{ap}^M -
\Re\left(\mathbb{F}_a \mathbf{F}_-^*\right) R_{am}^M -
\Re\left(\mathbb{F}_b \mathbb{F}_c^*\right) R_{bc}^M  \nonumber\\%
& \quad -
\Re\left(\mathbb{F}_b \mathbf{F}_+^*\right) R_{bp}^M 
-                                                    
\Re\left(\mathbb{F}_b \mathbf{F}_-^*\right) R_{bm}^M - \Re\left(\mathbb{F}_c \mathbf{F}_-^*\right) R_{cm}^M \nonumber\\*%
& \quad +
\Re\left(\mathbb{F}'_a \mathbb{F}_b^{\prime *}\right) R_{a'b'}^M +
\Re\left(\mathbb{F}'_a \mathbb{F}_c^{\prime *}\right) R_{a'c'}^M +
\Re\left(\mathbb{F}'_a \mathbf{F}_+^{\prime *}\right) R_{a'p'}^M 
\nonumber\\%
&  \quad +
\Re\left(\mathbb{F}'_a \mathbf{F}_-^{\prime *}\right) R_{a'm'}^M
 +
\Re\left(\mathbb{F}'_b \mathbb{F}_c^{\prime *}\right) R_{b'c'}^M + \Re\left(\mathbb{F}'_b \mathbf{F}_+^{\prime *}\right)
R_{b'p'}^M \nonumber\\%
& \quad + \Re\left(\mathbb{F}'_b \mathbf{F}_-^{\prime *}\right)
R_{b'm'}^M + \Re\left(\mathbb{F}'_c \mathbf{F}_-^{\prime *}\right)
R_{c'm'}^M \nonumber\\*%
& \quad - \Re\left(\mathbf{F}_+ \mathbf{F}_-^*\right) R_{pm}^M +
\Re\left(\mathbf{F}'_+ \mathbf{F}_-^{\prime *}\right) R_{p'm'}^M
\nonumber\\%
&  \quad - \Im\left(\mathbb{F}_a \mathbb{F}_b^*\right) I_{ab}^M
- \Im\left(\mathbb{F}_a \mathbb{F}_c^*\right) I_{ac}^M -
\Im\left(\mathbb{F}_a \mathbf{F}_-^*\right) I_{am}^M \nonumber\\*%
& \quad -
\Im\left(\mathbb{F}_b \mathbb{F}_c^*\right) I_{bc}^M -
\Im\left(\mathbb{F}_c \mathbf{F}_+^*\right) I_{cp}^M -
\Im\left(\mathbb{F}_c \mathbf{F}_-^*\right) I_{cm}^M \nonumber\\%
&  \quad + \Im\left(\mathbb{F}'_a \mathbb{F}_b^{\prime *}\right)
I_{a'b'}^M + \Im\left(\mathbb{F}'_a \mathbb{F}_c^{\prime *}\right)
I_{a'c'}^M + \Im\left(\mathbb{F}'_a \mathbf{F}_-^{\prime *}\right)
I_{a'm'}^M \nonumber\\*%
& \quad + \Im\left(\mathbb{F}'_b \mathbb{F}_c^{\prime *}\right)
I_{b'c'}^M + \Im\left(\mathbb{F}'_c \mathbf{F}_+^{\prime *}\right)
I_{c'p'}^M + \Im\left(\mathbb{F}'_c \mathbf{F}_-^{\prime *}\right)
I_{c'm'}^M  \nonumber\\%
&  \quad + m_\nu^2 \bigg( \Re\left(\mathbb{F}_a
\mathbb{F}_a^{\prime *}\right) R_{aa'}^M + \Re\left(\mathbb{F}_a
\mathbb{F}_b^{\prime *}\right) R_{ab'}^M +  \Re\left(\mathbb{F}_a
\mathbf{F}_+^{\prime *}\right) R_{ap'}^M \nonumber\\%
& \qquad\qquad + \Re\left(\mathbb{F}_b
\mathbb{F}_a^{\prime *}\right) R_{ba'}^M + \Re\left(\mathbb{F}_b
\mathbb{F}_b^{\prime *}\right) R_{bb'}^M + \Re\left(\mathbb{F}_b
\mathbb{F}_c^{\prime *}\right) R_{bc'}^M  \nonumber\\%
& \qquad\qquad + \Re\left(\mathbb{F}_b \mathbf{F}_+^{\prime
*}\right) R_{bp'}^M + \Re\left(\mathbb{F}_b \mathbf{F}_-^{\prime
*}\right) R_{bm'}^M + \Re\left(\mathbb{F}_c \mathbb{F}_b^{\prime
*}\right) R_{cb'}^M \nonumber\\%
& \qquad\qquad + \Re\left(\mathbb{F}_c \mathbb{F}_c^{\prime
*}\right) R_{cc'}^M + \Re\left(\mathbb{F}_c \mathbf{F}_+^{\prime
*}\right) R_{cp'}^M + \Re\left(\mathbb{F}_c \mathbf{F}_-^{\prime
*}\right) R_{cm'}^M \nonumber\\%
& \qquad\qquad + \Re\left(\mathbb{F}'_a \mathbf{F}_+^*\right)
R_{a'p}^M + \Re\left(\mathbb{F}'_a \mathbf{F}_-^*\right) R_{a'm}^M +
\Re\left(\mathbb{F}'_b \mathbf{F}_+^*\right) R_{b'p}^M \nonumber\\%
& \qquad\qquad +
\Re\left(\mathbb{F}'_b \mathbf{F}_-^*\right) R_{b'm}^M +
\Re\left(\mathbb{F}'_c \mathbf{F}_+^*\right) R_{c'p}^M +
\Re\left(\mathbb{F}'_c \mathbf{F}_-^*\right) R_{c'm}^M   \nonumber\\%
& \qquad\qquad + \Im\left(\mathbb{F}_a \mathbb{F}_c^{\prime
*}\right) I_{ac'}^M + \Im\left(\mathbb{F}_a \mathbf{F}_-^{\prime
*}\right) I_{am'}^M + \Im\left(\mathbb{F}_b \mathbf{F}_+^{\prime
*}\right) I_{bp'}^M \nonumber\\%
& \qquad\qquad + \Im\left(\mathbb{F}_b \mathbf{F}_-^{\prime
*}\right) I_{bm'}^M + \Im\left(\mathbb{F}_c \mathbb{F}_a^{\prime
*}\right) I_{ca'}^M + \Im\left(\mathbb{F}_c \mathbb{F}_c^{\prime
*}\right) I_{cc'}^M \nonumber\\%
& \qquad\qquad + \Im\left(\mathbb{F}_c \mathbf{F}_+^{\prime
*}\right) I_{cp'}^M + \Im\left(\mathbb{F}_c \mathbf{F}_-^{\prime
*}\right) I_{cm'}^M + \Im\left(\mathbb{F}'_b \mathbf{F}_+^*\right)
I_{b'p}^M \nonumber\\%
& \qquad\qquad + \Im\left(\mathbb{F}'_b \mathbf{F}_-^*\right) I_{b'm}^M +
\Im\left(\mathbb{F}'_c \mathbf{F}_+^*\right) I_{c'p}^M +
\Im\left(\mathbb{F}'_c \mathbf{F}_-^*\right) I_{c'm}^M  \bigg)
\Bigg).\label{eq:DiffDecayRate-DmM}
\end{align}

In absence of analytical expressions for all the form factors, we note
the following important features that can be easily observed in
Eq.~\eqref{eq:DiffDecayRate-DmM}.
\begin{enumerate}
\item There are direct and exchange terms which are related to one
another by $p_1 \leftrightarrow p_2$ exchange. In
Eq.~\eqref{eq:DiffDecayRate-DmM} all the direct terms appear with
negative sign. And the corresponding exchange terms have positive
sign. Therefore, these terms after integration over $\cos\theta_n$ and
$\phi$ should vanish, as they would have equal and opposite
contributions.%
\item There are interference terms which are invariant under the $p_1
\leftrightarrow p_2$ exchange. All these interference terms are
explicitly found to be proportional to $m_\nu^2$. These terms would
survive after integration over $\cos\theta_n$ and $\phi$, simply
because there is no way to cancel them, unless the integral itself
vanishes. For example, if one were to assume the form factors to be
constants, then all the $I_{j^{(\prime)}k^{(\prime)}}^M$ terms (for
$j,k=a,b,c,p,m$) would vanish.%
\item The integration has been carried out over $\cos\theta_n$ and
$\phi$, the variables necessary to describe the individual $\nu$ and
$\bar\nu$. The variable $m_{\nu\nu}^2$, also associated with the
$\nu,\bar\nu$ pair, is however unaffected by the $p_1 \leftrightarrow
p_2$ exchange. Essentially the integration over $\cos\theta_n$ and
$\phi$ wipes out the difference between the direct and the exchange
terms.
\end{enumerate}
Therefore, the difference between the Dirac and Majorana cases, as
shown in Eq.~\eqref{eq:DiffDecayRate-DmM} after integration over
neutrino pair variables, is now proportional to $m_\nu^2$:
\begin{equation}\label{eq:DM-confusion-theorem}
\frac{\mathrm{d}^3\Gamma^M}{\mathrm{d}m_{\mu\mu}^2 \; \mathrm{d}m_{\nu\nu}^2 \; \mathrm{d}\cos\theta_m} -
\frac{\mathrm{d}^3\Gamma^D}{\mathrm{d}m_{\mu\mu}^2 \; \mathrm{d}m_{\nu\nu}^2 \; \mathrm{d}\cos\theta_m}
\propto m_\nu^2,
\end{equation}
which proves DMCT in the present case. This is not our main point
since DMCT is a well known result. It would be interesting to see if
we may avoid the constraint imposed by the DMCT. We do this next, and
as a bonus we find the difference between Dirac and Majorana scenarios
is not just substantial, but {\it it eliminates the dependence on the
unknown neutrino mass} to a very good approximation. The corrections
coming from non-zero neutrino mass is negligible.

\subsection{Change of variables for back-to-back configuration}
                                                             
As shown in Sec.~\ref{sec:expectation} it is the back-to-back
configuration which holds the promise to probe the quantum statistics
of Majorana neutrinos most effectively. For this case, our choice of
kinematic variables is not helpful. We need to make change of
variables. Let us assume that the angle between the neutrino and
antineutrino in the rest frame of $B^0$ be $\Theta$. Then, in terms
of the neutrino energies $E_1$ and $E_2$ we have
\begin{equation}
\cos\Theta = \frac{Y^2 - E_1^2 - E_2^2 +
2m_\nu^2}{2\,\sqrt{E_1^2-m_\nu^2}\,\sqrt{E_2^2-m_\nu^2}}.
\end{equation}
When the neutrino and antineutrino are back-to-back in the $B^0$ rest
frame, we have $Y=0$ and $E_1 = E_2 = E_\nu$ (say). This implies that,
$\cos\Theta = -1$, as it should be for $\Theta=\pi$. It is easy to
show that,
\begin{align}
\mathrm{d}m_{\mu\mu}^2\,\mathrm{d}m_{\nu\nu}^2\,\mathrm{d}\cos\theta_n &= -
\frac{4\,m_B\,m_{\nu\nu}}{Y\,Y_n} \sqrt{\left(E_1^2 -
m_\nu^2\right)\,\left(E_2^2 - m_\nu^2\right)}\nonumber\\*%
&\quad \times \mathrm{d}E_1\,\mathrm{d}E_2\,\mathrm{d}\cos\Theta,
\end{align}
where
\begin{subequations}
\begin{align}
m_{\nu\nu}^2 &= 2\,m_\nu^2 + 2\,E_1\,E_2 - 2 \sqrt{\left(E_1^2 -
m_\nu^2\right)\,\left(E_2^2 -
m_\nu^2\right)}\,\cos\Theta,\label{eq:mnunuSq}\\%
m_{\mu\mu}^2 &= m_B^2 + 2\,m_\nu^2 - 2 m_B \left(E_1 + E_2\right) +
2\,E_1\,E_2 \nonumber\\*%
&\quad - 2 \sqrt{\left(E_1^2 - m_\nu^2\right)\,\left(E_2^2 -
m_\nu^2\right)}\,\cos\Theta.\label{eq:mmumuSq}
\end{align}
\end{subequations}
Therefore,
\begin{align}
\frac{\mathrm{d}^5\Gamma^{D/M}}{\mathrm{d}E_1\,\mathrm{d}E_2\,\mathrm{d}\cos\Theta\,\mathrm{d}\cos\theta_m\,\mathrm{d}\phi}
&= -
\frac{Y_m\,\braketSq{\mathscr{M}^{D/M}}}{4^5\,\pi^6\,m_B\,m_{\mu\mu}}
\nonumber\\*%
&\quad \times \sqrt{\left(E_1^2 - m_\nu^2\right)\,\left(E_2^2 -
m_\nu^2\right)}
\end{align}
It should be noted that the differential decay rate for back-to-back
configuration is obtained from the full five variable differential
decay rate as shown in Eq.~\eqref{eq:diff-decay-rate-1} without any
integration and after making the suitable change of variables
mentioned above.
                                             
\subsection{Addressing the back-to-back case}

For back-to-back case, with $E_1 = E_2 = E_\nu$ (say) and
$\Theta=\pi$, we get the following from Eqs.~\eqref{eq:mnunuSq} and
\eqref{eq:mmumuSq},
\begin{subequations}
\begin{align}
m_{\nu\nu}^2 &= 4\,E_\nu^2\;,\\%
m_{\mu\mu}^2 &= \left(m_B - 2\,E_\nu\right)^2,%
\end{align}
\end{subequations}
which correctly implies $Y = 0$, meaning that the di-muon and
di-neutrino systems are at rest in the $B^0$ rest frame, as they
should be. Moreover, for the back-to-back case we have
\begin{subequations}
\begin{align}
Y_m &= \sqrt{\left(\frac{m_B}{2} - E_\nu\right)^2 - m_\mu^2 }\; ,\\%
Y_n &= \sqrt{E_\nu^2 - m_\nu^2}\;.
\end{align}
\end{subequations}
It can be shown that, in general,
\begin{equation}
\cos\theta_n = \frac{m_{\nu\nu} \left(E_1 - E_2\right)}{2\,Y\,Y_n}.
\end{equation}
Whenever $E_1 = E_2$ for any value of the angle $\Theta$ between the
neutrino and antineutrino we get $\cos\theta_n=0$. This would
therefore hold true for the back-to-back case. Moreover, in the
back-to-back case we have both the back-to-back muons and the
back-to-back neutrino antineutrino pair, in one single plane. This
implies that for the back-to-back case we have $\phi = 0$. These
choices put the orientation of the coordinate axes in such a way that
the back-to-back neutrino and antineutrino fly away defining the
$x$-axis. The $xz$-plane in Fig.~\ref{fig:kinematics} is the one in
which the $3$-momenta of muons lie, and now the back-to-back neutrino
antineutrino define the $x$-direction. The direction perpendicular to
the neutrino direction is the $z$-direction. If we define the angle
between the neutrino and muon directions to be $\theta$, then
$\theta_m = \pi/2 - \theta$. This implies that
\begin{equation}
\cos\theta_m = \sin\theta.
\end{equation}
Finally, we note that the energy of neutrino $E_\nu$ in the
back-to-back case can be easily known from the experimentally measured
energy of either of the back-to-back muons $E_\mu$ via
Eq.~\eqref{eq:B2B-energies-relation}. The muon energy $E_\mu$, in the
back-to-back case, can vary in the range $\left[ m_\mu, m_B/2 - m_\nu
\right]$. It is easy to show that for the back-to-back configuration,
\begin{subequations}\label{eq:dot-products-B2B}
\begin{align}
p_1 \cdot p_\pm &= E_\mu \left(\frac{m_B}{2} - E_\mu\right) \mp
Y_m\,Y_n\cos\theta,\\%
p_2 \cdot p_\pm &= E_\mu \left(\frac{m_B}{2} - E_\mu\right) \pm
Y_m\,Y_n\cos\theta,
\end{align}
\end{subequations}
with $Y_m = \sqrt{E_\mu^2-m_\mu^2}$ and $Y_n =
\sqrt{\left(m_B/2-E_\mu\right)^2-m_\nu^2}$.

The differential decay rate in the back-to-back case is therefore
given by,
\begin{equation}\label{eq:diff-decay-rate-3}
\frac{\mathrm{d}^3\Gamma_{\leftrightarrow}^{D/M}}{\mathrm{d}E_\mu^2\,\mathrm{d}\sin\theta} =
\frac{2\,\sqrt{E_\mu^2 - m_\mu^2}}{\left(4\,\pi\right)^6\,m_B\,E_\mu}
\left( \left(\frac{m_B}{2} - E_\mu\right)^2 - m_\nu^2 \right)
\braketSq{\mathscr{M}^{D/M}_{\leftrightarrow}},
\end{equation}
where $\braketSq{\mathscr{M}^{D/M}_{\leftrightarrow}}$ is same as $\braketSq{\mathscr{M}^{D/M}}$ with the necessary dot product substitutions as shown in Eq.~\eqref{eq:dot-products-B2B}. In the expression for $\braketSq{\mathscr{M}^{D/M}_{\leftrightarrow}}$
we have form factors which are functions of $q_\pm^2$ or
$q_\pm^{\prime 2}$ and it is easy to show that,
\begin{subequations}
\begin{align}
q_\pm^2 &= m_\mu^2 + m_\nu^2 + E_\mu \left(m_B-2\,E_\mu\right) +
2\,Y_m\,Y_n\,\cos\theta,\\%
q_\pm^{\prime 2} &= m_\mu^2 + m_\nu^2 + E_\mu
\left(m_B-2\,E_\mu\right) - 2\,Y_m\,Y_n\,\cos\theta.%
\end{align}
\end{subequations}

\subsection{The difference between Dirac and Majorana cases in back-to-back configuration}

We are interested in whether there is any difference between Dirac and
Majorana cases in the back-to-back configuration which would be
independent of the mass $m_\nu$ which can be practically neglected in
comparison with other masses and the energy $E_\mu$. The difference
between the decay rates for Dirac and Majorana cases can be obtained
using Eq.~\eqref{eq:diff-decay-rate-3}. We find it convenient to
express the difference in differential decay rates for back-to-back
c
ase, after neglecting the neutrino mass in comparison with other
masses, as follows,                                                                    
\begin{align}
&\frac{\mathrm{d}^3\Gamma^D_{\leftrightarrow}}{\mathrm{d}E_\mu^2\,\mathrm{d}\sin\theta} -
\frac{\mathrm{d}^3\Gamma^M_{\leftrightarrow}}{\mathrm{d}E_\mu^2\,\mathrm{d}\sin\theta} =
\frac{G_F^4\,\sqrt{E_\mu^2 -
m_\mu^2}}{\left(4\,\pi\right)^6\,m_B\,E_\mu} \left(\frac{m_B}{2} -
E_\mu\right)^2 \nonumber\\%
&\quad \times \Bigg(\left(\modulus{\mathbb{F}_a}^2 -
\modulus{\mathbb{F}'_a}^2 \right) \Delta_{aa} +
\left(\modulus{\mathbb{F}_b}^2 - \modulus{\mathbb{F}'_b}^2 \right)
\Delta_{bb} + \left(\modulus{\mathbb{F}_c}^2 -
\modulus{\mathbb{F}'_c}^2 \right) \Delta_{cc} \nonumber\\%
& \qquad + \left(\modulus{\mathbf{F}_+}^2 - \modulus{\mathbf{F}'_+}^2
\right) \Delta_{pp} + \left(\modulus{\mathbf{F}_-}^2 -
\modulus{\mathbf{F}'_-}^2 \right) \Delta_{mm} \nonumber\\%
&\qquad + \bigg(
\Re\left(\mathbb{F}_a \mathbb{F}_b^*\right) - \Re\left(\mathbb{F}'_a
\mathbb{F}_b^{\prime *}\right) \bigg) \Delta_{ab} \nonumber\\%
& \qquad + \bigg( \Re\left(\mathbb{F}_a \mathbf{F}_+^*\right) -
\Re\left(\mathbb{F}'_a \mathbf{F}_+^{\prime *}\right) \bigg)
\Delta_{ap} \nonumber\\%
&\qquad + \bigg( \Re\left(\mathbb{F}_b \mathbf{F}_+^*\right) -
\Re\left(\mathbb{F}'_b \mathbf{F}_+^{\prime *}\right) \bigg)
\Delta_{bp} \nonumber\\%
& \qquad + \bigg( \Re\left(\mathbb{F}_a \mathbf{F}_-^*\right) -
\Re\left(\mathbb{F}'_a \mathbf{F}_-^{\prime *}\right) \bigg)
\Delta_{am} \nonumber\\%
&\qquad + \bigg( \Re\left(\mathbb{F}_b \mathbf{F}_-^*\right) -
\Re\left(\mathbb{F}'_b \mathbf{F}_-^{\prime *}\right) \bigg)
\Delta_{bm} \nonumber\\%
& \qquad + \bigg( \Re\left(\mathbb{F}_c \mathbf{F}_-^*\right) -
\Re\left(\mathbb{F}'_c \mathbf{F}_-^{\prime *}\right) \bigg)
\Delta_{cm} \nonumber\\%
&\qquad + \bigg( \Re\left(\mathbf{F}_+ \mathbf{F}_-^*\right) -
\Re\left(\mathbf{F}'_+ \mathbf{F}_-^{\prime *}\right) \bigg)
\Delta_{pm} \nonumber\\%
&\qquad + \cos\theta\,\Bigg( \left(\modulus{\mathbb{F}_a}^2 +
\modulus{\mathbb{F}'_a}^2 \right) \Sigma_{aa} +
\left(\modulus{\mathbb{F}_b}^2 + \modulus{\mathbb{F}'_b}^2 \right)
\Sigma_{bb} \nonumber\\%
&\hspace{2cm} + \left(\modulus{\mathbf{F}_+}^2 +
\modulus{\mathbf{F}'_+}^2 \right) \Sigma_{pp} +
\left(\modulus{\mathbf{F}_-}^2 + \modulus{\mathbf{F}'_-}^2 \right)
\Sigma_{mm} \nonumber\\%
&\hspace{2cm} + \bigg( \Re\left(\mathbb{F}_a \mathbb{F}_b^*\right) +
\Re\left(\mathbb{F}'_a \mathbb{F}_b^{\prime *}\right) \bigg)
\Sigma_{ab} \nonumber\\%
&\hspace{2cm} + \bigg( \Re\left(\mathbb{F}_a \mathbf{F}_-^*\right) +
\Re\left(\mathbb{F}'_a \mathbf{F}_-^{\prime *}\right) \bigg)
\Sigma_{am} \nonumber\\%
&\hspace{2cm} + \bigg( \Re\left(\mathbb{F}_b \mathbf{F}_-^*\right) +
\Re\left(\mathbb{F}'_b \mathbf{F}_-^{\prime *}\right) \bigg)
\Sigma_{bm} \nonumber\\%
&\hspace{2cm} + \bigg( \Re\left(\mathbf{F}_+ \mathbf{F}_-^*\right) +
\Re\left(\mathbf{F}'_+ \mathbf{F}_-^{\prime *}\right) \bigg)
\Sigma_{pm} \Bigg)\Bigg),\label{eq:DmM-B2B}
\end{align}
where the various non-vanishing $\Delta_{ij}$ and $\Sigma_{ij}$ terms,
with $i,j \in \{a,b,c,p\equiv +,m\equiv -\}$, are given in Appendix
\ref{app:sigma-delta}. It is interesting to note that, all the
$\Sigma_{ij}$ terms are directly proportional to $\cos\theta$, and
therefore do not contribute when $\theta=\pi/2$, i.e.\ when the
back-to-back neutrino antineutrino pair is perpendicular to the
back-to-back muons. It is also true that for this very special case of
$\theta=\pi/2$, we have $q_\pm^2 = q_\pm^{\prime 2}$ which implies
that both the primed and unprimed form factors are equal in this case.
This implies that when the muons fly perpendicular to the neutrino
antineutrino pair in the back-to-back case, there is no difference
between the Dirac and Majorana cases (when the neutrino mass is
neglected in comparison with other masses). For other values of
$\theta$, the difference between Dirac and Majorana cases is non-zero,
in general.

\subsection{A simple case study}\label{subsec:simple-case-study}

\begin{figure*}[hbtp]
\centering%
\subfloat[Three dimensional view of the differential decay rate for
Dirac case with an appropriate normalization as
mentioned.\label{fig:3D-D}]{\includegraphics[width=0.495\linewidth,keepaspectratio]{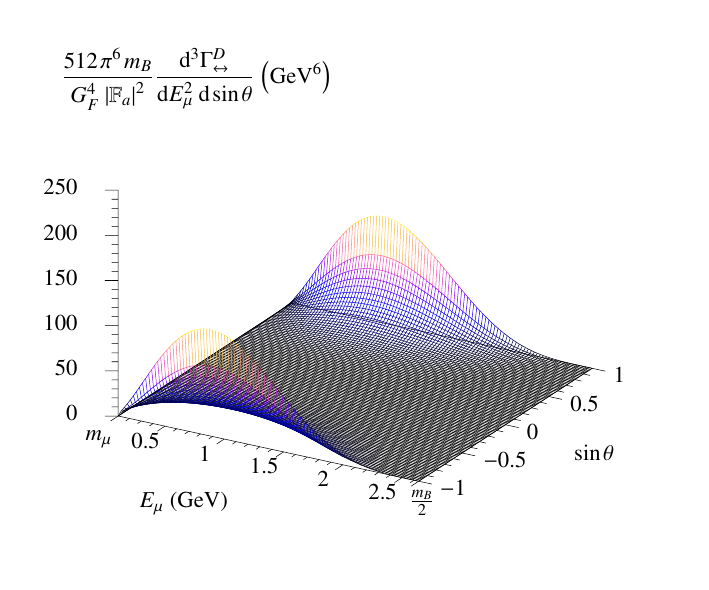}} \hfill%
\subfloat[Three dimensional view of the differential decay rate for
Majorana case with an appropriate normalization as
mentioned.\label{fig:3D-M}]{\includegraphics[width=0.495\linewidth,keepaspectratio]{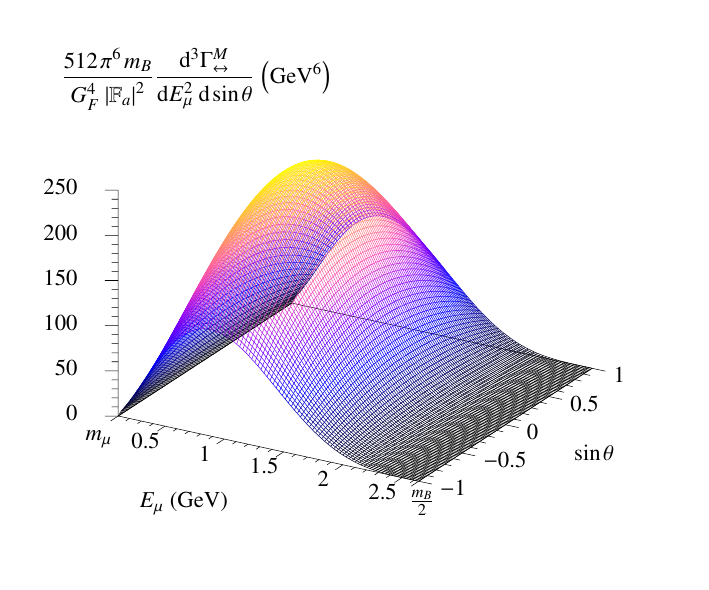}}\\%
\subfloat[Comparison of muon energy distributions between Dirac and
Majorana cases in the back-to-back
scenario.\label{fig:energy-dist}]{\includegraphics[width=0.495\linewidth, keepaspectratio]{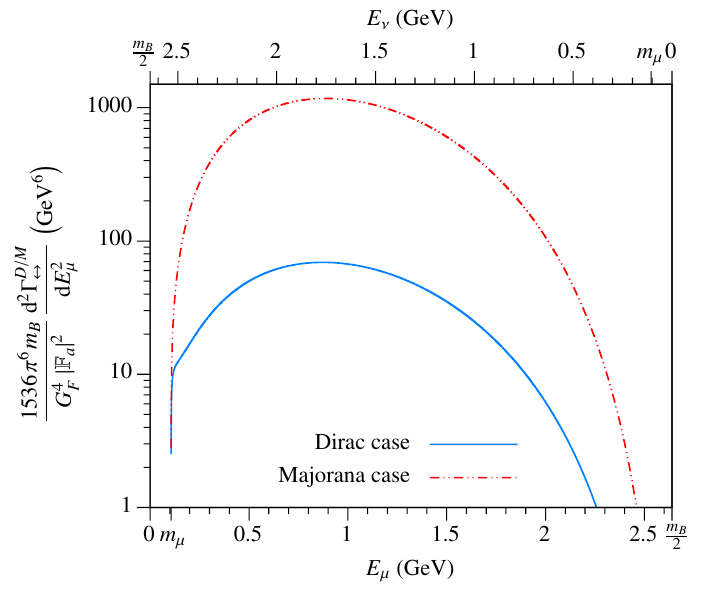}} \hfill%
\subfloat[Comparison of $\sin\theta$ distribution alone between Dirac
and Majorana cases. Compare with
Fig.~\ref{fig:B2B-angular-distribution}.\label{fig:Sth-dist}]{\includegraphics[width=0.495\linewidth,keepaspectratio]{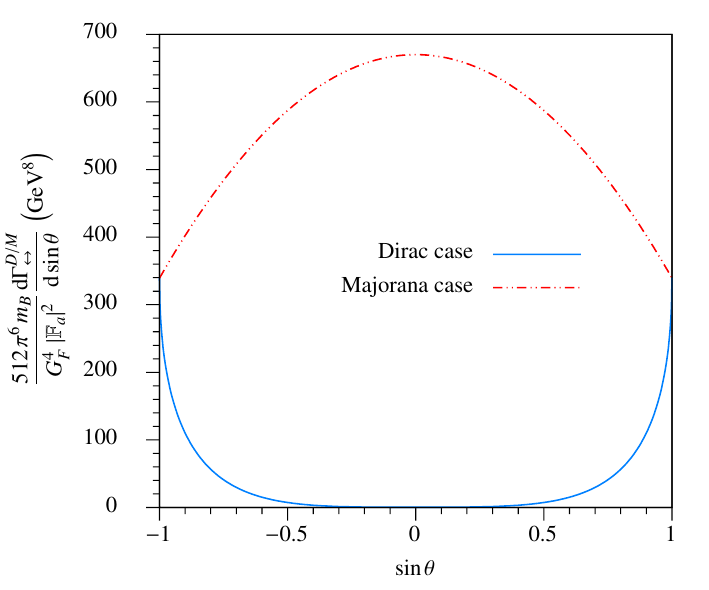}}%
\caption{Comparison of Dirac and Majorana cases via angular
distribution as given in Eq.~\eqref{eq:angular-distributions-b2b}
after being normalized appropriately.}%
\label{fig:angular-distribution}
\end{figure*}
To make a simple numerical estimate of our findings, we now neglect
the masses of muons and neutrinos in comparison with the mass of $B^0$
as well as the energies. We also consider only the non-resonant
contributions in the first approximation (only the $\Sigma_{aa}$,
$\Sigma_{bb}$ and $\Sigma_{ab}$ terms survive when the muon and
neutrino mass dependencies are neglected in comparison with other
terms, see Appendix~\ref{app:sigma-delta}). As an example, let us
consider the dominant contribution that arises from the form factors
$\mathbb{F}^{(\prime)}_a$ alone. For simplicity we also assume it to
be a constant form factor. The full differential back-to-back decay
rates are then given by,
\begin{subequations}\label{eq:angular-distributions-b2b}
\begin{align}
\frac{\mathrm{d}^3\Gamma^{D}_\leftrightarrow}{\mathrm{d}E_\mu^2\,\mathrm{d}\sin\theta} &= \frac{G_F^4 \modulus{\mathbb{F}_a}^2 \left( m_B-2\,E_\mu \right)^4 K_\mu}{512\,\pi^6\,m_B\,E_\mu} \left( E_\mu - K_\mu\,\cos\theta \right)^2,\label{eq:angular-distribution-D-b2b}\\%
\frac{\mathrm{d}^3\Gamma^M_\leftrightarrow}{\mathrm{d}E_\mu^2\,\mathrm{d}\sin\theta} &= \frac{G_F^4 \modulus{\mathbb{F}_a}^2 \left( m_B-2\,E_\mu \right)^4 K_\mu}{512\,\pi^6\,m_B\,E_\mu} \left( E_\mu^2 + K_\mu^2\cos^2\theta \right),\label{eq:angular-distribution-M-b2b}
\end{align}
\end{subequations}
where $K_\mu = \sqrt{E_\mu^2 -  m_\mu^2}$ is the magnitude of the
3-momentum of the back-to-back muons. There are no $m_\nu$ dependent
terms here. These distributions are shown in
Fig.~\ref{fig:angular-distribution}. In Figs.~\ref{fig:3D-D} and
\ref{fig:3D-M} the full distributions are shown. The one dimensional
muon energy distribution obtained after integrating over $\sin\theta$
is shown in \ref{fig:energy-dist}, which is also the one dimensional
neutrino energy distribution (see
Eq.~\eqref{eq:B2B-energies-relation}), and the angular distribution
with respect to $\sin\theta$ alone is shown in
Fig.~\ref{fig:B2B-angular-distribution}. If we neglect $m_\mu$ as
well, it is easy to see from Eq.~\eqref{eq:angular-distributions-b2b}
that         
\begin{subequations}
\begin{align}
\frac{\mathrm{d}^3\Gamma^{D}_\leftrightarrow}{\mathrm{d}E_\mu^2\,\mathrm{d}\sin\theta} &=\frac{G_F^4\,\modulus{\mathbb{F}_a}^2\,\left( m_B-2\,E_\mu \right)^4\,E_\mu^2}{512\,\pi^6\,m_B} \Big(1-\cos\theta\Big)^2,\\%
\frac{\mathrm{d}^3\Gamma^{M}_\leftrightarrow}{\mathrm{d}E_\mu^2\,\mathrm{d}\sin\theta} &=\frac{G_F^4\,\modulus{\mathbb{F}_a}^2\,\left( m_B-2\,E_\mu \right)^4\,E_\mu^2}{512\,\pi^6\,m_B} \Big(1+\cos^2\theta\Big),
\end{align}
\end{subequations}
confirming our expectation in Eqs.~\eqref{eq:B2B-helicity-dist-D},
\eqref{eq:B2B-helicity-dist-M}. The similarity between
Fig.~\ref{fig:B2B-angular-distribution} and Fig.~\ref{fig:Sth-dist} is
unmissable.

Integrating over the currently unobservable angle $\theta$ in
Eq.~\eqref{eq:angular-distributions-b2b} we get the muon energy
distributions for Dirac and Majorana cases,
\begin{subequations}\label{eq:DM-b2b}        
\begin{align}                                
\frac{\mathrm{d}^2\GammaDB}{\mathrm{d}E_\mu^2} &=
\frac{G_F^4\,\modulus{\mathbb{F}_a}^2}{1536\,\pi^6\,m_B\,E_\mu}
\left(m_B-2\,E_\mu\right)^4 K_\mu \nonumber\\%
&\quad \times \left(10\,E_\mu^2-3\,\pi\,E_\mu\,K_\mu-4\,m_\mu^2\right),\label{eq:D-b2b}\\%
\frac{\mathrm{d}^2\GammaMB}{\mathrm{d}E_\mu^2} &=
\frac{G_F^4\,\modulus{\mathbb{F}_a}^2}{1536\,\pi^6\,m_B\,E_\mu}
\left(m_B-2\,E_\mu\right)^4 K_\mu \,
\left(10\,E_\mu^2-4\,m_\mu^2\right),\label{eq:M-b2b}
\end{align}
\end{subequations}
which are shown in Fig.~\ref{fig:energy-dist}. It is also clear that
there is still non-zero difference between muon energy distributions
for Dirac and Majorana cases. Thus, this back-to-back muon energy
distribution can be explored to distinguish between Dirac and Majorana
neutrinos, and this difference is a direct consequence of the
antisymmetrization of the decay amplitude for Majorana neutrinos.
Therefore, the back-to-back muon energy distribution also probes the
quantum statistics of the Majorana neutrinos. While observing the
neutrino energy distribution we are not using
Eq.~\eqref{eq:general-DMCT}, instead we are utilizing
Eq.~\eqref{eq:DTneqET} directly to distinguish between Dirac and
Majorana neutrinos. Also note that the available phase space, as shown
in Fig.~\ref{fig:angular-distribution} is nonnegligible and the effect
of $m_\nu$ on the phase space can be neglected.

\paragraph{Branching ratio and experimental feasibility:}%
The branching ratios of the back-to-back configuration for Dirac and Majorana
cases are estimated to be,
\begin{subequations}\label{eq:branching-ratios}
\begin{align}
\mathcal{B}^D_\leftrightarrow &= \GammaDB/\Gamma_B \approx 1.1 \times
10^{-12} \, \textrm{GeV}^{-2} \times \modulus{\mathbb{F}_a}^2 ,\\%
\mathcal{B}^M_\leftrightarrow &= \GammaMB/\Gamma_B \approx 1.8 \times
10^{-11} \, \textrm{GeV}^{-2} \times \modulus{\mathbb{F}_a}^2,
\end{align}
\end{subequations}
where $\mathbb{F}_a$ has mass dimension 1 (expressed in GeV), and
$\Gamma_B$ is the total decay rate of the $B^0$ meson. Adding the
$\bar{B}^0$ mode would double the statistics. The branching ratios in
Eq.~\eqref{eq:branching-ratios} are very small, and at present with
about $4.8\times 10^8$ fully reconstructable $B$ decays at Belle~II
\cite{Kou:2018nap} it is not possible to observe these back-to-back
events. If in addition to the muon mode $B^0 \,\big(\bar{B}^0\big) \to
\mu^-\,\mu^+\,\nu_\mu\,\bar{\nu}_\mu$ one also considers the electron
mode $B^0 \,\big(\bar{B}^0\big) \to e^+\,e^-\,\nu_e\,\bar{\nu}_e$, \textit{the
statistics could be increased four fold}, such that the next generation
of $B$ factories might start to investigate these. One would probably
require a very high-luminosity $B$ factory to experimentally probe
this back-to-back configuration. Note that the $B$ decay considered
here is only one out of many possible modes that can be exploited
which we discuss in more detail in Sec.~\ref{sec:other-decay}.
Therefore, the apparent experimental difficulty of observing our
example $B$ decay must be considered in this context.

\paragraph{Background processes:} Since flavor changing neutral
current is absent at tree-level in the SM, we do not have background
events that are of the same order for the $B^0$ decay under our
consideration. Nevertheless, there are two possible $B^0$ decays that
can mimic the final state experimental signature of
$\mu^-\,\mu^+\,+\text{``missing momentum''}$:
\begin{enumerate}
\item $B^0 \to \tau^+\,\nu_\tau\,\mu^-\,\bar{\nu}_\mu \to \mu^-\,\mu^+\,\nu_\mu\,\bar{\nu}_\mu\,\nu_\tau\,\bar{\nu}_\tau$, and
\item $B^0 \to \tau^+\,\tau^- \to \mu^-\,\mu^+\,\nu_\mu\,\bar{\nu}_\mu\,\nu_\tau\,\bar{\nu}_\tau$.
\end{enumerate}
Both these decays involve (i) additional weak vertices, and (ii) phase
space suppression due to six final particles when compared with the
signal that has four final particles. Thus these decays are further
suppressed in comparison with the doubly weak signal decay mode. One
can therefore safely neglect these background processes.

\paragraph{For the future:}%
Though at present, it is not possible to detect and measure the
$4$-momentum of neutrinos at their place of origin, one might consider
a future where technological advancements could make this feasible.
Such futuristic detectors dedicated to neutrino detection might as
well follow the trend of additional detector set-ups such as FASER
\cite{Feng:2017uoz, Ariga:2018zuc, Ariga:2018uku, Ariga:2018pin,
Abreu:2019yak} or CODEX-b \cite{Aielli:2019ivi} at the LHC, or the
proposed MATHUSLA \cite{Chou:2016lxi, Alpigiani:2018fgd,
Curtin:2018mvb, Lubatti:2019vkf, Alpigiani:2020tva} and SHiP
\cite{Bonivento:2013jag, Anelli:2015pba, Alekhin:2015byh} detectors at
the high luminosity LHC, or the proposed GAZELLE \cite{Gazelle}
detector at Belle II. Such futuristic detectors could enable us to
directly probe the angular distribution of Fig.~\ref{fig:Sth-dist},
which dramatically shows the difference between Dirac and Majorana
neutrinos. Finally we note that, if the angle $\theta$ could be measured, then in addition
to the difference in back-to-back branching ratios as well as the muon
energy distributions of Fig.~\ref{fig:energy-dist} for Dirac and Majorana cases, one can also probe
whether the number of events increases away from $\theta=0$ or not.
From Fig.~\ref{fig:Sth-dist} the angular distribution for Majorana (or
Dirac) case exhibits a down-ward (or up-ward) trend while going away
from $\theta=0$.

\section{Discussion on other possible decay modes}\label{sec:other-decay}

From the discussion following Eq.~\eqref{eq:branching-ratios} it is
clear that the study of the back-to-back kinematics of the decay
$B^0\to \mu^-\,\mu^+\,\nu_\mu\,\bar{\nu}_\mu$ will have to wait for
future experimental advancement (along with additional theoretical
knowledge about the various form factors). Thus, it might be helpful
to identify some other potential decay modes which could exhibit
similar signatures as what we have found in the $B^0$ decay here.

One fine possibility would be to consider the Higgs decay $H \to
W^{(*)} W^* \to \mu^- \mu^+ \nu_\mu \bar{\nu}_\mu$. This could help us
avoid the consideration of the unknown form factors all together.
Nevertheless, a careful consideration of the Higgs decay mode presents
the following challenges.

\begin{enumerate}
\item \textit{Initial 4-momentum of the Higgs boson:} The initial
4-momentum of the Higgs must be known before. This is probably
achievable in an $e^- e^+$ collider tuned to produce the Higgs boson
at rest. Therefore, the study of the Higgs decay under consideration
is not feasible in any ongoing experiment such as the LHC where the
initial 4-momentum of the Higgs bosons varies.%
\item \textit{Background processes:} The final state $\mu^-\mu^+
\nu_\mu \bar{\nu}_\mu$ can also arise from other Higgs decays, such as
the sequential decay involving two $Z$ bosons, $H \to Z^{(*)}Z^* \to
\mu^-\mu^+ \nu_\mu \bar{\nu}_\mu$. Moreover, since the two neutrinos
are missing, one indeed needs to distinguish the signal events from
the dominant background coming from decays of $\tau$ arising from
Higgs decay, $H \to \tau^- \tau^+ \to \mu^- \mu^+ \nu_\tau
\bar{\nu}_\tau \nu_\mu \bar{\nu}_\mu$ which also has the final
signature of $\mu^- \mu^+ +$~``missing''. It is easy to throw away
on-shell $Z$ contributions by studying the invariant mass square of
the final muon pair or neutrino pair. However, there is no such
strategy to throw away the $H \to \tau^- \tau^+$ mediated events,
though such background is expected to be low in comparison with the
signal decay. The major background events from two off-shell $Z$
bosons would imply that additional Feynman diagrams must be taken into
consideration, and it is not be possible to obtain analytical results
for the Higgs decay by making simple substitutions in
Eq.~\eqref{eq:DiffDecayRate-DmM}. It must be noted that the $B$ decay
we have considered before is free from such background processes in
the SM due to absence of flavor changing neutral currents at tree
level.
\end{enumerate}

From these considerations, we therefore conclude that it is not
straightforward to apply our results from Sec.~\ref{sec:B-decay} in
the case of the Higgs decay $H \to \mu^- \mu^+ \nu_\mu \bar{\nu}_\mu$.
Since the Higgs decay can probe a much larger set of heavy neutrino
scenarios than the $B$ decay of Sec.~\ref{sec:B-decay}, it would be
interesting to study the Higgs decay. However, a detailed study of the
Higgs decay mode to differentiate Majorana neutrinos from Dirac
neutrinos is beyond the scope of this paper and is thus reserved for a
future work.

In addition to the $B$ meson and Higgs ($H$) decays to $\mu^- \mu^+
\nu_\mu \bar{\nu}_\mu$ final state, one can also consider some other
decay modes, such as $D \to \mu^- \mu^+ \nu_\mu \bar{\nu}_\mu$ (with
dominant $K$ pole contribution), $J/\psi \to \mu^- \mu^+ \nu_\mu
\bar{\nu}_\mu$ (involving the $WW\gamma$ vertex), and $\psi(2S) \to
\pi^+\pi^-\nu_\tau \bar{\nu}_\tau$ (with dominant $\tau$ pole
contributions). The $D$ decay can be analyzed in exactly the same
fashion as the $B$ decay we have considered in Sec.~\ref{sec:B-decay}.
Analogous calculations can be undertaken for the decays of $J/\psi$
and $\psi(2S)$ with the later probing the Majorana nature of
tau-neutrino.

A very interesting possibility from the experimental perspective could
be the kaon decay $K^0 \to \mu^-\,\mu^+\,\nu_\mu\,\bar{\nu}_\mu$. If
we were to consider the special case of contribution from the form
factors $\mathbb{F}_a^{(\prime)}$ alone, then we would get the
following branching ratios for the decays of $K^0_S$ and $K^0_L$ for
the back-to-back muons configuration in the kaon rest frame:
\begin{subequations}\label{eq:kaon-BR-B2B}     
\begin{align}
\mathcal{B}_{\leftrightarrow}^D \Big( K^0_S \to
\mu^-\,\mu^+\,\nu_\mu\,\bar{\nu}_\mu \Big) &= 1.6 \times 10^{-18} \,
\textrm{GeV}^{-2} \times \modulus{\mathbb{F}_a}^2,\\*%
\mathcal{B}_{\leftrightarrow}^M \Big( K^0_S \to
\mu^-\,\mu^+\,\nu_\mu\,\bar{\nu}_\mu \Big) &= 7.0 \times 10^{-18} \,
\textrm{GeV}^{-2} \times \modulus{\mathbb{F}_a}^2,\\*%
\mathcal{B}_{\leftrightarrow}^D \Big( K^0_L \to
\mu^-\,\mu^+\,\nu_\mu\,\bar{\nu}_\mu \Big) &= 9.2 \times 10^{-16} \,
\textrm{GeV}^{-2} \times \modulus{\mathbb{F}_a}^2,\\*%
\mathcal{B}_{\leftrightarrow}^M \Big( K^0_L \to
\mu^-\,\mu^+\,\nu_\mu\,\bar{\nu}_\mu \Big) &= 4.0 \times 10^{-15} \,
\textrm{GeV}^{-2} \times \modulus{\mathbb{F}_a}^2.%
\end{align}
\end{subequations}
The branching ratios given in Eq.~\eqref{eq:kaon-BR-B2B} are much
smaller than those for the $B^0$ decay as given in
Eq.~\eqref{eq:branching-ratios} because of the much reduced
phase-space for the kaon decay. Nevertheless, kaons are both
relatively much easier to produce in experiments and with extremely
larger numbers than the $B^0$ mesons. This might make the kaon decays
experimentally more accessible in near future.

Thus, we have found that one can think of many other decay modes which
can potentially be tapped in a manner similar to the $B^0$ mode we
have studied in this paper to probe the Dirac or Majorana nature of
the neutrinos, light as well as heavy, if they exist. One could, in
principle, extend our formalism to explore the Majorana nature of
heavy neutrinos, supersymmetric neutralinos, or any other exotic
electrically neutral fermions.

\section{Conclusion}\label{sec:conclusion}

In this paper, we have presented a technique, which is complimentary
to lepton number violating processes, to probe the Majorana nature of
neutrino. It is based on the idea of implementing the Fermi-Dirac
statistics and hence requires presence of a neutrino antineutrino pair
in the final state. We consider specifically the $B$ meson decay $B
\to \mu^- \mu^+ \nu_\mu \bar{\nu}_\mu$, taking both resonant and
non-resonant contributions simultaneously, in a very generalized
manner. We consider the most general vertex factor for the $B \to
W^*W^*$ vertex, involving three presently unknown, complex, transition
form factors. The differential decay rates for Dirac and Majorana
cases are expressed in terms of five independent variables: two mass
squares and three angles. If we integrate over the neutrino and     
antineutrino momenta completely, the difference between the
differential decay rates for Dirac and Majorana cases is
\textit{non-zero} albeit being directly proportional to the square of
the neutrino mass. Nevertheless, this difference is in agreement with
the `practical Dirac Majorana confusion theorem' which states that the
difference between Dirac and Majorana cases would vanish when the mass
of neutrino goes to zero. This mass dependence would, nevertheless,
favour heavy neutrino scenarios more than the active neutrinos, if
they exist in the kinematically allowed mass range.

We have demonstrated that it is possible that there can exist striking
difference between Dirac and Majorana cases which do not depend on the
mass of the neutrino, if we consider the special kinematic
configuration of back-to-back muons in the $B^0$ rest frame. Thus we
can measure directly the neutrino energy distribution to see the
difference between Dirac and Majorana neutrinos. The unknown
non-resonant transition form factors imply that a proper numerical
study of the $B$ decay process  must be carried out. This is under our
active consideration. Finally we note that the study of a similar
decay of the Higgs $H \to \mu^- \mu^+ \nu_\mu \bar{\nu}_\mu$ is much
more complicated with contributions from $W$ mediated and $Z$ mediated
channels as well as background contributions from $\tau$ decays that
arise from $H \to \tau^- \tau^+$ mode. This puts meson decays such as
the decays of $B,D,J/\psi,\psi(2S)$ at a unique position that these
are free from such background processes in the SM. Our approach
proposed in this paper is important from the point of view that our
methodology probes the Majorana nature of neutrinos by exploiting
their quantum statistics which is a fundamental property.

\section*{Acknowledgments}%
The work of CSK is supported by NRF of Korea (NRF-2021R1A4A2001897,
NRF-2022R1I1A1A01055643). MVNM thanks G Rajasekaran, Rahul Sinha and
the members of the Journal Club for comments and criticisms. DS is
thankful to Manimala Mitra for discussions, and to Minakshi Nayak for
bringing the GAZELLE detector to our attention. The authors are
grateful to Claudio Dib, Kaoru Hagiwara, Anjan Joshipura and Alexei
Smirnov for their critical comments.

\appendix

\section{The various terms of Eqs.~\eqref{eq:DAmpSq} and \eqref{eq:MAmpSq}}\label{appendix:terms}
             
In Eqs.~\eqref{eq:DAmpSq} and \eqref{eq:MAmpSq} the following terms vanish
\begin{align}
R^M_{ca'} &= R^M_{ac'} = R^{D/M}_{cp} = R^M_{c'p'} = I^M_{aa'} =
I^M_{ba'} = I^M_{ab'} = I^M_{bb'} \nonumber\\%
&= I^M_{cb'} = I^M_{bc'} = I^{D/M}_{ap} = I^{D/M}_{bp} = I^M_{a'p} =
I^{D/M}_{bm} = I^M_{a'm} \nonumber\\%
&= I^{D/M}_{pm} = I^M_{ap'} = I^M_{a'p'} = I^M_{b'p'} = I^M_{p'm'} =
I^M_{am'} = I^M_{b'm'} = 0.\label{eq:vanishing-terms}
\end{align}
The remaining $20$ non-vanishing terms for Dirac case as well as the
$70$ non-vanishing terms for Majorana case can be expressed by using
the following expressions involving the 4-momenta of the final
particles in the rest frame of the parent $B$ meson and in terms of
$m_{\mu\mu}^2$, $m_{\nu\nu}^2$, $\theta_m$, $\theta_n$ and $\phi$,
\begin{align}                                      
p_1 \cdot p_\pm &= \mp Y_m\,Y_n\,\sin\theta_m\,\sin\theta_n\,\cos
\phi \nonumber\\*%
&\quad \pm \frac{Y_m\,Y_n\,\left(m_B^2-
m_{\mu\mu}^2-m_{\nu\nu}^2\right)}{2\,m_{\nu\nu}\,m_{\mu\mu}}\,\cos\theta_m\,\cos\theta_n \nonumber\\*%
&\quad
+\frac{Y\,Y_n\,m_B}{2\,m_{\nu\nu}}\cos\theta_n \pm \frac{Y\,Y_m\,m_B}{2\,m_{\mu\mu}}\cos\theta_m \nonumber\\*%
&\quad +\frac{1}{8}\left(m_B^2-m_{\mu\mu}^2- m_{\nu\nu}^2\right),\\%
p_2 \cdot p_\pm &= \pm Y_m\,Y_n\,\sin\theta_m\,\sin\theta_n\,\cos
\phi \nonumber\\*%
&\quad \mp \frac{Y_m\,Y_n\,\left(m_B^2-
m_{\mu\mu}^2-m_{\nu\nu}^2\right)}{2\,m_{\nu\nu}\,m_{\mu\mu}}\,\cos\theta_m\,\cos\theta_n \nonumber\\*%
&\quad
-\frac{Y\,Y_n\,m_B}{2\,m_{\nu\nu}}\cos\theta_n \pm \frac{Y\,Y_m\,m_B}{2\,m_{\mu\mu}}\cos\theta_m \nonumber\\*%
&\quad +\frac{1}{8}\left(m_B^2-m_{\mu\mu}^2- m_{\nu\nu}^2\right),\\%
\mathbb{O} &= -Y\,Y_m\,Y_n\,m_B\,\sin\theta_m\,\sin\theta_n \,\sin
\phi.
\end{align}
Using these, the $20$ non-vanishing terms common to both
Eq.~\eqref{eq:DAmpSq} and Eq.~\eqref{eq:MAmpSq} are given by,
\begin{widetext}
\begin{align}
S^{D/M}_{aa} &= 64\,\Big( p_1 \cdot p_+ \Big)\,\Big( p_2 \cdot p_- \Big),\label{eq:term-first}\\%
S^{D/M}_{bb} &= 4\,\Bigg(2\,\left(2\,\Big(p_1 \cdot p_+\Big)\, \Big(p_2 \cdot
p_-\Big)+2\,\Big(p_1 \cdot p_-\Big)\, \Big(p_2 \cdot p_-\Big)+2\,\Big(p_1 \cdot
p_-\Big)\, \Big(p_1 \cdot p_+\Big)+2\,\Big(p_1 \cdot p_-\Big)^2-m_B^2\, \Big(p_1
\cdot p_-\Big)\right)\nonumber\\*%
&\qquad +2\,m_{\nu\nu}^2\,\left( \Big(p_2 \cdot p_-\Big)+\Big(p_1 \cdot
p_-\Big)\right)+2\,m_{\mu\mu} ^2\,\left(\Big(p_1 \cdot p_+\Big)+\Big(p_1 \cdot
p_-\Big)\right)+ m_{\nu\nu}^2\,m_{\mu\mu}^2\Bigg)\nonumber\\*%
&\quad \times \Bigg(2\,\left(2\, \Big(p_2 \cdot p_+\Big)^2+2\,\Big(p_2 \cdot
p_-\Big)\, \Big(p_2 \cdot p_+\Big)+2\,\Big(p_1 \cdot p_+\Big)\, \Big(p_2 \cdot
p_+\Big)-m_B^2\,\Big(p_2 \cdot p_+\Big)+2\, \Big(p_1 \cdot p_+\Big)\,\Big(p_2
\cdot p_-\Big)\right)\nonumber\\*%
&\qquad +2\, m_{\mu\mu}^2\,\left(\Big(p_2 \cdot p_+\Big)+\Big(p_2 \cdot p_-\Big)
\right)+2\,m_{\nu\nu}^2\,\left(\Big(p_2 \cdot p_+\Big)+ \Big(p_1 \cdot
p_+\Big)\right)+m_{\nu\nu}^2\,m_{\mu\mu}^2\Bigg),\\%
S^{D/M}_{cc} &= -4\,\Bigg(8\,m_\nu^2\,\bigg(\Big(p_1 \cdot p_-\Big)\, \Big(p_2 \cdot
p_+\Big)^2-2\,\Big(p_2 \cdot p_-\Big)^2\, \Big(p_2 \cdot p_+\Big)+\Big(p_1 \cdot
p_+\Big)\, \Big(p_2 \cdot p_-\Big)\,\Big(p_2 \cdot p_+\Big)\nonumber\\%
&\hspace{2cm} + \Big(p_1 \cdot p_-\Big)^2\,\Big(p_2 \cdot p_+\Big)+ \Big(p_1
\cdot p_-\Big)\,\Big(p_1 \cdot p_+\Big)\, \Big(p_2 \cdot p_-\Big)-2\,\Big(p_1
\cdot p_-\Big)\, \Big(p_1 \cdot p_+\Big)^2\bigg)\nonumber\\%
&\hspace{1cm} +8\,m_\mu^2\,\bigg( \Big(p_1 \cdot p_-\Big)\,\Big(p_2 \cdot
p_+\Big)^2+ \Big(p_1 \cdot p_+\Big)\,\Big(p_2 \cdot p_-\Big)\, \Big(p_2 \cdot
p_+\Big)-2\,\Big(p_1 \cdot p_+\Big)^2\, \Big(p_2 \cdot p_+\Big) \nonumber\\%
&\hspace{2cm} +\Big(p_1 \cdot p_-\Big)^2\, \Big(p_2 \cdot p_+\Big)-2\,\Big(p_1
\cdot p_-\Big)\, \Big(p_2 \cdot p_-\Big)^2+\Big(p_1 \cdot p_-\Big)\, \Big(p_1
\cdot p_+\Big)\,\Big(p_2 \cdot p_-\Big)\bigg)\nonumber\\%
&\hspace{1cm} -8\,m_\mu^2\, m_\nu^2\,\bigg(3\,\Big(p_2 \cdot p_-\Big)\,\Big(p_2
\cdot p_+\Big)+3 \,\Big(p_1 \cdot p_+\Big)\,\Big(p_2 \cdot p_+\Big)+2\, \Big(p_1
\cdot p_-\Big)\,\Big(p_2 \cdot p_+\Big)+3\, \Big(p_2 \cdot
p_-\Big)^2\nonumber\\%
&\hspace{3cm} +4\,\Big(p_1 \cdot p_+\Big)\, \Big(p_2 \cdot p_-\Big)+3\,\Big(p_1
\cdot p_-\Big)\, \Big(p_2 \cdot p_-\Big)+3\,\Big(p_1 \cdot p_+\Big)^2+3\,
\Big(p_1 \cdot p_-\Big)\,\Big(p_1 \cdot p_+\Big)\bigg)\nonumber\\%
&\hspace{1cm}+4\, m_{\mu\mu}^2\,m_\nu^2\,\bigg(2\,\Big(p_2 \cdot p_-\Big)\,
\Big(p_2 \cdot p_+\Big)+\Big(p_1 \cdot p_+\Big)\, \Big(p_2 \cdot
p_+\Big)+4\,\Big(p_1 \cdot p_-\Big)\, \Big(p_2 \cdot p_+\Big)\nonumber\\%
&\hspace{3cm} +4\,\Big(p_1 \cdot p_+\Big)\, \Big(p_2 \cdot p_-\Big)+\Big(p_1
\cdot p_-\Big)\, \Big(p_2 \cdot p_-\Big)+2\,\Big(p_1 \cdot p_-\Big)\, \Big(p_1
\cdot p_+\Big)\bigg)\nonumber\\%
&\qquad +4\,m_{\nu\nu}^2\,m_\mu^2\,\bigg( \Big(p_2 \cdot p_-\Big)\,\Big(p_2
\cdot p_+\Big)+2\, \Big(p_1 \cdot p_+\Big)\,\Big(p_2 \cdot p_+\Big)+4\, \Big(p_1
\cdot p_-\Big)\,\Big(p_2 \cdot p_+\Big)\nonumber\\%
&\hspace{3cm}+4\, \Big(p_1 \cdot p_+\Big)\,\Big(p_2 \cdot p_-\Big)+2\, \Big(p_1
\cdot p_-\Big)\,\Big(p_2 \cdot p_-\Big)+ \Big(p_1 \cdot p_-\Big)\,\Big(p_1 \cdot
p_+\Big)\bigg)\nonumber\\%
&\qquad -4\, m_{\nu\nu}^2\,m_\nu^2\,\bigg(\Big(p_2 \cdot p_-\Big)\, \Big(p_2
\cdot p_+\Big)+\Big(p_1 \cdot p_-\Big)\, \Big(p_1 \cdot
p_+\Big)\bigg)\nonumber\\%
&\qquad -4\,m_{\mu\mu}^2\,m_\mu^2\,\bigg( \Big(p_1 \cdot p_+\Big)\,\Big(p_2
\cdot p_+\Big)+ \Big(p_1 \cdot p_-\Big)\,\Big(p_2 \cdot
p_-\Big)\bigg)\nonumber\\%
&\qquad +16\,\left( \Big(p_1 \cdot p_-\Big)\,\Big(p_2 \cdot p_+\Big)- \Big(p_1
\cdot p_+\Big)^2\right)\,\left(\Big(p_1 \cdot p_-\Big)\, \Big(p_2 \cdot
p_+\Big)-\Big(p_2 \cdot p_-\Big)^2\right)\nonumber\\%
&\qquad -8\, m_{\nu\nu}^2\,m_{\mu\mu}^2\,\left(\Big(p_1 \cdot p_-\Big)\,
\Big(p_2 \cdot p_+\Big)+\Big(p_1 \cdot p_+\Big)\, \Big(p_2 \cdot
p_-\Big)\right)\nonumber\\%
&\qquad +4\,m_{\mu\mu}^2\,m_\nu^4\,\left( \Big(p_2 \cdot p_+\Big)+\Big(p_2 \cdot
p_-\Big)+ \Big(p_1 \cdot p_+\Big)+\Big(p_1 \cdot p_-\Big)\right)\nonumber\\%
&\qquad +4\,m_{\nu\nu} ^2\,m_\mu^4\,\left(\Big(p_2 \cdot p_+\Big)+\Big(p_2 \cdot
p_-\Big)+ \Big(p_1 \cdot p_+\Big)+\Big(p_1 \cdot p_-\Big)\right)\nonumber\\%
&\qquad +8\,m_\mu^4\, \left(\Big(p_1 \cdot p_+\Big)+\Big(p_1 \cdot
p_-\Big)\right)\,\left( \Big(p_2 \cdot p_+\Big)+\Big(p_2 \cdot
p_-\Big)\right)\nonumber\\%
&\qquad +4\,m_{\mu\mu} ^2\,m_\mu^2\,m_\nu^2\,\left(\Big(p_2 \cdot p_+\Big)-
\Big(p_2 \cdot p_-\Big)-\Big(p_1 \cdot p_+\Big)+ \Big(p_1 \cdot
p_-\Big)\right)\nonumber\\%
&\qquad +4\,m_{\nu\nu}^2\,m_\mu^2\,m_\nu^2\, \left(\Big(p_2 \cdot
p_+\Big)-\Big(p_2 \cdot p_-\Big)- \Big(p_1 \cdot p_+\Big)+\Big(p_1 \cdot
p_-\Big)\right)\nonumber\\%
&\qquad +8\,m_\nu^4\, \left(\Big(p_2 \cdot p_-\Big)+\Big(p_1 \cdot
p_-\Big)\right)\,\left( \Big(p_2 \cdot p_+\Big)+\Big(p_1 \cdot
p_+\Big)\right)\nonumber\\%
&\qquad -2\,m_{\nu\nu} ^2\,m_{\mu\mu}^2\,m_\nu^2\,\left(\Big(p_2 \cdot p_+\Big)+
\Big(p_1 \cdot p_-\Big)\right)-2\,m_{\nu\nu}^2\,m_{\mu\mu}^2\,m_\mu^
2\,\left(\Big(p_2 \cdot p_+\Big)+\Big(p_1 \cdot p_-\Big)\right)\nonumber\\%
&\qquad +2\,
m_{\mu\mu}^4\,m_\nu^4+12\,m_{\nu\nu}^2\,m_{\mu\mu}^2\,m_\mu^2\,m_\nu
^2-4\,m_{\nu\nu}^2\,m_{\mu\mu}^4\,m_\nu^2+2\,m_{\nu\nu}^4\,m_\mu^4-4
\,m_{\nu\nu}^4\,m_{\mu\mu}^2\,m_\mu^2+m_{\nu\nu}^4\,m_{\mu\mu}^4 \Bigg),\\%
S^{D/M}_{pp} &= 16\,\left(m_\nu^2\,\Big(p_1 \cdot p_-\Big)+m_\mu^2\, \Big(p_1 \cdot
p_-\Big)+2\,m_\mu^2\,m_\nu^2\right)\,\left(m_\nu^2\, \Big(p_2 \cdot
p_+\Big)+m_\mu^2\,\Big(p_2 \cdot p_+\Big)+2\,m_\mu^2 \,m_\nu^2\right),\\%
S^{D/M}_{mm} &= -8\,\left(m_\nu^2\,\Big(p_1 \cdot p_-\Big)+m_\mu^2\, \Big(p_1 \cdot
p_-\Big)+2\,m_\mu^2\,m_\nu^2\right)\,\bigg(4\,\left( \Big(p_1 \cdot
p_-\Big)\,\Big(p_2 \cdot p_+\Big)- \Big(p_1 \cdot p_+\Big)\,\Big(p_2 \cdot
p_-\Big)\right)\nonumber\\*%
&\qquad +2\,m_\mu^2\, \left(\Big(p_2 \cdot p_+\Big)+2\,\Big(p_2 \cdot
p_-\Big)\right)+2\, m_\nu^2\,\left(\Big(p_2 \cdot p_+\Big)+2\,\Big(p_1 \cdot
p_+\Big) \right)-2\,m_{\mu\mu}^2\,\Big(p_2 \cdot p_-\Big)-2\,m_{\nu\nu}^2\,
\Big(p_1 \cdot p_+\Big)\nonumber\\*%
&\qquad -4\,m_\mu^2\,m_\nu^2+2\,m_{\mu\mu}^2\,m_\nu^2
+2\,m_{\nu\nu}^2\,m_\mu^2-m_{\nu\nu}^2\,m_{\mu\mu}^2\bigg),\\%
R^{D/M}_{ab} &= -8\,\Bigg(4\,\bigg(2\,\Big(p_1 \cdot p_-\Big)\, \Big(p_2 \cdot
p_+\Big)^2+2\,\Big(p_1 \cdot p_+\Big)\, \Big(p_2 \cdot p_-\Big)\,\Big(p_2 \cdot
p_+\Big)+4\, \Big(p_1 \cdot p_-\Big)\,\Big(p_2 \cdot p_-\Big)\, \Big(p_2 \cdot
p_+\Big)\nonumber\\%
&\qquad +4\,\Big(p_1 \cdot p_-\Big)\, \Big(p_1 \cdot p_+\Big)\,\Big(p_2 \cdot
p_+\Big)+2\, \Big(p_1 \cdot p_-\Big)^2\,\Big(p_2 \cdot p_+\Big)-m_B^2\, \Big(p_1
\cdot p_-\Big)\,\Big(p_2 \cdot p_+\Big)\nonumber\\%
&\qquad +2\, \Big(p_1 \cdot p_-\Big)\,\Big(p_1 \cdot p_+\Big)\, \Big(p_2 \cdot
p_-\Big)-m_B^2\,\Big(p_1 \cdot p_+\Big)\, \Big(p_2 \cdot
p_-\Big)\bigg)\nonumber\\%
&\qquad +4\,m_{\mu\mu}^2\,\left( \Big(p_1 \cdot p_-\Big)\,\Big(p_2 \cdot
p_+\Big)- \Big(p_1 \cdot p_+\Big)\,\Big(p_2 \cdot p_-\Big)\right)+4\,
m_{\nu\nu}^2\,\left(\Big(p_1 \cdot p_-\Big)\,\Big(p_2 \cdot p_+\Big) -\Big(p_1
\cdot p_+\Big)\,\Big(p_2 \cdot p_-\Big)\right)\nonumber\\%
&\qquad -m_{\nu\nu}^ 2\,m_{\mu\mu}^2\,\left(2\,\Big(p_2 \cdot p_+\Big)+4\,
\Big(p_2 \cdot p_-\Big)+4\,\Big(p_1 \cdot p_+\Big)+2\, \Big(p_1 \cdot
p_-\Big)-m_B^2\right)\nonumber\\%
&\qquad +2\,m_{\mu\mu}^2\,m_\nu^2\, \left(2\,\Big(p_2 \cdot p_+\Big)+2\,\Big(p_2
\cdot p_-\Big)+2\, \Big(p_1 \cdot p_+\Big)+2\,\Big(p_1 \cdot
p_-\Big)-m_B^2\right)\nonumber\\%
&\qquad +2\, m_{\nu\nu}^2\,m_\mu^2\,\left(2\,\Big(p_2 \cdot p_+\Big)+2\,
\Big(p_2 \cdot p_-\Big)+2\,\Big(p_1 \cdot p_+\Big)+2\, \Big(p_1 \cdot
p_-\Big)-m_B^2\right)\nonumber\\%
&\qquad +8\,m_\mu^2\,\left( \Big(p_1 \cdot p_+\Big)+\Big(p_1 \cdot
p_-\Big)\right)\,\left( \Big(p_2 \cdot p_+\Big)+\Big(p_2 \cdot
p_-\Big)\right)\nonumber\\%
&\qquad +8\,m_\nu^2\, \left(\Big(p_2 \cdot p_-\Big)+\Big(p_1 \cdot
p_-\Big)\right)\,\left( \Big(p_2 \cdot p_+\Big)+\Big(p_1 \cdot
p_+\Big)\right)\nonumber\\%
&\qquad +4\,m_\mu^2\,
m_\nu^2\,m_B^2+2\,m_{\mu\mu}^4\,m_\nu^2+2\,m_{\nu\nu}^4\,m_\mu^2-
m_{\nu\nu}^2\,m_{\mu\mu}^4-m_{\nu\nu}^4\,m_{\mu\mu}^2\Bigg),\\%
R^{D/M}_{ac} &= -16\,\Bigg(4\,m_\nu^2\,\left(\Big(p_2 \cdot p_-\Big)\, \Big(p_2
\cdot p_+\Big)-\Big(p_1 \cdot p_-\Big)\, \Big(p_1 \cdot
p_+\Big)\right)-4\,m_\mu^2\,\left( \Big(p_1 \cdot p_+\Big)\,\Big(p_2 \cdot
p_+\Big)- \Big(p_1 \cdot p_-\Big)\,\Big(p_2 \cdot p_-\Big)\right)\nonumber\\%
&\hspace{1cm} +4\,\left( \Big(p_2 \cdot p_-\Big)-\Big(p_1 \cdot
p_+\Big)\right)\,\left( \Big(p_1 \cdot p_-\Big)\,\Big(p_2 \cdot p_+\Big)+
\Big(p_1 \cdot p_+\Big)\,\Big(p_2 \cdot p_-\Big)\right)\nonumber\\%
&\hspace{1cm} +2\, m_{\mu\mu}^2\,m_\nu^2\,\left(\Big(p_2 \cdot p_-\Big)-
\Big(p_1 \cdot p_+\Big)\right)+2\,m_{\nu\nu}^2\,m_\mu^2\,\left( \Big(p_2 \cdot
p_-\Big)-\Big(p_1 \cdot p_+\Big)\right)\nonumber\\%
&\hspace{1cm} -m_{\nu\nu}^2 \,m_{\mu\mu}^2\,\left(\Big(p_2 \cdot p_-\Big)-
\Big(p_1 \cdot p_+\Big)\right)\Bigg),\\%
R^{D/M}_{bc} &= 8\,\Bigg(4\,m_\nu^2\,\bigg(2\,\Big(p_2 \cdot p_-\Big)\, \Big(p_2
\cdot p_+\Big)^2+4\,\Big(p_2 \cdot p_-\Big)^2\, \Big(p_2 \cdot
p_+\Big)+4\,\Big(p_1 \cdot p_-\Big)\, \Big(p_2 \cdot p_-\Big)\,\Big(p_2 \cdot
p_+\Big)\nonumber\\%
&\hspace{15mm} -m_B^2\, \Big(p_2 \cdot p_-\Big)\,\Big(p_2 \cdot p_+\Big)-4\,
\Big(p_1 \cdot p_-\Big)\,\Big(p_1 \cdot p_+\Big)\, \Big(p_2 \cdot
p_+\Big)+2\,\Big(p_1 \cdot p_+\Big)\, \Big(p_2 \cdot p_-\Big)^2\nonumber\\%
&\hspace{15mm} -2\,\Big(p_1 \cdot p_+\Big)^2\, \Big(p_2 \cdot
p_-\Big)-4\,\Big(p_1 \cdot p_-\Big)\, \Big(p_1 \cdot p_+\Big)^2-2\,\Big(p_1
\cdot p_-\Big)^2\, \Big(p_1 \cdot p_+\Big)+m_B^2\,\Big(p_1 \cdot p_-\Big)\,
\Big(p_1 \cdot p_+\Big)\bigg)\nonumber\\%
&\qquad -4\,m_\mu^2\,\bigg(2\, \Big(p_1 \cdot p_+\Big)\,\Big(p_2 \cdot
p_+\Big)^2-4\, \Big(p_1 \cdot p_-\Big)\,\Big(p_2 \cdot p_-\Big)\, \Big(p_2 \cdot
p_+\Big)+4\,\Big(p_1 \cdot p_+\Big)^2\, \Big(p_2 \cdot p_+\Big)\nonumber\\%
&\hspace{2cm}+4\,\Big(p_1 \cdot p_-\Big)\, \Big(p_1 \cdot p_+\Big)\,\Big(p_2
\cdot p_+\Big)-m_B^2\, \Big(p_1 \cdot p_+\Big)\,\Big(p_2 \cdot p_+\Big)-2\,
\Big(p_1 \cdot p_+\Big)\,\Big(p_2 \cdot p_-\Big)^2\nonumber\\%
&\hspace{2cm}-4\, \Big(p_1 \cdot p_-\Big)\,\Big(p_2 \cdot p_-\Big)^2+2\,
\Big(p_1 \cdot p_+\Big)^2\,\Big(p_2 \cdot p_-\Big)-2\, \Big(p_1 \cdot
p_-\Big)^2\,\Big(p_2 \cdot p_-\Big)\nonumber\\%
&\hspace{2cm}+m_B^2\, \Big(p_1 \cdot p_-\Big)\,\Big(p_2 \cdot
p_-\Big)\bigg)\nonumber\\%
&\qquad +8\,\left( \Big(p_2 \cdot p_-\Big)-\Big(p_1 \cdot
p_+\Big)\right)\,\bigg(2\, \Big(p_1 \cdot p_-\Big)\,\Big(p_2 \cdot
p_+\Big)^2+3\, \Big(p_1 \cdot p_-\Big)\,\Big(p_2 \cdot p_-\Big)\, \Big(p_2 \cdot
p_+\Big)\nonumber\\%
&\hspace{2cm}+3\,\Big(p_1 \cdot p_-\Big)\, \Big(p_1 \cdot p_+\Big)\,\Big(p_2
\cdot p_+\Big)+2\, \Big(p_1 \cdot p_-\Big)^2\,\Big(p_2 \cdot p_+\Big)-m_B^2\,
\Big(p_1 \cdot p_-\Big)\,\Big(p_2 \cdot p_+\Big)\nonumber\\%
&\hspace{2cm}- \Big(p_1 \cdot p_+\Big)\,\Big(p_2 \cdot p_-\Big)^2- \Big(p_1
\cdot p_+\Big)^2\,\Big(p_2 \cdot p_-\Big)\bigg)\nonumber\\%
&\qquad-2\, m_{\nu\nu}^2\,m_\mu^2\,\bigg(2\,\Big(p_2 \cdot p_+\Big)^2+4\,
\Big(p_2 \cdot p_-\Big)\,\Big(p_2 \cdot p_+\Big)+2\, \Big(p_1 \cdot
p_+\Big)\,\Big(p_2 \cdot p_+\Big)-m_B^2\, \Big(p_2 \cdot p_+\Big)+2\,\Big(p_2
\cdot p_-\Big)^2\nonumber\\%
&\hspace{2cm}-2\, \Big(p_1 \cdot p_-\Big)\,\Big(p_2 \cdot p_-\Big)-2\, \Big(p_1
\cdot p_+\Big)^2-4\,\Big(p_1 \cdot p_-\Big)\, \Big(p_1 \cdot
p_+\Big)-2\,\Big(p_1 \cdot p_-\Big)^2+m_B^2\, \Big(p_1 \cdot
p_-\Big)\bigg)\nonumber\\%
&\qquad +2\,m_{\mu\mu}^2\,m_\nu^2\,\bigg(2\, \Big(p_2 \cdot
p_+\Big)^2+2\,\Big(p_2 \cdot p_-\Big)\, \Big(p_2 \cdot p_+\Big)+4\,\Big(p_1
\cdot p_+\Big)\, \Big(p_2 \cdot p_+\Big)-m_B^2\,\Big(p_2 \cdot p_+\Big)-2\,
\Big(p_2 \cdot p_-\Big)^2\nonumber\\%
&\hspace{25mm} -4\,\Big(p_1 \cdot p_-\Big)\, \Big(p_2 \cdot p_-\Big)+2\,\Big(p_1
\cdot p_+\Big)^2-2\, \Big(p_1 \cdot p_-\Big)\,\Big(p_1 \cdot p_+\Big)-2\,
\Big(p_1 \cdot p_-\Big)^2+m_B^2\,\Big(p_1 \cdot p_-\Big)\bigg)\nonumber\\%
&\qquad +4\, m_{\nu\nu}^2\,m_\nu^2\,\left(\Big(p_2 \cdot p_-\Big)\, \Big(p_2
\cdot p_+\Big)-\Big(p_1 \cdot p_-\Big)\, \Big(p_1 \cdot
p_+\Big)\right)\nonumber\\%
&\qquad -4\,m_{\mu\mu}^2\,m_\mu^2\,\left( \Big(p_1 \cdot p_+\Big)\,\Big(p_2
\cdot p_+\Big)- \Big(p_1 \cdot p_-\Big)\,\Big(p_2 \cdot
p_-\Big)\right)\nonumber\\%
&\qquad +4\, \left(m_{\mu\mu}^2+m_{\nu\nu}^2\right)\,\left(\Big(p_2 \cdot
p_-\Big)-\Big(p_1 \cdot p_+\Big) \right)\,\left(\Big(p_1 \cdot
p_-\Big)\,\Big(p_2 \cdot p_+\Big)- \Big(p_1 \cdot p_+\Big)\,\Big(p_2 \cdot
p_-\Big)\right)\nonumber\\%
&\qquad -2\, m_{\nu\nu}^2\,m_{\mu\mu}^2\,m_\mu^2\,\left(\Big(p_2 \cdot
p_+\Big)+2 \,\Big(p_2 \cdot p_-\Big)-2\,\Big(p_1 \cdot p_+\Big)- \Big(p_1 \cdot
p_-\Big)\right)\nonumber\\%
&\qquad +16\,m_\mu^2\,m_\nu^2\,\left( \Big(p_2 \cdot p_-\Big)-\Big(p_1 \cdot
p_+\Big)\right)\,\left( \Big(p_2 \cdot p_+\Big)+\Big(p_2 \cdot p_-\Big)+
\Big(p_1 \cdot p_+\Big)+\Big(p_1 \cdot p_-\Big)\right)\nonumber\\%
&\qquad -2\,m_{\nu\nu} ^4\,m_\mu^2\,\left(\Big(p_2 \cdot p_+\Big)+\Big(p_2 \cdot
p_-\Big)- \Big(p_1 \cdot p_+\Big)-\Big(p_1 \cdot p_-\Big)\right)\nonumber\\%
&\qquad +2\,m_{\mu\mu} ^4\,m_\nu^2\,\left(\Big(p_2 \cdot p_+\Big)-\Big(p_2 \cdot
p_-\Big)+ \Big(p_1 \cdot p_+\Big)-\Big(p_1 \cdot p_-\Big)\right) \nonumber\\%
&\qquad +2\,m_{\nu\nu} ^2\,m_{\mu\mu}^2\,m_\nu^2\,\left(\Big(p_2 \cdot
p_+\Big)-2\, \Big(p_2 \cdot p_-\Big)+2\,\Big(p_1 \cdot p_+\Big)- \Big(p_1 \cdot
p_-\Big)\right)\nonumber\\%
&\qquad +2\,m_{\nu\nu}^2\,m_{\mu\mu}^2\,\left( \Big(p_2 \cdot p_-\Big)-\Big(p_1
\cdot p_+\Big)\right)\,\left( \Big(p_2 \cdot p_-\Big)+\Big(p_1 \cdot
p_+\Big)\right)\nonumber\\%
&\qquad + \left(8\,m_\mu^2\,m_\nu^2 +
m_{\nu\nu}^2\,m_{\mu\mu}^2\right)\,\left(m_{\mu\mu}^2+m_{\nu\nu}^2\right)\,\left(\Big(p_2 \cdot p_-\Big)- \Big(p_1 \cdot p_+\Big)\right)\Bigg),\\%
R^{D/M}_{ap} &= 16\,\left(2\,m_\mu^2\,m_\nu^2\,\left(\Big(p_2 \cdot p_-\Big)+
\Big(p_1 \cdot p_+\Big)\right)-2\,m_\mu^2\,m_\nu^4+m_{\mu\mu}^2\,
m_\nu^4-2\,m_\mu^4\,m_\nu^2+m_{\nu\nu}^2\,m_\mu^4\right),\\%
R^{D/M}_{bp} &= 8\,\left(2\,m_\nu^2\,\left(\Big(p_2 \cdot p_-\Big)+ \Big(p_1 \cdot
p_-\Big)\right)+2\,m_\mu^2\,\left( \Big(p_1 \cdot p_+\Big)+\Big(p_1 \cdot
p_-\Big)\right)+m_{\mu\mu}^2 \,m_\nu^2+m_{\nu\nu}^2\,m_\mu^2\right)\nonumber\\%
&\quad \times \left(2\,m_\mu^2\,\left( \Big(p_2 \cdot p_+\Big)+\Big(p_2 \cdot
p_-\Big)\right)+2\,m_\nu^2\, \left(\Big(p_2 \cdot p_+\Big)+\Big(p_1 \cdot
p_+\Big)\right)+ m_{\mu\mu}^2\,m_\nu^2+m_{\nu\nu}^2\,m_\mu^2\right),\\%
R^{D/M}_{am} &= -8\,\Bigg(4\,m_\nu^2\,\left(\Big(p_1 \cdot p_-\Big)\, \Big(p_2 \cdot
p_+\Big)-\Big(p_1 \cdot p_+\Big)\, \Big(p_2 \cdot
p_-\Big)\right)+4\,m_\mu^2\,\left( \Big(p_1 \cdot p_-\Big)\,\Big(p_2 \cdot
p_+\Big)- \Big(p_1 \cdot p_+\Big)\,\Big(p_2 \cdot p_-\Big)\right)\nonumber\\%
&\hspace{1cm} +8\,m_\mu^2\, m_\nu^2\,\left(\Big(p_2 \cdot p_+\Big)+\Big(p_2
\cdot p_-\Big)+ \Big(p_1 \cdot p_+\Big)\right)-4\,m_{\mu\mu}^2\,m_\nu^2\,
\Big(p_2 \cdot p_-\Big)-4\,m_{\nu\nu}^2\,m_\mu^2\, \Big(p_1 \cdot
p_+\Big)-4\,m_\mu^2\,m_\nu^4\nonumber\\%
&\hspace{1cm} +2\,m_{\mu\mu}^2\,m_\nu^4
-4\,m_\mu^4\,m_\nu^2+2\,m_{\mu\mu}^2\,m_\mu^2\,m_\nu^2+2\,m_{\nu\nu}
^2\,m_\mu^2\,m_\nu^2-m_{\nu\nu}^2\,m_{\mu\mu}^2\,m_\nu^2+2\,
m_{\nu\nu}^2\,m_\mu^4-m_{\nu\nu}^2\,m_{\mu\mu}^2\,m_\mu^2\Bigg),\\%
R^{D/M}_{bm} &= -8\,\left(2\,m_\nu^2\,\left(\Big(p_2 \cdot p_-\Big)+ \Big(p_1 \cdot
p_-\Big)\right)+2\,m_\mu^2\,\left( \Big(p_1 \cdot p_+\Big)+\Big(p_1 \cdot
p_-\Big)\right)+m_{\mu\mu}^2 \,m_\nu^2+m_{\nu\nu}^2\,m_\mu^2\right)\nonumber\\%
&\quad \times \bigg(4\,\left( \Big(p_1 \cdot p_-\Big)\,\Big(p_2 \cdot p_+\Big)-
\Big(p_1 \cdot p_+\Big)\,\Big(p_2 \cdot p_-\Big)\right)+2\,m_\mu^2\,
\left(\Big(p_2 \cdot p_+\Big)+\Big(p_2 \cdot p_-\Big)\right)\nonumber\\%
&\qquad +2\, m_\nu^2\,\left(\Big(p_2 \cdot p_+\Big)+\Big(p_1 \cdot p_+\Big)
\right)-2\,m_{\mu\mu}^2\,\Big(p_2 \cdot p_-\Big)-2\,m_{\nu\nu}^2\, \Big(p_1
\cdot p_+\Big)+m_{\mu\mu}^2\,m_\nu^2+m_{\nu\nu}^2\,m_\mu^2-
m_{\nu\nu}^2\,m_{\mu\mu}^2\bigg),\\%
R^{D/M}_{cm} &= 4\,\left(m_\mu^2-m_\nu^2\right)\,\bigg(-4\,\left( m_{\mu\mu}^2 +
m_{\nu\nu}^2\right)\,\left(\Big(p_1 \cdot p_-\Big)\,\Big(p_2 \cdot p_+\Big)
+\Big(p_1 \cdot p_+\Big)\,\Big(p_2 \cdot p_-\Big)\right)\nonumber\\%
&\hspace{25mm} +8\,\left( \Big(p_2 \cdot p_-\Big)+\Big(p_1 \cdot
p_+\Big)\right)\,\left( \Big(p_1 \cdot p_-\Big)\,\Big(p_2 \cdot p_+\Big)-
\Big(p_1 \cdot p_+\Big)\,\Big(p_2 \cdot p_-\Big)\right)\nonumber\\%
&\hspace{25mm}+16\,m_\mu^2 \,m_\nu^2\,\left(\Big(p_2 \cdot p_+\Big)+\Big(p_2
\cdot p_-\Big)+ \Big(p_1 \cdot p_+\Big)+\Big(p_1 \cdot
p_-\Big)\right)\nonumber\\%
&\hspace{25mm}-4\,\left(m_{\mu\mu} ^2\,m_\nu^2+m_{\nu\nu}
^2\,m_\mu^2\right)\,\left(\Big(p_2 \cdot p_+\Big)+\Big(p_2 \cdot p_-\Big)+
\Big(p_1 \cdot p_+\Big)+\Big(p_1 \cdot p_-\Big)\right)\nonumber\\%
&\hspace{25mm}+8\,m_\mu^2\, \left(\Big(p_1 \cdot p_+\Big)+\Big(p_1 \cdot
p_-\Big)\right)\,\left( \Big(p_2 \cdot p_+\Big)+\Big(p_2 \cdot
p_-\Big)\right)\nonumber\\%
&\hspace{25mm}+8\,m_\nu^2\, \left(\Big(p_2 \cdot p_-\Big)+\Big(p_1 \cdot
p_-\Big)\right)\,\left( \Big(p_2 \cdot p_+\Big)+\Big(p_1 \cdot
p_+\Big)\right)\nonumber\\%
&\hspace{25mm}+2\,m_{\nu\nu} ^2\,m_{\mu\mu}^2\,\left(\Big(p_2 \cdot p_-\Big)+
\Big(p_1 \cdot p_+\Big)\right)+8\,m_{\mu\mu}^2\,m_\mu^2\,m_\nu^2+8\,
m_{\nu\nu}^2\,m_\mu^2\,m_\nu^2-2\,m_{\mu\mu}^4\,m_\nu^2\nonumber\\%
&\hspace{25mm}-4\,
m_{\nu\nu}^2\,m_{\mu\mu}^2\,m_\nu^2-4\,m_{\nu\nu}^2\,m_{\mu\mu}^2\,
m_\mu^2-2\,m_{\nu\nu}^4\,m_\mu^2+m_{\nu\nu}^2\,m_{\mu\mu}^4+
m_{\nu\nu}^4\,m_{\mu\mu}^2\bigg),\\%
R^{D/M}_{pm} &= 16\,\left(m_\nu^2\,\Big(p_1 \cdot p_-\Big)+m_\mu^2\, \Big(p_1 \cdot
p_-\Big)+2\,m_\mu^2\,m_\nu^2\right)\,\bigg(2\,m_\mu^2 \,\Big(p_2 \cdot
p_-\Big)+2\,m_\nu^2\,\Big(p_1 \cdot p_+\Big)\nonumber\\%
&\hspace{8cm} -4\,
m_\mu^2\,m_\nu^2+m_{\mu\mu}^2\,m_\nu^2+m_{\nu\nu}^2\,m_\mu^2\bigg),\\%
I^{D/M}_{ab} &= 32\,\mathbb{O}\,\left(2\,\Big(p_2 \cdot p_+\Big)+4\, \Big(p_1 \cdot
p_+\Big)+2\,\Big(p_1 \cdot p_-\Big)-m_B^2+m_{\mu\mu}^ 2+m_{\nu\nu}^2\right),\\%
I^{D/M}_{ac} &= -64\,\mathbb{O}\,\left(\Big(p_2 \cdot p_-\Big)+ \Big(p_1 \cdot
p_+\Big)\right),\\%
I^{D/M}_{bc} &= 32\,\mathbb{O}\,\bigg(2\,\left(2\,\Big(p_1 \cdot p_-\Big)\, \Big(p_2
\cdot p_+\Big)-\Big(p_2 \cdot p_-\Big)^2- \Big(p_1 \cdot
p_+\Big)^2\right)\nonumber\\%
&\hspace{15mm} +2\,\left(m_\mu^2+m_\nu^2\right)\,\left( \Big(p_2 \cdot p_+\Big)+\Big(p_2 \cdot
p_-\Big)+ \Big(p_1 \cdot p_+\Big)+\Big(p_1 \cdot p_-\Big)\right)\nonumber\\%
&\hspace{15mm}-\left(m_{\mu\mu}^2+m_{\nu\nu}^2\right) \,\left(\Big(p_2 \cdot p_-\Big)+\Big(p_1 \cdot
p_+\Big)\right)+\left(m_{\mu\mu}^2+m_{\nu\nu}^2\right)\left(m_\mu^2+m_\nu^2\right)-m_{\nu\nu}^2\,m_{\mu\mu}^2\bigg),\\%
I^{D/M}_{cp} &= -32\,\mathbb{O}\,\left(m_\mu^2-m_\nu^2\right)^2,\\%
I^{D/M}_{am} &= -32\,\mathbb{O}\,\left(m_\mu^2-m_\nu^2\right),\\%
I^{D/M}_{cm} &= 16\,\mathbb{O}\,\left(m_\mu^2-m_\nu^2\right)\,\left(2\,\left(\Big(p_2 \cdot p_-\Big)-\Big(p_1 \cdot p_+\Big)
 \right)-2\,m_\nu^2+2\,m_\mu^2-m_{\mu\mu}^2+m_{\nu\nu}^2\right).
\end{align}
The rest of the $50$ non-vanishing terms exclusive to Majorana case
and appearing in Eq.~\eqref{eq:MAmpSq} are given by,
\begin{align}
S^M_{a'a'} &= 64\,\Big(p_1 \cdot p_-\Big)\,\Big(p_2 \cdot p_+\Big),\\%
S^M_{b'b'} &= 4\,\bigg(2\,\left(2\,\Big(p_1 \cdot p_+\Big)\, \Big(p_2 \cdot
p_+\Big)+2\,\Big(p_1 \cdot p_-\Big)\, \Big(p_2 \cdot p_+\Big)+2\,\Big(p_1 \cdot
p_+\Big)^2+2\, \Big(p_1 \cdot p_-\Big)\,\Big(p_1 \cdot p_+\Big)-m_B^2\, \Big(p_1
\cdot p_+\Big)\right)\nonumber\\%
&\qquad +2\,m_{\nu\nu}^2\,\left( \Big(p_2 \cdot p_+\Big)+\Big(p_1 \cdot
p_+\Big)\right)+2\,m_{\mu\mu} ^2\,\left(\Big(p_1 \cdot p_+\Big)+\Big(p_1 \cdot
p_-\Big)\right)+ m_{\nu\nu}^2\,m_{\mu\mu}^2\bigg)\nonumber\\%
&\quad \times \bigg(2\,\left(2\, \Big(p_2 \cdot p_-\Big)\,\Big(p_2 \cdot
p_+\Big)+2\, \Big(p_1 \cdot p_-\Big)\,\Big(p_2 \cdot p_+\Big)+2\, \Big(p_2 \cdot
p_-\Big)^2+2\,\Big(p_1 \cdot p_-\Big)\, \Big(p_2 \cdot p_-\Big)-m_B^2\,\Big(p_2
\cdot p_-\Big)\right)\nonumber\\%
&\qquad +2\, m_{\mu\mu}^2\,\left(\Big(p_2 \cdot p_+\Big)+\Big(p_2 \cdot p_-\Big)
\right)+2\,m_{\nu\nu}^2\,\left(\Big(p_2 \cdot p_-\Big)+ \Big(p_1 \cdot
p_-\Big)\right)+m_{\nu\nu}^2\,m_{\mu\mu}^2\bigg),\\%
S^M_{c'c'} &= 4\,\bigg(8\,m_\nu^2\,\bigg(2\,\Big(p_2 \cdot p_-\Big)\, \Big(p_2
\cdot p_+\Big)^2-\Big(p_1 \cdot p_-\Big)\, \Big(p_2 \cdot p_-\Big)\,\Big(p_2
\cdot p_+\Big)- \Big(p_1 \cdot p_-\Big)\,\Big(p_1 \cdot p_+\Big)\, \Big(p_2
\cdot p_+\Big)\nonumber\\%
&\hspace{2cm} -\Big(p_1 \cdot p_+\Big)\, \Big(p_2 \cdot p_-\Big)^2-\Big(p_1
\cdot p_+\Big)^2\, \Big(p_2 \cdot p_-\Big)+2\,\Big(p_1 \cdot p_-\Big)^2\,
\Big(p_1 \cdot p_+\Big)\bigg)\nonumber\\%
&\qquad +8\,m_\mu^2\,\bigg(2\, \Big(p_1 \cdot p_+\Big)\,\Big(p_2 \cdot
p_+\Big)^2- \Big(p_1 \cdot p_-\Big)\,\Big(p_2 \cdot p_-\Big)\, \Big(p_2 \cdot
p_+\Big)-\Big(p_1 \cdot p_-\Big)\, \Big(p_1 \cdot p_+\Big)\,\Big(p_2 \cdot
p_+\Big)\nonumber\\%
&\hspace{2cm} - \Big(p_1 \cdot p_+\Big)\,\Big(p_2 \cdot p_-\Big)^2- \Big(p_1
\cdot p_+\Big)^2\,\Big(p_2 \cdot p_-\Big)+2\, \Big(p_1 \cdot
p_-\Big)^2\,\Big(p_2 \cdot p_-\Big)\bigg) \nonumber\\%
&\qquad +8\,m_\mu^2 \,m_\nu^2\,\bigg(3\,\Big(p_2 \cdot p_+\Big)^2+3\, \Big(p_2
\cdot p_-\Big)\,\Big(p_2 \cdot p_+\Big)+3\, \Big(p_1 \cdot p_+\Big)\,\Big(p_2
\cdot p_+\Big)+4\, \Big(p_1 \cdot p_-\Big)\,\Big(p_2 \cdot p_+\Big)\nonumber\\%
&\hspace{25mm} +2\, \Big(p_1 \cdot p_+\Big)\,\Big(p_2 \cdot p_-\Big)+3\,
\Big(p_1 \cdot p_-\Big)\,\Big(p_2 \cdot p_-\Big)+3\, \Big(p_1 \cdot
p_-\Big)\,\Big(p_1 \cdot p_+\Big)+3\, \Big(p_1 \cdot
p_-\Big)^2\bigg)\nonumber\\%
&\qquad +16\,\left(\Big(p_1 \cdot p_+\Big)\, \Big(p_2 \cdot p_-\Big)-\Big(p_1
\cdot p_-\Big)^2\right)\,\left( \Big(p_2 \cdot p_+\Big)^2-\Big(p_1 \cdot
p_+\Big)\, \Big(p_2 \cdot p_-\Big)\right)\nonumber\\%
&\qquad -4\,m_{\mu\mu}^2\,m_\nu^2\,\bigg(2\, \Big(p_2 \cdot p_-\Big)\,\Big(p_2
\cdot p_+\Big)+ \Big(p_1 \cdot p_+\Big)\,\Big(p_2 \cdot p_+\Big)+4\, \Big(p_1
\cdot p_-\Big)\,\Big(p_2 \cdot p_+\Big)\nonumber\\%
&\hspace{25mm}+4\, \Big(p_1 \cdot p_+\Big)\,\Big(p_2 \cdot p_-\Big)+ \Big(p_1
\cdot p_-\Big)\,\Big(p_2 \cdot p_-\Big)+2\, \Big(p_1 \cdot p_-\Big)\,\Big(p_1
\cdot p_+\Big)\bigg)\nonumber\\%
&\qquad -4\, m_{\nu\nu}^2\,m_\mu^2\,\bigg(\Big(p_2 \cdot p_-\Big)\, \Big(p_2
\cdot p_+\Big)+2\,\Big(p_1 \cdot p_+\Big)\, \Big(p_2 \cdot p_+\Big)+4\,\Big(p_1
\cdot p_-\Big)\, \Big(p_2 \cdot p_+\Big)\nonumber\\%
&\hspace{25mm}+4\,\Big(p_1 \cdot p_+\Big)\, \Big(p_2 \cdot p_-\Big)+2\,\Big(p_1
\cdot p_-\Big)\, \Big(p_2 \cdot p_-\Big)+\Big(p_1 \cdot p_-\Big)\, \Big(p_1
\cdot p_+\Big)\bigg)\nonumber\\%
&\qquad +4\,m_{\nu\nu}^2\,m_\nu^2\,\left( \Big(p_2 \cdot p_-\Big)\,\Big(p_2
\cdot p_+\Big)+ \Big(p_1 \cdot p_-\Big)\,\Big(p_1 \cdot p_+\Big)\right)
\nonumber\\%
&\qquad +4\, m_{\mu\mu}^2\,m_\mu^2\,\left(\Big(p_1 \cdot p_+\Big)\, \Big(p_2
\cdot p_+\Big)+\Big(p_1 \cdot p_-\Big)\, \Big(p_2 \cdot
p_-\Big)\right)\nonumber\\%
&\qquad +8\,m_{\nu\nu}^2\,m_{\mu\mu}^2\,\left( \Big(p_1 \cdot p_-\Big)\,\Big(p_2
\cdot p_+\Big)+ \Big(p_1 \cdot p_+\Big)\,\Big(p_2 \cdot
p_-\Big)\right)\nonumber\\%
&\qquad -4\, \left(m_{\mu\mu}^2+m_{\nu\nu}^2\right)\,m_\nu^4\,\left(\Big(p_2
\cdot p_+\Big)+ \Big(p_2 \cdot p_-\Big)+\Big(p_1 \cdot p_+\Big)+ \Big(p_1 \cdot
p_-\Big)\right)\nonumber\\%
&\qquad -8\,m_\mu^4\, \left(\Big(p_1 \cdot p_+\Big)+\Big(p_1 \cdot
p_-\Big)\right)\,\left( \Big(p_2 \cdot p_+\Big)+\Big(p_2 \cdot p_-\Big)\right)
\nonumber\\%
&\qquad +4\,\left(m_{\mu\mu}
^2+m_{\nu\nu}^2\right)\,m_\mu^2\,m_\nu^2\,\left(\Big(p_2 \cdot p_+\Big)-
\Big(p_2 \cdot p_-\Big)-\Big(p_1 \cdot p_+\Big)+ \Big(p_1 \cdot
p_-\Big)\right)\nonumber\\%
&\qquad -8\,m_\nu^4\, \left(\Big(p_2 \cdot p_-\Big)+\Big(p_1 \cdot
p_-\Big)\right)\,\left( \Big(p_2 \cdot p_+\Big)+\Big(p_1 \cdot p_+\Big)\right)
\nonumber\\%
&\qquad +2\,m_{\nu\nu}
^2\,m_{\mu\mu}^2\,\left(m_\mu^2+m_\nu^2\right)\,\left(\Big(p_2 \cdot p_-\Big)+
\Big(p_1 \cdot p_+\Big)\right)-2\,
m_{\mu\mu}^4\,m_\nu^4-12\,m_{\nu\nu}^2\,m_{\mu\mu}^2\,m_\mu^2\,m_\nu
^2+4\,m_{\nu\nu}^2\,m_{\mu\mu}^4\,m_\nu^2\nonumber\\%
&\qquad -2\,m_{\nu\nu}^4\,m_\mu^4+4
\,m_{\nu\nu}^4\,m_{\mu\mu}^2\,m_\mu^2-m_{\nu\nu}^4\,m_{\mu\mu}^4 \bigg),\\%
S^M_{p'p'} &= 16\,\left(\left(m_\mu^2+m_\nu^2\right)\,\Big(p_1 \cdot
p_+\Big)+2\,m_\mu^2\,m_\nu^2\right)\,\left(\left(m_\mu^2+m_\nu^2\right)\,
\Big(p_2 \cdot p_-\Big)+2\,m_\mu^2 \,m_\nu^2\right),\\%
S^M_{m'm'} &= 8\,\left(m_\nu^2\,\Big(p_2 \cdot p_-\Big)+m_\mu^2\, \Big(p_2 \cdot
p_-\Big)+2\,m_\mu^2\,m_\nu^2\right)\nonumber\\%
&\quad \times \bigg(4\,\left( \Big(p_1 \cdot p_-\Big)\,\Big(p_2 \cdot p_+\Big)-
\Big(p_1 \cdot p_+\Big)\,\Big(p_2 \cdot p_-\Big)\right)-2\,m_\nu^2\,
\left(2\,\Big(p_2 \cdot p_+\Big)+\Big(p_1 \cdot p_+\Big)\right)\nonumber\\%
&\hspace{1cm} +2\, m_{\nu\nu}^2\,\Big(p_2 \cdot p_+\Big)-2\,m_\mu^2\,\left(
\Big(p_1 \cdot p_+\Big)+2\,\Big(p_1 \cdot p_-\Big)\right)+2\,
m_{\mu\mu}^2\,\Big(p_1 \cdot p_-\Big)\nonumber\\%
&\hspace{1cm} +4\,m_\mu^2\,m_\nu^2-2\,
m_{\mu\mu}^2\,m_\nu^2-2\,m_{\nu\nu}^2\,m_\mu^2+m_{\nu\nu}^2\,
m_{\mu\mu}^2\bigg),\\%
R^M_{aa'} &= 32\,\left(2\,m_\mu^2-m_{\mu\mu}^2\right),\\%
R^M_{ba'} &= -16\,\left(2\,\left(\Big(p_2 \cdot p_-\Big)+ \Big(p_1
\cdot p_-\Big)\right)+m_{\mu\mu}^2\right)\,\left(2\,\left( \Big(p_2 \cdot
p_+\Big)+\Big(p_1 \cdot p_+\Big)\right)+m_{\mu\mu}^2 \right),\\%
R^M_{ab'} &= -16\,\left(2\,\left(\Big(p_2 \cdot p_-\Big)+ \Big(p_1
\cdot p_-\Big)\right)+m_{\mu\mu}^2\right)\,\left(2\,\left( \Big(p_2 \cdot
p_+\Big)+\Big(p_1 \cdot p_+\Big)\right)+m_{\mu\mu}^2 \right),\\%
R^M_{bb'} &= 8\,m_B^4\,\left(2\,m_\mu^2-m_{\mu\mu}^2\right),\\%
R^M_{cb'} &= -16\,\bigg(4\,\bigg(\Big(p_1 \cdot p_-\Big)\, \Big(p_2
\cdot p_+\Big)^2-\Big(p_2 \cdot p_-\Big)^2\, \Big(p_2 \cdot p_+\Big)-2\,\Big(p_1
\cdot p_-\Big)\, \Big(p_2 \cdot p_-\Big)\,\Big(p_2 \cdot p_+\Big)\nonumber\\%
&\hspace{25mm} +2\, \Big(p_1 \cdot p_-\Big)\,\Big(p_1 \cdot p_+\Big)\, \Big(p_2
\cdot p_+\Big)-\Big(p_1 \cdot p_-\Big)^2\, \Big(p_2 \cdot p_+\Big)+\Big(p_1
\cdot p_-\Big)\, \Big(p_1 \cdot p_+\Big)^2\bigg)\nonumber\\%
&\hspace{2cm}-2\,\left(m_{\mu\mu}^2+m_{\nu\nu}^2\right)\,\left( \Big(p_2 \cdot
p_-\Big)\,\Big(p_2 \cdot p_+\Big)- \Big(p_1 \cdot p_-\Big)\,\Big(p_1 \cdot
p_+\Big)\right)\nonumber\\%
&\hspace{2cm}+4\,m_\mu^2 \,\left(\left(\Big(p_2 \cdot p_+\Big)+ \Big(p_1 \cdot
p_+\Big)\right)^2-\left(\Big(p_1 \cdot p_-\Big)+\Big(p_2 \cdot
p_-\Big)\right)^2\right)\nonumber\\%
&\hspace{2cm}+2\,\left(m_{\mu\mu}^2+m_{\nu\nu}^2\right)\,m_\mu^2\,\left(\Big(p_2
\cdot p_+\Big)-\Big(p_2 \cdot p_-\Big)+ \Big(p_1 \cdot p_+\Big)-\Big(p_1 \cdot
p_-\Big)\right)\nonumber\\%
&\hspace{2cm}-m_{\nu\nu}^2 \,m_{\mu\mu}^2\,\left(\Big(p_2 \cdot p_+\Big)-
\Big(p_1 \cdot p_-\Big)\right)\bigg),\\%
R^M_{a'b'} &= -8\,\bigg(4\,\bigg(4\,\Big(p_1 \cdot p_+\Big)\, \Big(p_2 \cdot
p_-\Big)\,\Big(p_2 \cdot p_+\Big)+2\, \Big(p_1 \cdot p_-\Big)\,\Big(p_2 \cdot
p_-\Big)\, \Big(p_2 \cdot p_+\Big)+2\,\Big(p_1 \cdot p_-\Big)\, \Big(p_1 \cdot
p_+\Big)\,\Big(p_2 \cdot p_+\Big)\nonumber\\%
&\hspace{2cm}-m_B^2\, \Big(p_1 \cdot p_-\Big)\,\Big(p_2 \cdot p_+\Big)+2\,
\Big(p_1 \cdot p_+\Big)\,\Big(p_2 \cdot p_-\Big)^2+2\, \Big(p_1 \cdot
p_+\Big)^2\,\Big(p_2 \cdot p_-\Big)\nonumber\\%
&\hspace{2cm}+4\, \Big(p_1 \cdot p_-\Big)\,\Big(p_1 \cdot p_+\Big)\, \Big(p_2
\cdot p_-\Big)-m_B^2\,\Big(p_1 \cdot p_+\Big)\, \Big(p_2 \cdot
p_-\Big)\bigg)\nonumber\\%
&\hspace{15mm}-4\,\left(m_{\mu\mu}^2+m_{\nu\nu}^2\right)\,\left( \Big(p_1 \cdot
p_-\Big)\,\Big(p_2 \cdot p_+\Big)- \Big(p_1 \cdot p_+\Big)\,\Big(p_2 \cdot
p_-\Big)\right)\nonumber\\%
&\hspace{15mm}-m_{\nu\nu}^ 2\,m_{\mu\mu}^2\,\left(4\,\Big(p_2 \cdot p_+\Big)+2\,
\Big(p_2 \cdot p_-\Big)+2\,\Big(p_1 \cdot p_+\Big)+4\, \Big(p_1 \cdot
p_-\Big)-m_B^2\right) \nonumber\\%
&\hspace{15mm}+2\,\left(m_{\mu\mu}^2\,m_\nu^2+m_{\nu\nu}^2\,m_\mu^2\right)\,
\left(2\,\Big(p_2 \cdot p_+\Big)+2\,\Big(p_2 \cdot p_-\Big)+2\, \Big(p_1 \cdot
p_+\Big)+2\,\Big(p_1 \cdot p_-\Big)-m_B^2\right)\nonumber\\%
&\hspace{15mm}+8\,m_\mu^2\,\left( \Big(p_1 \cdot
p_+\Big)+\Big(p_1 \cdot p_-\Big)\right)\,\left( \Big(p_2 \cdot p_+\Big)+\Big(p_2
\cdot p_-\Big)\right)\nonumber\\%
&\hspace{15mm}+8\,m_\nu^2\,\left( \Big(p_1 \cdot
p_+\Big)+\Big(p_2 \cdot p_+\Big)\right)\,\left( \Big(p_1 \cdot p_-\Big)+\Big(p_2
\cdot p_-\Big)\right)\nonumber\\%
&\hspace{15mm}+4\,m_\mu^2\,
m_\nu^2\,m_B^2+2\,m_{\mu\mu}^4\,m_\nu^2+2\,m_{\nu\nu}^4\,m_\mu^2-
m_{\nu\nu}^2\,m_{\mu\mu}^4-m_{\nu\nu}^4\,m_{\mu\mu}^2\bigg),\\%
R^M_{bc'} &= -16\,\Bigg(4\,\bigg(\Big(p_2 \cdot p_-\Big)\, \Big(p_2
\cdot p_+\Big)^2+2\,\Big(p_1 \cdot p_+\Big)\, \Big(p_2 \cdot p_-\Big)\,\Big(p_2
\cdot p_+\Big)- \Big(p_1 \cdot p_+\Big)\,\Big(p_2 \cdot p_-\Big)^2\nonumber\\*%
&\hspace{2cm}+ \Big(p_1 \cdot p_+\Big)^2\,\Big(p_2 \cdot p_-\Big)-2\, \Big(p_1
\cdot p_-\Big)\,\Big(p_1 \cdot p_+\Big)\, \Big(p_2 \cdot p_-\Big)-\Big(p_1 \cdot
p_-\Big)^2\, \Big(p_1 \cdot p_+\Big)\bigg)\nonumber\\*%
&\hspace{15mm}+2\,\left(m_{\mu\mu}^2+m_{\nu\nu}^2\right)\,\left( \Big(p_2 \cdot  
p_-\Big)\,\Big(p_2 \cdot p_+\Big)- \Big(p_1 \cdot p_-\Big)\,\Big(p_1 \cdot       
p_+\Big)\right)\nonumber\\*%
&\hspace{15mm}+4\,m_\mu^2 \,\left(\left(\Big(p_2 \cdot p_+\Big)+ \Big(p_1 \cdot
p_+\Big)\right)^2-\left(\Big(p_1 \cdot p_-\Big)+\Big(p_2 \cdot                   
p_-\Big)\right)^2\right) \nonumber\\*%
&\hspace{15mm}+2\,\left(m_{\mu\mu}^2+m_{\nu\nu}^2\right)\,m_\mu^2\,\left(\Big(p_2
\cdot p_+\Big)-\Big(p_2 \cdot p_-\Big)+ \Big(p_1 \cdot p_+\Big)-\Big(p_1 \cdot
p_-\Big)\right)\nonumber\\*%
&\hspace{15mm}+m_{\nu\nu}^2 \,m_{\mu\mu}^2\,\left(\Big(p_2 \cdot p_-\Big)-
\Big(p_1 \cdot p_+\Big)\right)\Bigg),\\%
R^M_{cc'} &= 8\,\Bigg(4\,m_\mu^2\,\bigg(\Big(p_2 \cdot p_+\Big)^2+2\,
\Big(p_2 \cdot p_-\Big)\,\Big(p_2 \cdot p_+\Big)+ \Big(p_2 \cdot
p_-\Big)^2\nonumber\\%
&\hspace{2cm}+\Big(p_1 \cdot p_+\Big)^2+2\, \Big(p_1 \cdot p_-\Big)\,\Big(p_1
\cdot p_+\Big)+ \Big(p_1 \cdot p_-\Big)^2\bigg)\nonumber\\%
&\hspace{1cm}-4\,m_{\mu\mu}^2\,\left( \Big(p_2 \cdot p_-\Big)\,\Big(p_2 \cdot
p_+\Big)+ \Big(p_1 \cdot p_-\Big)\,\Big(p_1 \cdot p_+\Big)\right)\nonumber\\%
&\hspace{1cm}+4\, m_{\nu\nu}^2\,\left(\Big(p_1 \cdot p_-\Big)\,\Big(p_2 \cdot
p_+\Big) +\Big(p_1 \cdot p_+\Big)\,\Big(p_2 \cdot p_-\Big)\right)\nonumber\\%
&\hspace{1cm}+4\,\left( \Big(p_2 \cdot p_+\Big)-\Big(p_2 \cdot p_-\Big)-
\Big(p_1 \cdot p_+\Big)+\Big(p_1 \cdot p_-\Big)\right)\,\left( \Big(p_1 \cdot
p_-\Big)\,\Big(p_2 \cdot p_+\Big)- \Big(p_1 \cdot p_+\Big)\,\Big(p_2 \cdot
p_-\Big)\right)\nonumber\\%
&\hspace{1cm}-16\,m_\mu^2 \,m_\nu^2\,\left(\Big(p_2 \cdot p_+\Big)+\Big(p_2
\cdot p_-\Big)+ \Big(p_1 \cdot p_+\Big)+\Big(p_1 \cdot
p_-\Big)\right)\nonumber\\%
&\hspace{1cm}+4\,\left(m_{\mu\mu} ^2\,m_\nu^2+m_{\nu\nu}
^2\,m_\mu^2\right)\,\left(\Big(p_2 \cdot p_+\Big)+\Big(p_2 \cdot p_-\Big)+
\Big(p_1 \cdot p_+\Big)+\Big(p_1 \cdot p_-\Big)\right)\nonumber\\%
&\hspace{1cm}-m_{\nu\nu}^2 \,m_{\mu\mu}^2\,\left(\Big(p_2 \cdot p_+\Big)+
\Big(p_2 \cdot p_-\Big)+\Big(p_1 \cdot p_+\Big)+ \Big(p_1 \cdot
p_-\Big)\right)\nonumber\\%
&\hspace{1cm}-8\,m_\nu^2\,\left( \Big(p_2 \cdot p_-\Big)+\Big(p_1 \cdot
p_-\Big)\right)\,\left( \Big(p_2 \cdot p_+\Big)+\Big(p_1 \cdot
p_+\Big)\right)\nonumber\\%
&\hspace{1cm}-8\,m_{\mu\mu}
^2\,m_\mu^2\,m_\nu^2-8\,m_{\nu\nu}^2\,m_\mu^2\,m_\nu^2+2\,m_{\mu\mu}
^4\,m_\nu^2+4\,m_{\nu\nu}^2\,m_{\mu\mu}^2\,m_\nu^2+2\,m_{\nu\nu}^4\,
m_\mu^2-m_{\nu\nu}^4\,m_{\mu\mu}^2\Bigg),\\%
R^M_{a'c'} &= 16\,\Bigg(4\,m_\nu^2\,\left(\Big(p_2 \cdot p_-\Big)\, \Big(p_2
\cdot p_+\Big)-\Big(p_1 \cdot p_-\Big)\, \Big(p_1 \cdot
p_+\Big)\right)+4\,m_\mu^2\,\left( \Big(p_1 \cdot p_+\Big)\,\Big(p_2 \cdot
p_+\Big)- \Big(p_1 \cdot p_-\Big)\,\Big(p_2 \cdot p_-\Big)\right)\nonumber\\%
&\hspace{1cm}+4\,\left( \Big(p_2 \cdot p_+\Big)-\Big(p_1 \cdot
p_-\Big)\right)\,\left( \Big(p_1 \cdot p_-\Big)\,\Big(p_2 \cdot p_+\Big)+
\Big(p_1 \cdot p_+\Big)\,\Big(p_2 \cdot p_-\Big)\right)\nonumber\\%
&\hspace{1cm}+
\left(2\,\left(m_{\mu\mu}^2\,m_\nu^2+m_{\nu\nu}^2\,m_\mu^2\right)-m_{\nu\nu}^2
\,m_{\mu\mu}^2\right)\,\left(\Big(p_2 \cdot p_+\Big)- \Big(p_1 \cdot
p_-\Big)\right)\Bigg),\\%
R^M_{b'c'} &= 8\,\Bigg(-4\,m_\nu^2\,\bigg(4\,\Big(p_2 \cdot p_-\Big)\, \Big(p_2
\cdot p_+\Big)^2+2\,\Big(p_1 \cdot p_-\Big)\, \Big(p_2 \cdot
p_+\Big)^2+2\,\Big(p_2 \cdot p_-\Big)^2\, \Big(p_2 \cdot p_+\Big)\nonumber\\%
&\hspace{25mm}+4\,\Big(p_1 \cdot p_+\Big)\, \Big(p_2 \cdot p_-\Big)\,\Big(p_2
\cdot p_+\Big)-m_B^2\, \Big(p_2 \cdot p_-\Big)\,\Big(p_2 \cdot p_+\Big)-2\,
\Big(p_1 \cdot p_-\Big)^2\,\Big(p_2 \cdot p_+\Big)\nonumber\\%
&\hspace{25mm}-4\, \Big(p_1 \cdot p_-\Big)\,\Big(p_1 \cdot p_+\Big)\, \Big(p_2
\cdot p_-\Big)-2\,\Big(p_1 \cdot p_-\Big)\, \Big(p_1 \cdot
p_+\Big)^2-4\,\Big(p_1 \cdot p_-\Big)^2\, \Big(p_1 \cdot p_+\Big)\nonumber\\%
&\hspace{25mm}+m_B^2\,\Big(p_1 \cdot p_-\Big)\, \Big(p_1 \cdot
p_+\Big)\bigg)\nonumber\\%
&\hspace{1cm}-4\,m_\mu^2\,\bigg(4\, \Big(p_1 \cdot p_+\Big)\,\Big(p_2 \cdot
p_+\Big)^2+2\, \Big(p_1 \cdot p_-\Big)\,\Big(p_2 \cdot p_+\Big)^2+4\, \Big(p_1
\cdot p_+\Big)\,\Big(p_2 \cdot p_-\Big)\, \Big(p_2 \cdot p_+\Big)\nonumber\\%
&\hspace{25mm}+2\,\Big(p_1 \cdot p_+\Big)^2\, \Big(p_2 \cdot
p_+\Big)-m_B^2\,\Big(p_1 \cdot p_+\Big)\, \Big(p_2 \cdot p_+\Big)-2\,\Big(p_1
\cdot p_-\Big)^2\, \Big(p_2 \cdot p_+\Big)\nonumber\\%
&\hspace{25mm}-2\,\Big(p_1 \cdot p_-\Big)\, \Big(p_2 \cdot
p_-\Big)^2-4\,\Big(p_1 \cdot p_-\Big)\, \Big(p_1 \cdot p_+\Big)\,\Big(p_2 \cdot
p_-\Big)\nonumber\\%
&\hspace{25mm}-4\, \Big(p_1 \cdot p_-\Big)^2\,\Big(p_2 \cdot p_-\Big)+m_B^2\,
\Big(p_1 \cdot p_-\Big)\,\Big(p_2 \cdot p_-\Big)\bigg)\nonumber\\%
&\hspace{1cm}+8\,\left( \Big(p_2 \cdot p_+\Big)-\Big(p_1 \cdot
p_-\Big)\right)\,\bigg( \Big(p_1 \cdot p_-\Big)\,\Big(p_2 \cdot p_+\Big)^2-3\,
\Big(p_1 \cdot p_+\Big)\,\Big(p_2 \cdot p_-\Big)\, \Big(p_2 \cdot
p_+\Big)\nonumber\\%
&\hspace{55mm}+\Big(p_1 \cdot p_-\Big)^2\, \Big(p_2 \cdot p_+\Big)-2\,\Big(p_1
\cdot p_+\Big)\, \Big(p_2 \cdot p_-\Big)^2\nonumber\\%
&\hspace{55mm}-2\,\Big(p_1 \cdot p_+\Big)^2\, \Big(p_2 \cdot
p_-\Big)-3\,\Big(p_1 \cdot p_-\Big)\, \Big(p_1 \cdot p_+\Big)\,\Big(p_2 \cdot
p_-\Big)\nonumber\\%
&\hspace{55mm}+m_B^2\, \Big(p_1 \cdot p_+\Big)\,\Big(p_2 \cdot
p_-\Big)\bigg)\nonumber\\%
&\hspace{15mm}+2\, m_{\nu\nu}^2\,m_\mu^2\,\bigg(2\,\Big(p_2 \cdot p_+\Big)^2+4\,
\Big(p_2 \cdot p_-\Big)\,\Big(p_2 \cdot p_+\Big)-2\, \Big(p_1 \cdot
p_+\Big)\,\Big(p_2 \cdot p_+\Big)+2\, \Big(p_2 \cdot p_-\Big)^2\nonumber\\%
&\hspace{35mm}+2\,\Big(p_1 \cdot p_-\Big)\, \Big(p_2 \cdot
p_-\Big)-m_B^2\,\Big(p_2 \cdot p_-\Big)-2\, \Big(p_1 \cdot
p_+\Big)^2-4\,\Big(p_1 \cdot p_-\Big)\, \Big(p_1 \cdot p_+\Big)\nonumber\\%
&\hspace{35mm}+m_B^2\,\Big(p_1 \cdot p_+\Big)-2\, \Big(p_1 \cdot
p_-\Big)^2\bigg)\nonumber\\%
&\hspace{15mm}+2\,m_{\mu\mu}^2\,m_\nu^2\,\bigg(2\, \Big(p_2 \cdot
p_+\Big)^2-2\,\Big(p_2 \cdot p_-\Big)\, \Big(p_2 \cdot p_+\Big)+4\,\Big(p_1
\cdot p_+\Big)\, \Big(p_2 \cdot p_+\Big)-2\,\Big(p_2 \cdot
p_-\Big)^2\nonumber\\%
&\hspace{35mm}-4\, \Big(p_1 \cdot p_-\Big)\,\Big(p_2 \cdot p_-\Big)+m_B^2\,
\Big(p_2 \cdot p_-\Big)+2\,\Big(p_1 \cdot p_+\Big)^2+2\, \Big(p_1 \cdot
p_-\Big)\,\Big(p_1 \cdot p_+\Big)\nonumber\\%
&\hspace{35mm}-m_B^2\, \Big(p_1 \cdot p_+\Big)-2\,\Big(p_1 \cdot
p_-\Big)^2\bigg)\nonumber\\%
&\hspace{15mm}-4\, m_{\nu\nu}^2\,m_\nu^2\,\left(\Big(p_2 \cdot p_-\Big)\,
\Big(p_2 \cdot p_+\Big)-\Big(p_1 \cdot p_-\Big)\, \Big(p_1 \cdot
p_+\Big)\right)\nonumber\\%
&\hspace{15mm}-4\,m_{\mu\mu}^2\,m_\mu^2\,\left( \Big(p_1 \cdot
p_+\Big)\,\Big(p_2 \cdot p_+\Big)- \Big(p_1 \cdot p_-\Big)\,\Big(p_2 \cdot
p_-\Big)\right)\nonumber\\%
&\hspace{15mm}+4\, m_{\mu\mu}^2\,\left(\Big(p_2 \cdot p_+\Big)-\Big(p_1 \cdot
p_-\Big) \right)\,\left(\Big(p_1 \cdot p_-\Big)\,\Big(p_2 \cdot p_+\Big)-
\Big(p_1 \cdot p_+\Big)\,\Big(p_2 \cdot p_-\Big)\right)\nonumber\\%
&\hspace{15mm}+4\, m_{\nu\nu}^2\,\left(\Big(p_2 \cdot p_+\Big)-\Big(p_1 \cdot
p_-\Big) \right)\,\left(\Big(p_1 \cdot p_-\Big)\,\Big(p_2 \cdot p_+\Big)-
\Big(p_1 \cdot p_+\Big)\,\Big(p_2 \cdot p_-\Big)\right)\nonumber\\%
&\hspace{15mm}+2\, m_{\nu\nu}^2\,m_{\mu\mu}^2\,m_\mu^2\,\left(2\, \Big(p_2 \cdot
p_+\Big)+\Big(p_2 \cdot p_-\Big)- \Big(p_1 \cdot p_+\Big)-2\,\Big(p_1 \cdot
p_-\Big)\right)\nonumber\\%
&\hspace{15mm}+2\, m_{\nu\nu}^2\,m_{\mu\mu}^2\,m_\nu^2\,\left(2\, \Big(p_2 \cdot
p_+\Big)-\Big(p_2 \cdot p_-\Big)+ \Big(p_1 \cdot p_+\Big)-2\,\Big(p_1 \cdot
p_-\Big)\right)\nonumber\\%
&\hspace{15mm}-16\,m_\mu^ 2\,m_\nu^2\,\left(\Big(p_2 \cdot p_+\Big)-\Big(p_1
\cdot p_-\Big) \right)\,\left(\Big(p_2 \cdot p_+\Big)+\Big(p_2 \cdot p_-\Big)+
\Big(p_1 \cdot p_+\Big)+\Big(p_1 \cdot p_-\Big)\right)\nonumber\\%
&\hspace{15mm}+2\,m_{\nu\nu} ^4\,m_\mu^2\,\left(\Big(p_2 \cdot p_+\Big)+\Big(p_2
\cdot p_-\Big)- \Big(p_1 \cdot p_+\Big)-\Big(p_1 \cdot
p_-\Big)\right)\nonumber\\%
&\hspace{15mm}+2\,m_{\mu\mu} ^4\,m_\nu^2\,\left(\Big(p_2 \cdot p_+\Big)-\Big(p_2
\cdot p_-\Big)+ \Big(p_1 \cdot p_+\Big)-\Big(p_1 \cdot
p_-\Big)\right)\nonumber\\%
&\hspace{15mm}-2\,m_{\nu\nu} ^2\,m_{\mu\mu}^2\,\left(\Big(p_2 \cdot p_+\Big)-
\Big(p_1 \cdot p_-\Big)\right)\,\left(\Big(p_2 \cdot p_+\Big)+ \Big(p_1 \cdot
p_-\Big)\right)\nonumber\\%
&\hspace{15mm}-\left(m_{\mu\mu}^2+m_{\nu\nu}^2\right)\,\left(8\,m_\mu^2\,m_\nu^2
+m_{\mu\mu}^2m_{\nu\nu}^2\right)\, \left(\Big(p_2 \cdot
p_+\Big)-\Big(p_1 \cdot p_-\Big)\right)\Bigg),\\%
R^M_{a'p} &= -64\,\left(\Big(p_1 \cdot p_-\Big)+
m_\mu^2\right)\,\left(\Big(p_2 \cdot p_+\Big)+m_\mu^2\right),\\%
R^M_{b'p} &= 4\,\Bigg(4\,\bigg(4\,\Big(p_1 \cdot p_+\Big)\, \Big(p_2
\cdot p_-\Big)\,\Big(p_2 \cdot p_+\Big)+2\, \Big(p_1 \cdot p_-\Big)\,\Big(p_2
\cdot p_-\Big)\, \Big(p_2 \cdot p_+\Big)+2\,\Big(p_1 \cdot p_-\Big)\, \Big(p_1
\cdot p_+\Big)\,\Big(p_2 \cdot p_+\Big)\nonumber\\%
&\hspace{2cm}-m_B^2\, \Big(p_1 \cdot p_-\Big)\,\Big(p_2 \cdot p_+\Big)+2\,
\Big(p_1 \cdot p_+\Big)\,\Big(p_2 \cdot p_-\Big)^2+2\, \Big(p_1 \cdot
p_+\Big)^2\,\Big(p_2 \cdot p_-\Big)\nonumber\\%
&\hspace{2cm}+4\, \Big(p_1 \cdot p_-\Big)\,\Big(p_1 \cdot p_+\Big)\, \Big(p_2
\cdot p_-\Big)-m_B^2\,\Big(p_1 \cdot p_+\Big)\, \Big(p_2 \cdot
p_-\Big)\bigg)\nonumber\\%
&\hspace{15mm}+4\,m_\mu^2\,\bigg(2\, \Big(p_1 \cdot p_+\Big)\,\Big(p_2 \cdot
p_+\Big)+2\, \Big(p_1 \cdot p_-\Big)\,\Big(p_2 \cdot p_+\Big)-m_B^2\, \Big(p_2
\cdot p_+\Big)+2\,\Big(p_1 \cdot p_+\Big)\, \Big(p_2 \cdot p_-\Big)\nonumber\\%
&\hspace{3cm}+2\,\Big(p_1 \cdot p_-\Big)\, \Big(p_2 \cdot
p_-\Big)-m_B^2\,\Big(p_1 \cdot p_-\Big)\bigg)\nonumber\\%
&\hspace{15mm}-4\, \left(m_{\mu\mu}^2+m_{\nu\nu}^2\right)\,\left(\Big(p_1 \cdot
p_-\Big)\,\Big(p_2 \cdot p_+\Big) -\Big(p_1 \cdot p_+\Big)\,\Big(p_2 \cdot
p_-\Big)\right)\nonumber\\%
&\hspace{15mm}-m_{\nu\nu}^ 2\,m_{\mu\mu}^2\,\left(4\,\Big(p_2 \cdot p_+\Big)+2\,
\Big(p_2 \cdot p_-\Big)+2\,\Big(p_1 \cdot p_+\Big)+4\, \Big(p_1 \cdot
p_-\Big)-m_B^2\right)\nonumber\\%
&\hspace{15mm}+2\,\left(m_{\mu\mu}^2\,m_\nu^2+m_{\nu\nu}^2\,m_\mu^2\right)\,
\left(2\,\Big(p_2 \cdot p_+\Big)+2\,\Big(p_2 \cdot p_-\Big)+2\, \Big(p_1 \cdot
p_+\Big)+2\,\Big(p_1 \cdot p_-\Big)-m_B^2\right)\nonumber\\%
&\hspace{15mm}+8\,m_\nu^2\,\left( \Big(p_2 \cdot p_-\Big)+\Big(p_1 \cdot
p_-\Big)\right)\,\left( \Big(p_2 \cdot p_+\Big)+\Big(p_1 \cdot
p_+\Big)\right)\nonumber\\%
&\hspace{15mm}+4\,m_\mu^2\,
m_\nu^2\,m_B^2-4\,m_\mu^4\,m_B^2+2\,m_{\mu\mu}^4\,m_\nu^2+2\,
m_{\nu\nu}^4\,m_\mu^2-m_{\nu\nu}^2\,m_{\mu\mu}^4-m_{\nu\nu}^4\,
m_{\mu\mu}^2\Bigg),\\%
R^M_{c'p} &= -8\,\Bigg(4\,m_\mu^2\,\left(\Big(p_2 \cdot p_+\Big)^2+
\Big(p_1 \cdot p_+\Big)\,\Big(p_2 \cdot p_+\Big)- \Big(p_1 \cdot
p_-\Big)\,\Big(p_2 \cdot p_-\Big)- \Big(p_1 \cdot p_-\Big)^2\right)\nonumber\\%
&\hspace{15mm}+4\,m_\nu^2\,\left( \Big(p_2 \cdot p_-\Big)\,\Big(p_2 \cdot
p_+\Big)- \Big(p_1 \cdot p_-\Big)\,\Big(p_1 \cdot p_+\Big)\right)\nonumber\\%
&\hspace{15mm}+4\,\left( \Big(p_2 \cdot p_+\Big)-\Big(p_1 \cdot
p_-\Big)\right)\,\left( \Big(p_1 \cdot p_-\Big)\,\Big(p_2 \cdot p_+\Big)+
\Big(p_1 \cdot p_+\Big)\,\Big(p_2 \cdot p_-\Big)\right)\nonumber\\%
&\hspace{15mm}-4\,m_\mu^4\, \left(\Big(p_2 \cdot p_+\Big)+\Big(p_2 \cdot
p_-\Big)- \Big(p_1 \cdot p_+\Big)-\Big(p_1 \cdot p_-\Big)\right)\nonumber\\%
&\hspace{15mm}-4\,m_\mu^2\, m_\nu^2\,\left(\Big(p_2 \cdot p_+\Big)-\Big(p_2
\cdot p_-\Big)+ \Big(p_1 \cdot p_+\Big)-\Big(p_1 \cdot
p_-\Big)\right)\nonumber\\%
&\hspace{15mm}+\left(2\,m_{\mu\mu}^2\,m_\nu^2+2\,m_{\mu\mu}^2\,m_\mu^2+4\,m_{\nu\nu}^2\,m_\mu^2-m_{\nu\nu}^2 \,m_{\mu\mu}^2\right)\,\left(\Big(p_2 \cdot p_+\Big)-\Big(p_1 \cdot p_-\Big) \right)\Bigg),\\%
R^M_{a'm} &= -32\,\left(\Big(p_1 \cdot p_-\Big)+m_\mu^2\right)\,\left(2
\,\Big(p_1 \cdot p_+\Big)-2\,m_\mu^2+m_{\mu\mu}^2\right),\\%
R^M_{b'm} &= -4\,\Bigg(8\,\bigg(\Big(p_1 \cdot p_-\Big)\, \Big(p_2
\cdot p_-\Big)\,\Big(p_2 \cdot p_+\Big)- \Big(p_1 \cdot p_-\Big)\,\Big(p_1 \cdot
p_+\Big)\, \Big(p_2 \cdot p_+\Big)-\Big(p_1 \cdot p_+\Big)\, \Big(p_2 \cdot
p_-\Big)^2\nonumber\\%
&\hspace{2cm}-\Big(p_1 \cdot p_+\Big)^2\, \Big(p_2 \cdot p_-\Big)-2\,\Big(p_1
\cdot p_-\Big)\, \Big(p_1 \cdot p_+\Big)\,\Big(p_2 \cdot p_-\Big)-2\, \Big(p_1
\cdot p_-\Big)\,\Big(p_1 \cdot p_+\Big)^2\nonumber\\%
&\hspace{2cm}-2\, \Big(p_1 \cdot p_-\Big)^2\,\Big(p_1 \cdot p_+\Big)+m_B^2\,
\Big(p_1 \cdot p_-\Big)\,\Big(p_1 \cdot p_+\Big)\bigg)\nonumber\\%
&\hspace{15mm}-8\,m_\mu^2\, \bigg(\Big(p_1 \cdot p_+\Big)\,\Big(p_2 \cdot
p_+\Big)+ \Big(p_1 \cdot p_-\Big)\,\Big(p_2 \cdot p_+\Big)+ \Big(p_1 \cdot
p_+\Big)\,\Big(p_2 \cdot p_-\Big)+ \Big(p_1 \cdot p_-\Big)\,\Big(p_2 \cdot
p_-\Big)\nonumber\\%
&\hspace{3cmm}+2\, \Big(p_1 \cdot p_+\Big)^2+4\,\Big(p_1 \cdot p_-\Big)\,
\Big(p_1 \cdot p_+\Big)-m_B^2\,\Big(p_1 \cdot p_+\Big)+2\, \Big(p_1 \cdot
p_-\Big)^2\bigg)\nonumber\\%
&\hspace{15mm}-4\,m_{\nu\nu}^2\,\left( \Big(p_1 \cdot p_-\Big)\,\Big(p_2 \cdot
p_+\Big)+ \Big(p_1 \cdot p_+\Big)\,\Big(p_2 \cdot p_-\Big)+2\, \Big(p_1 \cdot
p_-\Big)\,\Big(p_1 \cdot p_+\Big)\right)\nonumber\\%
&\hspace{15mm}+4\, m_{\mu\mu}^2\,\left(\Big(p_1 \cdot p_-\Big)\,\Big(p_2 \cdot
p_+\Big) -\Big(p_1 \cdot p_+\Big)\,\Big(p_2 \cdot p_-\Big)+2\, \Big(p_1 \cdot
p_-\Big)\,\Big(p_2 \cdot p_-\Big)+2\, \Big(p_1 \cdot
p_-\Big)^2\right)\nonumber\\%
&\hspace{15mm}+2\,m_{\mu\mu}^2\,m_\nu^2\,\left(2\, \Big(p_2 \cdot
p_+\Big)+2\,\Big(p_2 \cdot p_-\Big)+2\, \Big(p_1 \cdot p_+\Big)+2\,\Big(p_1
\cdot p_-\Big)-m_B^2\right)\nonumber\\%
&\hspace{15mm}-4\, m_{\nu\nu}^2\,m_\mu^2\,\left(\Big(p_2 \cdot p_+\Big)+
\Big(p_2 \cdot p_-\Big)+3\,\Big(p_1 \cdot p_+\Big)+3\, \Big(p_1 \cdot
p_-\Big)\right)\nonumber\\%
&\hspace{15mm}+8\,m_\nu^2\,\left( \Big(p_2 \cdot p_-\Big)+\Big(p_1 \cdot
p_-\Big)\right)\,\left( \Big(p_2 \cdot p_+\Big)+\Big(p_1 \cdot
p_+\Big)\right)\nonumber\\%
&\hspace{15mm}+2\,m_{\nu\nu} ^2\,m_{\mu\mu}^2\,\left(\Big(p_2 \cdot p_-\Big)+
\Big(p_1 \cdot p_+\Big)+2\,\Big(p_1 \cdot p_-\Big)\right)\nonumber\\%
&\hspace{15mm}-2\, m_{\mu\mu}^2\,m_\mu^2\,\left(4\,\Big(p_1 \cdot p_+\Big)+4\,
\Big(p_1 \cdot p_-\Big)-m_B^2\right)+4\,m_{\mu\mu}^4\, \Big(p_1 \cdot
p_-\Big)+4\,m_\mu^2\,m_\nu^2\,m_B^2-4\,m_\mu^4\,m_B^2 \nonumber\\%
&\hspace{15mm}+2\,m_{\mu\mu}^4\,m_\nu^2-4\,m_{\nu\nu}^2\,m_{\mu\mu}^2\,m_\mu^2-2\, m_{\nu\nu}^4\,m_\mu^2+m_{\nu\nu}^2\,m_{\mu\mu}^4+m_{\nu\nu}^4\, m_{\mu\mu}^2\Bigg),\\%
R^M_{c'm} &= -4\,\Bigg(8\,\bigg(\Big(p_1 \cdot p_-\Big)\, \Big(p_1
\cdot p_+\Big)\,\Big(p_2 \cdot p_+\Big)+ \Big(p_1 \cdot p_-\Big)^2\,\Big(p_2
\cdot p_+\Big)+ \Big(p_1 \cdot p_+\Big)^2\,\Big(p_2 \cdot p_-\Big)\nonumber\\%
&\hspace{2cm}- \Big(p_1 \cdot p_-\Big)\,\Big(p_1 \cdot p_+\Big)\, \Big(p_2 \cdot
p_-\Big)-2\,\Big(p_1 \cdot p_-\Big)^2\, \Big(p_1 \cdot
p_+\Big)\bigg)\nonumber\\%
&\hspace{15mm}+8\,m_\mu^2\,\bigg( \Big(p_1 \cdot p_+\Big)\,\Big(p_2 \cdot
p_+\Big)-2\, \Big(p_1 \cdot p_+\Big)\,\Big(p_2 \cdot p_-\Big)\nonumber\\%
&\hspace{3cm}- \Big(p_1 \cdot p_-\Big)\,\Big(p_2 \cdot p_-\Big)+ \Big(p_1 \cdot
p_+\Big)^2-2\,\Big(p_1 \cdot p_-\Big)\, \Big(p_1 \cdot p_+\Big)-\Big(p_1 \cdot
p_-\Big)^2\bigg)\nonumber\\%
&\hspace{15mm}+8\,m_\nu^2 \,\left(\Big(p_1 \cdot p_-\Big)\,\Big(p_2 \cdot
p_+\Big)+ \Big(p_1 \cdot p_+\Big)\,\Big(p_2 \cdot p_-\Big)+2\, \Big(p_1 \cdot
p_-\Big)\,\Big(p_1 \cdot p_+\Big)\right)\nonumber\\%
&\hspace{15mm}+4\, m_{\mu\mu}^2\,\left(\Big(p_1 \cdot p_-\Big)\,\Big(p_2 \cdot
p_+\Big) +\Big(p_1 \cdot p_+\Big)\,\Big(p_2 \cdot p_-\Big)\right)\nonumber\\%
&\hspace{15mm}-8\,m_\mu^4 \,\left(\Big(p_2 \cdot p_+\Big)+\Big(p_2 \cdot
p_-\Big)+3\, \Big(p_1 \cdot p_+\Big)+3\,\Big(p_1 \cdot
p_-\Big)\right)\nonumber\\%
&\hspace{15mm}+8\,m_\mu^2 \,m_\nu^2\,\left(\Big(p_2 \cdot p_+\Big)-\Big(p_2
\cdot p_-\Big)+ \Big(p_1 \cdot p_+\Big)-\Big(p_1 \cdot
p_-\Big)\right)\nonumber\\%
&\hspace{15mm}+4\,m_{\mu\mu} ^2\,m_\mu^2\,\left(\Big(p_2 \cdot p_+\Big)+2\,
\Big(p_1 \cdot p_+\Big)+\Big(p_1 \cdot p_-\Big)\right)\nonumber\\%
&\hspace{15mm}+4\,m_{\mu\mu} ^2\,m_\nu^2\,\left(\Big(p_2 \cdot p_-\Big)+\Big(p_1
\cdot p_+\Big)+2 \,\Big(p_1 \cdot p_-\Big)\right)\nonumber\\%
&\hspace{15mm}-2\,m_{\nu\nu}^2\,m_{\mu\mu}^2\, \left(\Big(p_1 \cdot
p_+\Big)+\Big(p_1 \cdot p_-\Big)\right)-8\, m_{\nu\nu}^2\,\Big(p_1 \cdot
p_-\Big)\,\Big(p_1 \cdot p_+\Big)\nonumber\\%
&\hspace{15mm}+2\,
m_{\mu\mu}^4\,m_\nu^2-8\,m_{\mu\mu}^2\,m_\mu^4-8\,m_{\nu\nu}^2\,
m_\mu^4+2\,m_{\mu\mu}^4\,m_\mu^2+4\,m_{\nu\nu}^2\,m_{\mu\mu}^2\,
m_\mu^2-m_{\nu\nu}^2\,m_{\mu\mu}^4\Bigg),\\%
R^M_{ap'} &= -64\,\left(\Big(p_1 \cdot p_+\Big)+m_\mu^2\right)\,\left(
\Big(p_2 \cdot p_-\Big)+m_\mu^2\right),\\%
R^M_{bp'} &= 4\,\Bigg(4\,\bigg(2\,\Big(p_1 \cdot p_-\Big)\, \Big(p_2
\cdot p_+\Big)^2+2\,\Big(p_1 \cdot p_+\Big)\, \Big(p_2 \cdot p_-\Big)\,\Big(p_2
\cdot p_+\Big)+4\, \Big(p_1 \cdot p_-\Big)\,\Big(p_2 \cdot p_-\Big)\, \Big(p_2
\cdot p_+\Big)\nonumber\\%
&\hspace{2cm}+4\,\Big(p_1 \cdot p_-\Big)\, \Big(p_1 \cdot p_+\Big)\,\Big(p_2
\cdot p_+\Big)+2\, \Big(p_1 \cdot p_-\Big)^2\,\Big(p_2 \cdot p_+\Big)-m_B^2\,
\Big(p_1 \cdot p_-\Big)\,\Big(p_2 \cdot p_+\Big)\nonumber\\%
&\hspace{2cm}+2\, \Big(p_1 \cdot p_-\Big)\,\Big(p_1 \cdot p_+\Big)\, \Big(p_2
\cdot p_-\Big)-m_B^2\,\Big(p_1 \cdot p_+\Big)\, \Big(p_2 \cdot
p_-\Big)\bigg)\nonumber\\%
&\hspace{15mm}+4\,m_\mu^2\,\bigg(2\, \Big(p_1 \cdot p_+\Big)\,\Big(p_2 \cdot
p_+\Big)+2\, \Big(p_1 \cdot p_-\Big)\,\Big(p_2 \cdot p_+\Big)\nonumber\\%
&\hspace{3cm}+2\, \Big(p_1 \cdot p_+\Big)\,\Big(p_2 \cdot p_-\Big)+2\, \Big(p_1
\cdot p_-\Big)\,\Big(p_2 \cdot p_-\Big)-m_B^2\, \Big(p_2 \cdot
p_-\Big)-m_B^2\,\Big(p_1 \cdot p_+\Big)\bigg)\nonumber\\%
&\hspace{15mm}+4\, \left(m_{\mu\mu}^2+m_{\nu\nu}^2\right)\,\left(\Big(p_1 \cdot
p_-\Big)\,\Big(p_2 \cdot p_+\Big) -\Big(p_1 \cdot p_+\Big)\,\Big(p_2 \cdot
p_-\Big)\right)\nonumber\\%
&\hspace{15mm}-m_{\nu\nu}^ 2\,m_{\mu\mu}^2\,\left(2\,\Big(p_2 \cdot p_+\Big)+4\,
\Big(p_2 \cdot p_-\Big)+4\,\Big(p_1 \cdot p_+\Big)+2\, \Big(p_1 \cdot
p_-\Big)-m_B^2\right)\nonumber\\%
&\hspace{15mm}+2\,\left(m_{\mu\mu}^2\,m_\nu^2+m_{\nu\nu}^2\,m_\mu^2\right)\,
\left(2\,\Big(p_2 \cdot p_+\Big)+2\,\Big(p_2 \cdot p_-\Big)+2\, \Big(p_1 \cdot
p_+\Big)+2\,\Big(p_1 \cdot p_-\Big)-m_B^2\right)\nonumber\\%
&\hspace{15mm}+8\,m_\nu^2\,\left( \Big(p_2 \cdot p_-\Big)+\Big(p_1 \cdot
p_-\Big)\right)\,\left( \Big(p_2 \cdot p_+\Big)+\Big(p_1 \cdot
p_+\Big)\right)\nonumber\\%
&\hspace{15mm}+4\,m_\mu^2\,
m_\nu^2\,m_B^2-4\,m_\mu^4\,m_B^2+2\,m_{\mu\mu}^4\,m_\nu^2+2\,
m_{\nu\nu}^4\,m_\mu^2-m_{\nu\nu}^2\,m_{\mu\mu}^4-m_{\nu\nu}^4\,
m_{\mu\mu}^2\Bigg),\\%
R^M_{cp'} &= 8\,\Bigg(4\,m_\nu^2\,\left(\Big(p_2 \cdot p_-\Big)\,
\Big(p_2 \cdot p_+\Big)-\Big(p_1 \cdot p_-\Big)\, \Big(p_1 \cdot
p_+\Big)\right)\nonumber\\%
&\hspace{15mm}-4\,m_\mu^2\,\left( \Big(p_1 \cdot p_+\Big)\,\Big(p_2 \cdot
p_+\Big)- \Big(p_2 \cdot p_-\Big)^2-\Big(p_1 \cdot p_-\Big)\, \Big(p_2 \cdot
p_-\Big)+\Big(p_1 \cdot p_+\Big)^2\right)\nonumber\\%
&\hspace{15mm}+4\,\left( \Big(p_2 \cdot p_-\Big)-\Big(p_1 \cdot
p_+\Big)\right)\,\left( \Big(p_1 \cdot p_-\Big)\,\Big(p_2 \cdot p_+\Big)+
\Big(p_1 \cdot p_+\Big)\,\Big(p_2 \cdot p_-\Big)\right)\nonumber\\%
&\hspace{15mm}-4\,m_\mu^4\, \left(\Big(p_2 \cdot p_+\Big)+\Big(p_2 \cdot
p_-\Big)- \Big(p_1 \cdot p_+\Big)-\Big(p_1 \cdot p_-\Big)\right)\nonumber\\%
&\hspace{15mm}+4\,m_\mu^2\, m_\nu^2\,\left(\Big(p_2 \cdot p_+\Big)-\Big(p_2
\cdot p_-\Big)+ \Big(p_1 \cdot p_+\Big)-\Big(p_1 \cdot
p_-\Big)\right)\nonumber\\%
&\hspace{15mm}+\left(2\,m_{\mu\mu}
^2\,\left(m_\mu^2+m_\nu^2\right)+4m_{\nu\nu}^2\,m_\mu^2-m_{\nu\nu}^2\,
m_{\mu\mu}^2\right)\,\left(\Big(p_2 \cdot p_-\Big)-\Big(p_1 \cdot p_+\Big) \right)\Bigg),\\%
R^M_{a'p'} &= 16\,\left(2\,m_\mu^2\,m_\nu^2\,\left(\Big(p_2 \cdot p_+\Big)+
\Big(p_1 \cdot p_-\Big)\right)-2\,m_\mu^2\,m_\nu^4+m_{\mu\mu}^2\,
m_\nu^4-2\,m_\mu^4\,m_\nu^2+m_{\nu\nu}^2\,m_\mu^4\right),\\%
R^M_{b'p'} &= 8\,\left(2\,m_\nu^2\,\left(\Big(p_2 \cdot p_+\Big)+ \Big(p_1 \cdot
p_+\Big)\right)+2\,m_\mu^2\,\left( \Big(p_1 \cdot p_+\Big)+\Big(p_1 \cdot
p_-\Big)\right)+m_{\mu\mu}^2 \,m_\nu^2+m_{\nu\nu}^2\,m_\mu^2\right)\nonumber\\%
&\quad \times \left(2\,m_\mu^2\,\left( \Big(p_2 \cdot p_+\Big)+\Big(p_2 \cdot
p_-\Big)\right)+2\,m_\nu^2\, \left(\Big(p_2 \cdot p_-\Big)+\Big(p_1 \cdot
p_-\Big)\right)+ m_{\mu\mu}^2\,m_\nu^2+m_{\nu\nu}^2\,m_\mu^2\right),\\%
R^M_{p'm'} &= 16\,\left(m_\nu^2\,\Big(p_2 \cdot p_-\Big)+m_\mu^2\, \Big(p_2
\cdot p_-\Big)+2\,m_\mu^2\,m_\nu^2\right)\nonumber\\%
&\quad \times \left(2\,m_\nu^2 \,\Big(p_2 \cdot p_+\Big)+2\,m_\mu^2\,\Big(p_1
\cdot p_-\Big)-4\,
m_\mu^2\,m_\nu^2+m_{\mu\mu}^2\,m_\nu^2+m_{\nu\nu}^2\,m_\mu^2\right),\\%
R^M_{am'} &= -32\,\left(\Big(p_2 \cdot p_-\Big)+m_\mu^2\right)\,\left(2
\,\Big(p_2 \cdot p_+\Big)-2\,m_\mu^2+m_{\mu\mu}^2\right),\\%
R^M_{bm'} &= 4\,\Bigg(8\,\bigg(2\,\Big(p_2 \cdot p_-\Big)\, \Big(p_2
\cdot p_+\Big)^2+\Big(p_1 \cdot p_-\Big)\, \Big(p_2 \cdot p_+\Big)^2\nonumber\\%
&\hspace{2cm}+2\,\Big(p_2 \cdot p_-\Big)^2\, \Big(p_2 \cdot p_+\Big)+\Big(p_1
\cdot p_+\Big)\, \Big(p_2 \cdot p_-\Big)\,\Big(p_2 \cdot p_+\Big)+2\, \Big(p_1
\cdot p_-\Big)\,\Big(p_2 \cdot p_-\Big)\, \Big(p_2 \cdot p_+\Big)\nonumber\\%
&\hspace{2cm}-m_B^2\,\Big(p_2 \cdot p_-\Big)\, \Big(p_2 \cdot p_+\Big)+\Big(p_1
\cdot p_-\Big)^2\, \Big(p_2 \cdot p_+\Big)-\Big(p_1 \cdot p_-\Big)\, \Big(p_1
\cdot p_+\Big)\,\Big(p_2 \cdot p_-\Big)\bigg)\nonumber\\%
&\hspace{15mm}+8\,m_\mu^2\, \bigg(2\,\Big(p_2 \cdot p_+\Big)^2+4\,\Big(p_2 \cdot
p_-\Big)\, \Big(p_2 \cdot p_+\Big)+\Big(p_1 \cdot p_+\Big)\, \Big(p_2 \cdot
p_+\Big)+\Big(p_1 \cdot p_-\Big)\, \Big(p_2 \cdot p_+\Big)\nonumber\\%
&\hspace{3cm}-m_B^2\,\Big(p_2 \cdot p_+\Big)+2\, \Big(p_2 \cdot
p_-\Big)^2+\Big(p_1 \cdot p_+\Big)\, \Big(p_2 \cdot p_-\Big)+\Big(p_1 \cdot
p_-\Big)\, \Big(p_2 \cdot p_-\Big)\bigg)\nonumber\\%
&\hspace{15mm}+4\,m_{\nu\nu}^2\,\left(2\, \Big(p_2 \cdot p_-\Big)\,\Big(p_2
\cdot p_+\Big)+ \Big(p_1 \cdot p_-\Big)\,\Big(p_2 \cdot p_+\Big)+ \Big(p_1 \cdot
p_+\Big)\,\Big(p_2 \cdot p_-\Big)\right)\nonumber\\%
&\hspace{15mm}+4\, m_{\mu\mu}^2\,\left(\Big(p_1 \cdot p_-\Big)\,\Big(p_2 \cdot
p_+\Big) -2\,\Big(p_2 \cdot p_-\Big)^2-\Big(p_1 \cdot p_+\Big)\, \Big(p_2 \cdot
p_-\Big)-2\,\Big(p_1 \cdot p_-\Big)\, \Big(p_2 \cdot p_-\Big)\right)\nonumber\\%
&\hspace{15mm}+2\,m_{\mu\mu}^2\,m_\mu^2\,\left(4\, \Big(p_2 \cdot
p_+\Big)+4\,\Big(p_2 \cdot p_-\Big)-m_B^2\right)\nonumber\\%
&\hspace{15mm}+4\, m_{\nu\nu}^2\,m_\mu^2\,\left(3\,\Big(p_2 \cdot p_+\Big)+3\,
\Big(p_2 \cdot p_-\Big)+\Big(p_1 \cdot p_+\Big)+ \Big(p_1 \cdot
p_-\Big)\right)\nonumber\\%
&\hspace{15mm}-2\,m_{\mu\mu}^2\,m_\nu^2\,\left(2\, \Big(p_2 \cdot
p_+\Big)+2\,\Big(p_2 \cdot p_-\Big)+2\, \Big(p_1 \cdot p_+\Big)+2\,\Big(p_1
\cdot p_-\Big)-m_B^2\right)\nonumber\\%
&\hspace{15mm}-2\, m_{\nu\nu}^2\,m_{\mu\mu}^2\,\left(\Big(p_2 \cdot p_+\Big)+2\,
\Big(p_2 \cdot p_-\Big)+\Big(p_1 \cdot p_-\Big)\right)\nonumber\\%
&\hspace{15mm}-8\,m_\nu^2\, \left(\Big(p_2 \cdot p_-\Big)+\Big(p_1 \cdot
p_-\Big)\right)\,\left( \Big(p_2 \cdot p_+\Big)+\Big(p_1 \cdot
p_+\Big)\right)-4\,m_{\mu\mu} ^4\,\Big(p_2 \cdot p_-\Big)\nonumber\\%
&\hspace{15mm}-4\,m_\mu^2\,m_\nu^2\,m_B^2+4\,m_\mu^4\,
m_B^2-2\,m_{\mu\mu}^4\,m_\nu^2+4\,m_{\nu\nu}^2\,m_{\mu\mu}^2\,m_\mu^
2+2\,m_{\nu\nu}^4\,m_\mu^2-m_{\nu\nu}^2\,m_{\mu\mu}^4-m_{\nu\nu}^4\,
m_{\mu\mu}^2\Bigg),\\%
R^M_{cm'} &= -4\,\Bigg(8\,\bigg(\Big(p_1 \cdot p_-\Big)\, \Big(p_2
\cdot p_+\Big)^2-2\,\Big(p_2 \cdot p_-\Big)^2\, \Big(p_2 \cdot p_+\Big)+\Big(p_1
\cdot p_+\Big)\, \Big(p_2 \cdot p_-\Big)\,\Big(p_2 \cdot p_+\Big)\nonumber\\%
&\hspace{3cm}- \Big(p_1 \cdot p_-\Big)\,\Big(p_2 \cdot p_-\Big)\, \Big(p_2 \cdot
p_+\Big)+\Big(p_1 \cdot p_+\Big)\, \Big(p_2 \cdot p_-\Big)^2\bigg)\nonumber\\%
&\hspace{15mm}+8\,m_\mu^2\,\bigg( \Big(p_2 \cdot p_+\Big)^2-2\,\Big(p_2 \cdot
p_-\Big)\, \Big(p_2 \cdot p_+\Big)+\Big(p_1 \cdot p_+\Big)\, \Big(p_2 \cdot
p_+\Big)-2\,\Big(p_1 \cdot p_-\Big)\, \Big(p_2 \cdot p_+\Big)\nonumber\\%
&\hspace{3cm}-\Big(p_2 \cdot p_-\Big)^2- \Big(p_1 \cdot p_-\Big)\,\Big(p_2 \cdot
p_-\Big)\bigg)\nonumber\\%
&\hspace{15mm}+8\,m_\nu^2\, \left(2\,\Big(p_2 \cdot p_-\Big)\,\Big(p_2 \cdot
p_+\Big)+ \Big(p_1 \cdot p_-\Big)\,\Big(p_2 \cdot p_+\Big)+ \Big(p_1 \cdot
p_+\Big)\,\Big(p_2 \cdot p_-\Big)\right)\nonumber\\%
&\hspace{15mm}+4\, m_{\mu\mu}^2\,\left(\Big(p_1 \cdot p_-\Big)\,\Big(p_2 \cdot
p_+\Big) +\Big(p_1 \cdot p_+\Big)\,\Big(p_2 \cdot p_-\Big)\right)\nonumber\\%
&\hspace{15mm}-8\,m_\mu^4 \,\left(3\,\Big(p_2 \cdot p_+\Big)+3\,\Big(p_2 \cdot
p_-\Big)+ \Big(p_1 \cdot p_+\Big)+\Big(p_1 \cdot p_-\Big)\right)\nonumber\\%
&\hspace{15mm}+4\,m_{\mu\mu} ^2\,m_\mu^2\,\left(2\,\Big(p_2 \cdot p_+\Big)+
\Big(p_2 \cdot p_-\Big)+\Big(p_1 \cdot p_+\Big)\right)\nonumber\\%
&\hspace{15mm}+4\,m_{\mu\mu} ^2\,m_\nu^2\,\left(\Big(p_2 \cdot p_+\Big)+2\,
\Big(p_2 \cdot p_-\Big)+\Big(p_1 \cdot p_-\Big)\right)\nonumber\\%
&\hspace{15mm}-2\,m_{\nu\nu} ^2\,m_{\mu\mu}^2\,\left(\Big(p_2 \cdot p_+\Big)+
\Big(p_2 \cdot p_-\Big)\right)\nonumber\\%
&\hspace{15mm}+8\,m_\mu^2\,m_\nu^2\,\left( \Big(p_2 \cdot p_+\Big)-\Big(p_2
\cdot p_-\Big)+ \Big(p_1 \cdot p_+\Big)-\Big(p_1 \cdot
p_-\Big)\right)-8\,m_{\nu\nu} ^2\,\Big(p_2 \cdot p_-\Big)\,\Big(p_2 \cdot
p_+\Big)\nonumber\\%
&\hspace{15mm}+2\,m_{\mu\mu}^4
\,m_\nu^2-8\,m_{\mu\mu}^2\,m_\mu^4-8\,m_{\nu\nu}^2\,m_\mu^4+2\,
m_{\mu\mu}^4\,m_\mu^2+4\,m_{\nu\nu}^2\,m_{\mu\mu}^2\,m_\mu^2-
m_{\nu\nu}^2\,m_{\mu\mu}^4\Bigg),\\%
R^M_{a'm'} &= 8\,\Bigg(4\,m_\nu^2\,\left(\Big(p_1 \cdot p_-\Big)\, \Big(p_2
\cdot p_+\Big)-\Big(p_1 \cdot p_+\Big)\, \Big(p_2 \cdot
p_-\Big)\right)+4\,m_\mu^2\,\left( \Big(p_1 \cdot p_-\Big)\,\Big(p_2 \cdot
p_+\Big)- \Big(p_1 \cdot p_+\Big)\,\Big(p_2 \cdot p_-\Big)\right)\nonumber\\%
&\qquad-8\,m_\mu^2\, m_\nu^2\,\left(\Big(p_2 \cdot p_+\Big)+\Big(p_1 \cdot
p_+\Big)+ \Big(p_1 \cdot p_-\Big)\right)+4\,m_{\nu\nu}^2\,m_\mu^2\, \Big(p_2
\cdot p_+\Big)+4\,m_{\mu\mu}^2\,m_\nu^2\, \Big(p_1 \cdot
p_-\Big)+4\,m_\mu^2\,m_\nu^4\nonumber\\%
&\qquad -2\,m_{\mu\mu}^2\,m_\nu^4
+4\,m_\mu^4\,m_\nu^2-2\,m_{\mu\mu}^2\,m_\mu^2\,m_\nu^2-2\,m_{\nu\nu}
^2\,m_\mu^2\,m_\nu^2+m_{\nu\nu}^2\,m_{\mu\mu}^2\,m_\nu^2-2\,
m_{\nu\nu}^2\,m_\mu^4+m_{\nu\nu}^2\,m_{\mu\mu}^2\,m_\mu^2\Bigg),\\%
R^M_{b'm'} &= 8\,\left(2\,m_\mu^2\,\left(\Big(p_2 \cdot p_+\Big)+ \Big(p_2 \cdot
p_-\Big)\right)+2\,m_\nu^2\,\left( \Big(p_2 \cdot p_-\Big)+\Big(p_1 \cdot
p_-\Big)\right)+m_{\mu\mu}^2 \,m_\nu^2+m_{\nu\nu}^2\,m_\mu^2\right)\nonumber\\%
&\quad \times \bigg(4\,\left( \Big(p_1 \cdot p_-\Big)\,\Big(p_2 \cdot p_+\Big)-
\Big(p_1 \cdot p_+\Big)\,\Big(p_2 \cdot p_-\Big)\right)-2\,m_\nu^2\,
\left(\Big(p_2 \cdot p_+\Big)+\Big(p_1 \cdot p_+\Big)\right)+2\,
m_{\nu\nu}^2\,\Big(p_2 \cdot p_+\Big)\nonumber\\%
&\hspace{15mm}-2\,m_\mu^2\,\left( \Big(p_1 \cdot p_+\Big)+\Big(p_1 \cdot
p_-\Big)\right)+2\,m_{\mu\mu} ^2\,\Big(p_1 \cdot
p_-\Big)-m_{\mu\mu}^2\,m_\nu^2-m_{\nu\nu}^2\,
m_\mu^2+m_{\nu\nu}^2\,m_{\mu\mu}^2\bigg),\\%
R^M_{c'm'} &= 4\,\left(m_\nu^2-m_\mu^2\right)\,\Bigg(4\,
m_{\mu\mu}^2\,\left(\Big(p_1 \cdot p_-\Big)\,\Big(p_2 \cdot p_+\Big) +\Big(p_1
\cdot p_+\Big)\,\Big(p_2 \cdot p_-\Big)\right)\nonumber\\%
&\hspace{25mm}+4\, m_{\nu\nu}^2\,\left(\Big(p_1 \cdot p_-\Big)\,\Big(p_2 \cdot
p_+\Big) +\Big(p_1 \cdot p_+\Big)\,\Big(p_2 \cdot p_-\Big)\right)\nonumber\\%
&\hspace{25mm}+8\,\left( \Big(p_2 \cdot p_+\Big)+\Big(p_1 \cdot
p_-\Big)\right)\,\bigg( \Big(p_1 \cdot p_-\Big)\,\Big(p_2 \cdot p_+\Big)-
\Big(p_1 \cdot p_+\Big)\,\Big(p_2 \cdot p_-\Big)\bigg)\nonumber\\%
&\hspace{25mm}-16\,m_\mu^2 \,m_\nu^2\,\left(\Big(p_2 \cdot p_+\Big)+\Big(p_2
\cdot p_-\Big)+ \Big(p_1 \cdot p_+\Big)+\Big(p_1 \cdot
p_-\Big)\right)\nonumber\\%
&\hspace{25mm}+4\,m_{\mu\mu} ^2\,m_\nu^2\,\left(\Big(p_2 \cdot p_+\Big)+\Big(p_2
\cdot p_-\Big)+ \Big(p_1 \cdot p_+\Big)+\Big(p_1 \cdot
p_-\Big)\right)\nonumber\\%
&\hspace{25mm}+4\,m_{\nu\nu}^2\,m_\mu^2\,\left(\Big(p_2 \cdot p_+\Big)+\Big(p_2
\cdot p_-\Big)+ \Big(p_1 \cdot p_+\Big)+\Big(p_1 \cdot
p_-\Big)\right)\nonumber\\%
&\hspace{25mm}-8\,m_\mu^2\, \left(\Big(p_1 \cdot p_+\Big)+\Big(p_1 \cdot
p_-\Big)\right)\,\left( \Big(p_2 \cdot p_+\Big)+\Big(p_2 \cdot
p_-\Big)\right)\nonumber\\%
&\hspace{25mm}-8\,m_\nu^2\, \left(\Big(p_2 \cdot p_-\Big)+\Big(p_1 \cdot
p_-\Big)\right)\,\left( \Big(p_2 \cdot p_+\Big)+\Big(p_1 \cdot
p_+\Big)\right)\nonumber\\%
&\hspace{25mm}-2\,m_{\nu\nu}^2\,m_{\mu\mu}^2\,\left(\Big(p_2 \cdot p_+\Big)+
\Big(p_1 \cdot p_-\Big)\right)-8\,m_{\mu\mu}^2\,m_\mu^2\,m_\nu^2-8\,
m_{\nu\nu}^2\,m_\mu^2\,m_\nu^2+2\,m_{\mu\mu}^4\,m_\nu^2\nonumber\\%
&\hspace{25mm}+4\,
m_{\nu\nu}^2\,m_{\mu\mu}^2\,m_\nu^2+4\,m_{\nu\nu}^2\,m_{\mu\mu}^2\,
m_\mu^2+2\,m_{\nu\nu}^4\,m_\mu^2-m_{\nu\nu}^2\,m_{\mu\mu}^4-
m_{\nu\nu}^4\,m_{\mu\mu}^2\Bigg),\\%
I^M_{ca'} &= 64\,\mathbb{O},\\%
I^M_{a'b'} &= -32\,\mathbb{O}\,\left(4\,\Big(p_2 \cdot p_+\Big)+2\, \Big(p_2
\cdot p_-\Big)+2\,\Big(p_1 \cdot p_+\Big)-m_B^2+m_{\mu\mu}^
2+m_{\nu\nu}^2\right),\\%
I^M_{ac'} &= 64\,\mathbb{O},\\%
I^M_{cc'} &= 32\,\mathbb{O}\,\left(\Big(p_2 \cdot p_+\Big)- \Big(p_2
\cdot p_-\Big)+\Big(p_1 \cdot p_+\Big)- \Big(p_1 \cdot p_-\Big)\right),\\%
I^M_{a'c'} &= 64\,\mathbb{O}\,\left(\Big(p_2 \cdot p_+\Big)+ \Big(p_1 \cdot
p_-\Big)\right),\\%
I^M_{b'c'} &= 32\,\mathbb{O}\,\Bigg(2\,\left(\Big(p_2 \cdot p_+\Big)^2-2\,
\Big(p_1 \cdot p_+\Big)\,\Big(p_2 \cdot p_-\Big)+ \Big(p_1 \cdot
p_-\Big)^2\right)+ \left(m_{\mu\mu}^2 + m_{\nu\nu}^2\right)\,\left(\Big(p_2
\cdot p_+\Big)+\Big(p_1 \cdot p_-\Big) \right)\nonumber\\%
&\hspace{2cm}-2\,\left(m_\mu^2+m_\nu^2\right)\,\left( \Big(p_2 \cdot
p_+\Big)+\Big(p_2 \cdot p_-\Big)+ \Big(p_1 \cdot p_+\Big)+\Big(p_1 \cdot
p_-\Big)\right)\nonumber\\%
&\hspace{2cm}-m_{\mu\mu}^2\,m_\nu^2-m_{\nu\nu}^2\,m_\nu^2-m_{\mu\mu}^2\,
m_\mu^2-m_{\nu\nu}^2\,m_\mu^2+m_{\nu\nu}^2\,m_{\mu\mu}^2\Bigg),\\%
I^M_{b'p} &= -16\,\mathbb{O}\,\left(4\,\Big(p_2 \cdot p_+\Big)+2\,
\Big(p_2 \cdot p_-\Big)+2\,\Big(p_1 \cdot p_+\Big)-m_B^2+m_{\mu\mu}^
2+m_{\nu\nu}^2\right),\\%
I^M_{c'p} &= 32\,\mathbb{O}\,\left(\Big(p_2 \cdot p_+\Big)+ \Big(p_1
\cdot p_-\Big)+2\,m_\mu^2\right),\\%
I^M_{b'm} &= -16\,\mathbb{O}\,\left(2\,\left(\Big(p_2 \cdot p_-\Big)+
\Big(p_1 \cdot p_+\Big)+2\,\Big(p_1 \cdot p_-\Big)\right)+m_{\mu\mu}
^2+m_{\nu\nu}^2\right),\\%
I^M_{c'm} &= 16\,\mathbb{O}\,\left(2\,\left(\Big(p_1 \cdot p_+\Big)-
\Big(p_1 \cdot p_-\Big)\right)-4\,m_\mu^2+m_{\mu\mu}^2\right),\\%
I^M_{bp'} &= 16\,\mathbb{O}\,\left(2\,\Big(p_2 \cdot p_+\Big)+4\,
\Big(p_1 \cdot p_+\Big)+2\,\Big(p_1 \cdot p_-\Big)-m_B^2+m_{\mu\mu}^
2+m_{\nu\nu}^2\right),\\%
I^M_{cp'} &= -32\,\mathbb{O}\,\left(\Big(p_2 \cdot p_-\Big)+ \Big(p_1
\cdot p_+\Big)+2\,m_\mu^2\right),\\%
I^M_{c'p'} &= 32\,\mathbb{O}\,\left(m_\nu^2-m_\mu^2\right)^2,\\%
I^M_{bm'} &= 16\,\mathbb{O}\,\left(2\,\left(\Big(p_2 \cdot p_+\Big)+2\,
\Big(p_2 \cdot p_-\Big)+\Big(p_1 \cdot p_-\Big)\right)+m_{\mu\mu}^2+
m_{\nu\nu}^2\right),\\%
I^M_{cm'} &= -16\,\mathbb{O}\,\left(2\,\left(\Big(p_2 \cdot p_+\Big)-
\Big(p_2 \cdot p_-\Big)\right)-4\,m_\mu^2+m_{\mu\mu}^2\right),\\%
I^M_{a'm'} &=
32\,\mathbb{O}\,\left(m_\mu^2-m_\nu^2\right),\\%
I^M_{c'm'} &=
16\,\mathbb{O}\,\left(m_\mu^2-m_\nu^2\right)\,\left(2\,\left(\Big(p_2 \cdot p_+\Big)-\Big(p_1 \cdot p_-\Big)
\right)+2\,m_\nu^2-2\,m_\mu^2+m_{\mu\mu}^2-m_{\nu\nu}^2\right).\label{eq:term-last}
\end{align}

\end{widetext}

\section{Expressions for the various $\Sigma_{ij}$ and $\Delta_{ij}$ terms}\label{app:sigma-delta}

The $\Delta_{ij}$ terms appearing in Eq.~\eqref{eq:DmM-B2B} are given by
\begin{align}
\Delta_{aa} &=
-16\,\left(m_B-2\,E_\mu\right)^2\,\left(\left(m_\mu^2-E_\mu^2\right)\,\cos^2\theta-E_\mu^2\right),\\%
\Delta_{bb} &=
-4\,m_B^4\,\left(m_B-2\,E_\mu\right)^2\,\left(\left(m_\mu^2-E_\mu^2\right)\,\cos^2\theta-E_\mu^2\right),\\%
\Delta_{cc} &= -8\,m_\mu^2\,\left(m_\mu^2-E_\mu^2
\right)\,m_B^2\,\left(m_B-2\,E_\mu\right)^2\,\sin^2\theta,\\%
\Delta_{pp} &= -4
\,m_\mu^4\,\left(m_B-2\,E_\mu\right)^2\,\left(\left(m_\mu^2-E_\mu^2
\right)\,\cos^2\theta-E_\mu^2\right),\\%
\Delta_{mm} &= 4\,m_\mu^2\,\left(m_B-2\,E_\mu
\right)^2\,\Big(\left(m_\mu^2-E_\mu^2\right)
\,\left(m_B^2-m_\mu^2\right)\,\cos^2\theta \nonumber\\%
&\quad + E_\mu
\,\left(E_\mu\,m_B^2-2\,m_\mu^2\,m_B+E_\mu\,m_\mu^2\right)\Big),\\%
\Delta_{ab} &= -16\,m_B^2\,\left(m_B-2\,E_\mu\right)^2\nonumber\\*%
&\quad \times \left(\left(
m_\mu^2-E_\mu^2\right)\,\cos^2\theta-E_\mu^2 \right),\\%
\Delta_{ap} &= 16\,m_\mu^4 \,\left(m_B-2\,E_\mu\right)^2,\\%
\Delta_{bp} &= 8\,m_\mu^4\,m_B^2\,\left(m_B-2\,E_\mu\right)^2,\\%
\Delta_{am} &= 16
\,m_\mu^2\,\left(m_B-2\,E_\mu\right)^2\,\left(E_\mu\,m_B-m_\mu^2
\right),\\%
\Delta_{bm} &= 8\,m_\mu^2\,m_B^2\,\left(m_B-2\,E_\mu\right)^2
\,\left(E_\mu\,m_B-m_\mu^2\right),\\%
\Delta_{cm} &=
8\,m_\mu^2\,m_B^2\,\left(E_\mu^2-m_\mu^2\right)\,\left(m_B-2\,E_\mu\right)^2\,\sin^2\theta,\\%
\Delta_{pm} &= 8\,m_\mu^4\,\left(m_B-2\,E_\mu \right)^2\nonumber\\*%
&\quad \times \Big(\left(m_\mu^2-E_\mu^2\right) \,\cos^2\theta
+E_\mu\,\left(m_B-E_\mu\right)\Big),
\end{align}
and the $\Sigma_{ij}$ terms are given by,
\begin{align}
\Sigma_{aa} &= -32\,E_\mu\,
\sqrt{E_\mu^2-m_\mu^2}\,\left(m_B-2\,E_\mu\right)^2,\\%
\Sigma_{bb} &=
-8\,m_B^4\,E_\mu\,\sqrt{E_\mu^2- m_\mu^2}\,\left(m_B-2\,E_\mu\right)^2,\\%
\Sigma_{pp} &=
8\,E_\mu\,m_\mu^4\,\sqrt{E_\mu^2- m_\mu^2}\,\left(m_B-2\,E_\mu\right)^2,\\ \nonumber\\%
\Sigma_{mm} &=
-8\,m_\mu^4\,\sqrt{E_\mu^2-m_\mu^2}\,\left(m_B-2\,E_\mu\right) ^2\,\left(m_B-E_\mu\right),\\%
\Sigma_{ab} &= -32\,m_B^2\,E_\mu\,
\sqrt{E_\mu^2-m_\mu^2}\,\left(m_B-2\,E_\mu \right)^2,\\%
\Sigma_{am} &=
-16\,m_\mu^2\,m_B\,\sqrt{E_\mu^2-m_\mu^2
}\,\left(m_B-2\,E_\mu\right)^2,\\%
\Sigma_{bm} &=
-8\,m_\mu^2\,m_B^3\,\sqrt{E_\mu^2-m_\mu^2}\,\left(m_B-2 \,E_\mu\right)^2,\\%
\Sigma_{pm} &=
8\,m_\mu^4\,\sqrt{E_\mu^2-m_\mu^2}\,\left(m_B-2\,E_\mu\right)^3.
\end{align}

\end{document}